\documentclass[twocolumn,showpacs,preprintnumbers,amsmath,amssymb,APSl,prd,nofootinbib,superscriptaddress]{revtex4-2}
\usepackage{graphicx,color}
\usepackage{amsmath}
\usepackage{amssymb}
\usepackage{hyperref}
\usepackage[utf8]{inputenc}
\usepackage[english]{babel}
\usepackage{epsfig}
\usepackage{subfigure}
\usepackage{wasysym}
\usepackage{color,xcolor}
\usepackage{amsmath}
\usepackage{bm}
\usepackage{epsfig}
\usepackage{amsfonts}
\usepackage{dcolumn}
\usepackage{float}

\hypersetup{colorlinks=true, linkcolor=blue, citecolor=green}

\usepackage{dcolumn}
\usepackage{bm}
\usepackage{ifpdf}
\usepackage{hyperref}
\usepackage{float}
\usepackage{bm}
\usepackage{xcolor,color,graphicx,graphics}
\usepackage[OT1]{fontenc}
\usepackage{latexsym,amssymb,amsmath,amsfonts}
\usepackage{makeidx}
\usepackage{epsfig}
\usepackage{epstopdf}
\usepackage{mathrsfs}
\hypersetup{colorlinks=true, linkcolor=blue, citecolor=green}
\usepackage{enumerate}
\usepackage{xcolor}
 \usepackage{multirow}

\begin{document}

\title{Periodical orbits and waveforms with spontaneous Lorentz symmetry-breaking in Kalb-Ramond gravity}

	\author{Ednaldo L. B. Junior} \email{ednaldobarrosjr@gmail.com}
\affiliation{Faculdade de F\'{i}sica, Universidade Federal do Pará, Campus Universitário de Tucuruí, CEP: 68464-000, Tucuruí, Pará, Brazil}

     \author{José Tarciso S. S. Junior}
    \email{tarcisojunior17@gmail.com}
\affiliation{Faculdade de F\'{i}sica, Programa de P\'{o}s-Gradua\c{c}\~{a}o em F\'{i}sica, Universidade Federal do Par\'{a}, 66075-110, Bel\'{e}m, Par\'{a}, Brazill}

	\author{Francisco S. N. Lobo} \email{fslobo@ciencias.ulisboa.pt}
\affiliation{Instituto de Astrof\'{i}sica e Ci\^{e}ncias do Espa\c{c}o, Faculdade de Ci\^{e}ncias da Universidade de Lisboa, Edifício C8, Campo Grande, P-1749-016 Lisbon, Portugal}
\affiliation{Departamento de F\'{i}sica, Faculdade de Ci\^{e}ncias da Universidade de Lisboa, Edif\'{i}cio C8, Campo Grande, P-1749-016 Lisbon, Portugal}

    \author{\\Manuel E. Rodrigues} \email{esialg@gmail.com}
\affiliation{Faculdade de F\'{i}sica, Programa de P\'{o}s-Gradua\c{c}\~{a}o em F\'{i}sica, Universidade Federal do Par\'{a}, 66075-110, Bel\'{e}m, Par\'{a}, Brazill}
\affiliation{Faculdade de Ci\^{e}ncias Exatas e Tecnologia, Universidade Federal do Par\'{a}, Campus Universit\'{a}rio de Abaetetuba, 68440-000, Abaetetuba, Par\'{a}, Brazil}

 \author{Diego Rubiera-Garcia} \email{ drubiera@ucm.es}
\affiliation{Departamento de Física Téorica and IPARCOS, Universidad Complutense de Madrid, E-28040 Madrid, Spain}

     \author{Luís F. Dias da Silva} 
        \email{fc53497@alunos.fc.ul.pt}
\affiliation{Instituto de Astrof\'{i}sica e Ci\^{e}ncias do Espa\c{c}o, Faculdade de Ci\^{e}ncias da Universidade de Lisboa, Edifício C8, Campo Grande, P-1749-016 Lisbon, Portugal}

    \author{Henrique A. Vieira} \email{henriquefisica2017@gmail.com}
\affiliation{Faculdade de F\'{i}sica, Programa de P\'{o}s-Gradua\c{c}\~{a}o em F\'{i}sica, Universidade Federal do Par\'{a}, 66075-110, Bel\'{e}m, Par\'{a}, Brazill}

\begin{abstract}

In this paper, we study time-like geodesics around a spherically symmetric black hole in Kalb-Ramond (KR) gravity, characterized by the parameter $l$, which induces spontaneous Lorentz symmetry breaking. The geodesic equations and effective potential are derived to investigate the influence of $l$. We calculate the marginally bound orbits and innermost stable circular orbits, analyzing the parameter's impact. Periodic orbits are computed numerically and classified within the standard taxonomy, revealing significant effects of $l$ on their momentum and energy. Additionally, we explore an extreme mass ratio inspiral system under the adiabatic approximation to derive gravitational waveforms emitted by an object orbiting a supermassive black hole in KR gravity. These waveforms reflect the distinctive characteristics of periodic orbits and highlight the influence of $l$. With advancements in gravitational wave detection, these results offer insights into black holes influenced by Lorentz symmetry-breaking fields.

\end{abstract}

\date{\today}

\maketitle

\section{Introduction}

In 2015, the Laser Interferometer Gravitational-Wave Observatory (LIGO) Collaboration achieved a groundbreaking confirmation of Einstein's prediction regarding the emission of gravitational waves (GWs), detecting signals from the coalescence of two black holes located 1.3 billion light-years from Earth \cite{LIGO,LIGO2}. This milestone marked the beginning of gravitational wave astronomy. Subsequently, in 2019, the Event Horizon Telescope (EHT) succeeded in capturing the first direct image of the "shadow" of the supermassive object at the center of the M87 galaxy \cite{EHT1,EHT2,EHT3,EHT4,EHT5,EHT6}. This was followed in 2022 by the observation of a similar shadow at Sagittarius A* (Sgr A*), the supermassive object at the center of our Milky Way galaxy \cite{EHT7,EHT8,EHT9,EHT10,EHT11,EHT12,EHT13,EHT14,EHT15}. Both observations have been consistently interpreted as evidence of black holes, further solidifying our understanding of these enigmatic objects.
These advances in observational techniques have not only deepened our knowledge of the structure and dynamics of ultra-compact objects but also provided unique opportunities to test the behavior of gravitational fields under extreme conditions.

The case for the existence of supermassive black holes will be further reinforced by upcoming GW detections using next-generation space-based observatories such as Taiji \cite{Taiji}, TianQin \cite{TianQuin}, and the Laser Interferometer Space Antenna (LISA) \cite{LISA}. Within this framework, extreme mass ratio inspiral (EMRI) systems emerge as important sources of GWs for these advanced detectors. EMRI systems consist of a stellar-mass object, such as a neutron star or black hole, orbiting a supermassive black hole. The interaction within these systems generates low-frequency GWs over extended periods \cite{EMRI, EMRI2}, offering a unique window into the orbital dynamics and the space-time geometry surrounding the supermassive black hole \cite{EMRI3}.
Studying special orbital configurations around these black holes is therefore of paramount importance, as it enables the identification of distinct GW signatures within EMRI systems. These signatures carry detailed information about the strong-field gravitational regime, offering a powerful tool for probing the fundamental properties of black holes and testing the predictions of GR and alternative theories of gravity.

In particular, the zoom-whirl behavior exhibited by a test particle orbiting a supermassive black hole provides valuable insights into the underlying space-time geometry \cite{zw}. More specifically, periodic orbits, i.e., those that return to the same point after a finite time, serve as a notable example of such a behavior. These orbits are especially significant due to their capacity to display intricate zoom-whirl dynamics. An effective framework for studying periodic orbits was introduced in \cite{TAX}, where any generic bound orbit around a Schwarzschild or Kerr black hole can be approximated by a periodic orbit. These orbits are characterized by the triplet $(z, w, v)$: $z$, the zoom number, represents the number of complete leaves in the orbit; $w$, the whirl number, indicates the number of revolutions around the periastron before transitioning to the next zoom; and $v$, the vertex number, specifies the location of the subsequent apoastron. For each periodic orbit, the angular and orbital frequencies are associated with a rational number $q$, which encapsulates the orbit's characteristics.

The taxonomy of periodic orbits established in \cite{TAX} has been applied to study particle motion around various black hole spacetimes. Examples include investigations of Kerr black holes \cite{TAXKeer2, TAXKeer3, TAXKeer4, TAXKeer5}, the Reissner-Nordström solution \cite{TAXRN}, and quantum-corrected black holes \cite{TAXQC}. For the Schwarzschild geometry, an alternative approach focusing on energy and momentum has been proposed in \cite{SCNovo}. Additional research in this topic includes studies of orbit precession and periodic orbits of time-like particles in both vacuum \cite{P1} and charged \cite{P3} black-bounce spacetimes. Furthermore, the periodic orbits of massive particles were investigated in the context of charged black holes in the Einstein-Æther theory~\cite{AE}, revealing a similarity in the taxonomy to that of the Schwarzschild spacetime for certain choices of the Einstein-Æther parameter. Additionally, in~\cite{MOG}, the orbital motion of time-like geodesics was analyzed within a Schwarzschild-like metric that includes an extra parameter. It was shown that this parameter influences the space-time structure, effective potential, and orbital energy, with gravitational waves examined in relation to the orbital signatures.

In the study of EMRIs, various quantum and modified gravity models have been explored to analyze GW emission from periodic orbits. For instance, polymer black holes derived from loop quantum gravity (LQG) \cite{LQC} and black holes with quantum corrections \cite{base2} have been investigated for their impact on GW signatures. The geodesic motion of timelike particles around black holes in Einsteinian cubic gravity was analyzed in \cite{base5}, with results correlated to recent observational data, further enhancing our understanding of modified gravity effects on black hole dynamics.
Recent studies \cite{GWnovo1, GWnovo2} have also focused on the effective radial potential of test particles around black holes with hair, revealing significant effects of the hair parameter. As this parameter increases, key orbital features such as the marginally bound orbit (MBO) and innermost stable circular orbit (ISCO) are observed to decrease substantially. These analyses also consider GW emission in the context of these modified potentials, offering a deeper understanding of black hole properties.
In these investigations, EMRI systems are often modeled using the adiabatic approximation \cite{Adiabatic, Adiabatic2, Adiabatic3}, wherein the energy and angular momentum of the orbiting object remain nearly constant, allowing it to trace periodic orbits governed by the geodesic equations. These findings not only illuminate the dynamic properties of black holes but also provide critical insights into their existence and characteristics through the detection and analysis of gravitational waves.

The exploration of gravitational physics beyond GR has traditionally been driven by the fundamental incompatibility between GR and Quantum Mechanics. Among the most prominent attempts to address this issue are string theory \cite{corda1, corda2, corda3, corda4} and loop quantum gravity \cite{LQG1, LQG2, LQG3, LQG4}. In recent years, these frameworks have been joined by alternative approaches, including non-commutative field theory \cite{NC, NC1, NC2}, massive gravity \cite{MG1, MG2}, and $f(T)$ gravity \cite{fT1, fT2, fT3, fT4, fT5, fT6, fT7, fT8, fT9, fT10, fT11, fT12}, among others.
A notable class of gravitational theories within this broader context involves the breaking of Lorentz symmetry, which asserts the equivalence of all inertial frames. This idea gained prominence with the introduction of Horava-Lifshitz gravity \cite{Horava}, a seminal theory proposing deviations from Lorentz invariance. Lorentz symmetry breaking can occur in two principal forms. In explicit breaking, the Lagrangian itself lacks invariance under Lorentz transformations, often resulting in theoretical incompatibilities such as violations of the Bianchi identities \cite{Bianch}. Conversely, spontaneous symmetry breaking preserves the Lorentz-invariant form of the Lagrangian while allowing physical phenomena to exhibit symmetry-breaking behavior \cite{Lorentz1, Lorentz2, Lorentz3, Lorentz4, Lorentz5, Lorentz6, Lorentz7, bumblebee, bumblebee1}.  In \cite{Rajes}, possible Lorentz violations resulting in a source-dependent velocity of the GWs were considered. This made it possible to put a strict limit on such deviations using LIGO-Virgo observations for the first time.

In this work, we focus on a particularly promising theoretical framework involving the Kalb-Ramond (KR) field, a rank-two tensor field $B_{\mu\nu}$, which arises naturally in bosonic string theories when generalizing the description of electromagnetic fields from point-particles to strings \cite{KRoriginal}. Through a non-minimal coupling with the Ricci scalar, the KR field induces spontaneous Lorentz symmetry breaking, characterized by a non-zero vacuum expectation value \cite{VLorentz,  VLorentz3}. This field modifies the GR equations by introducing additional terms that affect the usual solutions. For black holes, the impact of this modification has been explored by considering observable phenomena such as gravitational lensing and perihelion precession \cite{KRteste}. Therefore, the corrections that arise in the solutions can lead to testable predictions for images of black holes observed by interferometry (such as those from the Event Horizon Telescope \cite{EHT7,EHT8,EHT9,EHT10,EHT11,EHT12,EHT13,EHT14,EHT15}), as well as new effects on the accretion disk and the X-ray emission profile, due to the presence of the extra field. \cite{Bailey, Capanelli, Jumaniyozov}

The potential role of KR fields in cosmic inflation has also been examined, specifically through models involving antisymmetric tensor fields coupled to gravity \cite{KRinfla}. Strong gravitational lensing effects in scenarios with extra dimensions incorporating the KR field were analyzed in \cite{KRlenteforte}, and subsequent research accounted for rotational effects \cite{KRKumar}. The influence of the KR parameter on the motion of massive and massless particles around black holes has been investigated in detail \cite{KRparticulas}. Additionally, the violation of gravitational parity by the 2-form KR field was studied in \cite{KRparity}, along with its potential role as a candidate for dark matter \cite{KRdarkmatter}, and its role in electromagnetic interactions \cite{KReletrico} and thermodynamic phenomena \cite{KRtermo}.

In \cite{KR}, the authors derived static and spherically symmetric solutions for Schwarzschild-like configurations, incorporating the KR field, both with and without a cosmological constant. Their formulation considers a self-interacting KR field within the framework of a non-minimally coupled Einstein-Hilbert action \cite{KRacoplado1, KRMaluf}. By varying the action with respect to the metric, they obtained field equations that closely resemble those of GR. In the absence of a cosmological constant, the resulting line element is expressed as:
\begin{eqnarray}
ds^2 &= & -\left(\frac{1}{1-l}-\frac{2M}{r}\right)dt^2+\left(\frac{1}{1-l}-\frac{2M}{r}\right)^{-1}dr^2
	\nonumber \\
&&+ r^2(d\theta^2+\sin^2\theta d\phi^2)\,,\label{KR0}
\end{eqnarray}
where $M$ is the mass parameter and $l$ is an additional dimensionless parameter that characterizes the spontaneous Lorentz symmetry-breaking.

In \cite{Guo} the authors investigated the effects on quasinormal modes and graybody factor of a black hole described by this KR metric.   We recently analyzed in \cite{nosso2} the gravitational lensing effects associated with the line element mentioned above, calculating the deflection angle and leveraging observational data from the Sgr A* images to determine key observables such as image position, luminosity, and delay time. The results indicated that these quantities could be measured within the weak-field regime, however, observations in the strong-field regime would require obervations with the next generation of interferometric instruments. We showed that both the deflection angle and the calculated observables depend explicitly on the symmetry-breaking parameter.
Additionally, in \cite{nosso}, we investigated time-like and light-like geodesics in Kalb-Ramond (KR) gravity around a black hole, focusing on constraining the Lorentz symmetry-breaking parameter. By examining the precession of the periastron of the S2 star orbiting Sgr A* and the geodesic precession around Earth, we provided a constraint on the spontaneous symmetry-breaking parameter within a specified interval, given by
\begin{equation} \label{eq:bound}
    -0.185022 \leq l \leq 0.0609 \,.
\end{equation}
We also showed that the shadow of the black hole depends strongly on the parameter $l$ and we constrain the value of this parameter by the leftover of Sgr A*.  These results therefore indicate possible implications for propagation in gravitational waves and periodic orbits.  Thus, this theory opens up new perspectives for understanding gravitational interaction and astrophysical phenomena.

The primary objective of this paper is to examine the behavior of periodic orbits and assess how the spontaneous Lorentz symmetry-breaking parameter $l$ in the KR metric, as derived in \cite{nosso}, influences these orbits, as well as the MBO and ISCO. To achieve this, we will numerically calculate the periodic orbits and consider the EMRI system within the adiabatic approximation. Specifically, we will model a black hole with a mass similar to that of Sgr A* and a secondary massive object with a mass comparable to that of S-stars. From this, we will obtain the corresponding waveforms and compare these results with those predicted by the Schwarzschild geometry. We will demonstrate that the waveforms reflect the characteristics of the periodic orbits and, as a result, their variations. With the ongoing advancements in gravitational wave detection, our findings may offer valuable insights into the gravitational structure described by the KR metric \eqref{KR0}.

This paper is organized in the following manner. In Sec. \ref{sec:two} we obtain the time-like geodesic equation and the effective potential for the KR black hole. In Sec.\,\ref{SEC3}  we analyze how the effective potential is affected by the $l$ parameter in specific MBO and ISCO orbits.  Sec.\,\ref{PO} is devoted to obtaining and classifying periodic orbits according to their taxonomy. In Sec.\,\ref{secGW} we examine the waveforms obtained from the periodic orbits for an EMRI system within the adiabatic approximation.   Finally, in Sec.\,\ref{Sec:Conclusion}, we summarize our work and conclude. Geometrized units ($G=1$ and $c=1$) where, $M$, $t$,  and $r$ have units of meters $\left[m\right]$  and the metric signature ($-,+,+,+$) are assumed throughout this work. All numerical results are obtained using \textit{Mathematica} software, using the \textsf{ Solve}, \textsf{NIntegrate} and \textsf{Plot} commands, with the standard precision for each.

\section{geodesics and effective potential: analytical results }\label{sec:two}


 The Lagrangian density that describes a test particle in GR is $\mathcal{L}=\frac{1}{2}g_{\mu\nu}\dot{x}^{\mu}\dot{x}^{\nu}$, where the overdot denotes differentiation with respect to an affine parameter with units $\left[m\right]$. Motion along geodesics must satisfy the condition $\mathcal{L}=\varepsilon$, where
 \begin{eqnarray}
 \varepsilon =
\begin{cases}
-1\,,\;\;\;\;\;\; \text{for light-like,}  \\
0\,, \;\;\;\;\;\;\;\;\; \text{for time-like,}
\end{cases}
 \end{eqnarray}
and therefore we have \cite{Inverno}
\begin{eqnarray}
g_{\mu\nu}\dot{x}^{\mu}\dot{x}^{\nu}=\varepsilon.
\label{gmn}
\end{eqnarray}

A general static and spherically symmetric metric can be written (in a certain set of coordinates) as
\begin{equation}
    ds^2 = -A(r) dt^2 +B(r)dr^2 + C(r)(d\theta^2 + \sin^2 \theta d\phi^2)\,,
    \label{eq:dsgeral}
\end{equation}
and we consider, without any loss of generality, a geodesic which is located in the equatorial plane $\theta= \pi /2$, with $\dot{\theta}=0$, such that \eqref{gmn} takes the form: 
\begin{equation}
    -A(r)\dot{t}^2+B(r)\dot{r}^2+C(r)\dot{\phi}^2= \varepsilon\,.
    \label{eq:dsmov}
\end{equation}
Furthermore, the Killing vectors associated to the time-reversal and rotation symmetry of the line element above entail the conservation of two quantities: $A(r)\dot{t}=E$ and $C(r)\dot{\phi}=L$, identified as the energy and angular momentum per unit mass, respectively. This way, if we multiply Eq.\,\eqref{eq:dsmov} by $A(r)$ and take into account the above quantities, then we obtain:
\begin{equation}
\dot{r}^2=\frac{1}{B(r)}\left[\varepsilon-\frac{L^2}{C(r)}+\frac{E^2}{A(r)}\right]\,,\label{rponto}
\end{equation}
where the effective potential $V_{\rm eff}$ is taken from \cite{Inverno} 
\begin{eqnarray}
\dot{r}^2=E^2-V_{\rm eff}\,.\label{rpontoquad}
\end{eqnarray}
Here $L$ has units $\left[m\right]$ while $E$ and $V_{\rm eff}$ are dimensionless.

\subsection{Light-like geodesics}

For light-like geodesics, the effective potential takes the form 
\begin{eqnarray}
V_{\rm eff}= A(r) \frac{L^2}{C(r)}
\end{eqnarray}  
and has a maximum in $\frac{dV_{\rm eff}}{d r}=0$ which for metric \eqref{KR0} leads to the equation for the radius of the photon sphere (i.e. circular orbits), $r_{\rm ph}=3M(1-l)$. This is an unstable orbit, meaning that small deviations will cause the photon to escape or fall into the black hole.  We studied the effects of the $l$ parameter on this type of orbit in \cite{nosso}. These results play a crucial role in understanding black hole shadows and strong-field tests of gravity.   Here the angular momentum can be arbitrary, unlike the case of a massive particle, where there is a minimum stable orbit, which we will deal with in the next subsection and in the rest of this paper. \\

\subsection{Time-like geodesic}

Let us assume a test particle that moves along a time-like geodesic with four-velocity $\dot{x}^{\mu}=dx^{\mu}/d\tau$. For the KR metric \eqref{KR0} the equation for the effective potential is
\begin{eqnarray}
V_{\rm eff}(r)=\left(\frac{1}{1-l}-\frac{2M}{r}\right)\left(1+\frac{L^2}{r^2}\right)\,.
\label{Veff}
\end{eqnarray}
This potential behaves asymptotically as $V_{\rm eff}(r)\rightarrow 1/(1-l)$, which is different from the Schwarzschild one except when  $l\rightarrow 0$.  This indicates that particles with energy $E^2>1/(1-l)$ can escape to infinity.  Therefore, bound orbits should only exist for energies  $E^2<1/(1-l)$, while $E^2=1/(1-l)$  is the maximum energy for particles in bound orbits.  The stability of the orbits is determined according to the sign of the second derivative of the effective potential, i.e. if $\frac{d^2V_{\rm eff}}{dr^2}> 0 $ then the orbit is stable while $\frac{d^2V_{\rm eff}}{dr^2}<0$ the orbit is unstable. 

For our analysis, we adopt values of $ l $ within the range (\ref{eq:bound})  obtained in \cite{nosso} through the orbital precession of the S2 star around Sgr A*. The spontaneous symmetry-breaking parameter can be assumed as follows 
\begin{eqnarray}
l=l_{\rm min}+\alpha\left(l_{\rm max}-l_{\rm min}\right)\,,\label{lcontrol}
\end{eqnarray}
where $0\leq \alpha \leq 1$,  $l_{\rm min}=-0.185022$ and $l_{\rm max}=0.060938$.

\section{MARGINALLY BOUND ORBITS AND INNERMOST STABLE CIRCULAR ORBITS: analytical results} \label{SEC3}

We are interested in analyzing the properties of periodic orbits around black holes described by the KR metric. These periodic orbits are contained in the class of bound orbits. A particle that falls freely into the black hole can be captured in an unstable circular orbit referred to as MBO  and has a maximum energy  $E$. Therefore, the MBO must meet the following conditions \cite{TAX, TAXKeer2, TAXKeer3, TAXKeer4, TAXKeer5, TAXRN, TAXQC, SCNovo, P1, P3}:
\begin{eqnarray}
V_{\rm eff}=\frac{1}{(1-l)}\,, \qquad 
 \frac{dV_{\rm eff}}{dr}=0\,.
\end{eqnarray}
We can obtain the radius $r_{\rm MBO}$ and the orbital angular momentum $L_{\rm MBO}$ for the marginally bound orbit of KR spacetime as
\begin{eqnarray}
r_{\rm MBO}=L_{\rm MBO}=4M(1-l)\,,
 \label{rMBOLMBO}
\end{eqnarray}
which for $l\rightarrow 0$ restores the right Schwarzschild behaviour \cite{TAX}.

Another particular periodic orbit is the ISCO, which has a minimum radius around the black hole. It can be obtained when the maximum and minimum points of the potential coincide \cite{TAX, TAXKeer2, TAXKeer3, TAXKeer4, TAXKeer5, TAXRN, TAXQC, SCNovo, P1, P3}, i.e.,
\begin{eqnarray}
V_{\rm eff}=E^2\,, \qquad 
 \frac{dV_{\rm eff}}{dr}=0\,,  \qquad 
 \frac{d^2V_{\rm eff}}{dr^2}=0\,,
\end{eqnarray}
from which we obtain the quantities
\begin{eqnarray}
r_{\rm ISCO}&=&6M(1-l)\,, \nonumber \\ 
 L_{\rm ISCO}&=&2 \sqrt{3} M \sqrt{l^2-2 l+1}\,, 
\nonumber 
	\\ 
E_{\rm ISCO} &=&\frac{2 \sqrt{2}}{3 \sqrt{1-l}}\,, \label{ISCOS}
\end{eqnarray}
which are the radius, the orbital angular momentum and energy of ISCO. Again, the results for Schwarzschild are recovered in limit $l\rightarrow 0 $ \cite{TAX}.

\begin{figure}[t!]
\centering
{\includegraphics[width=8.75cm]{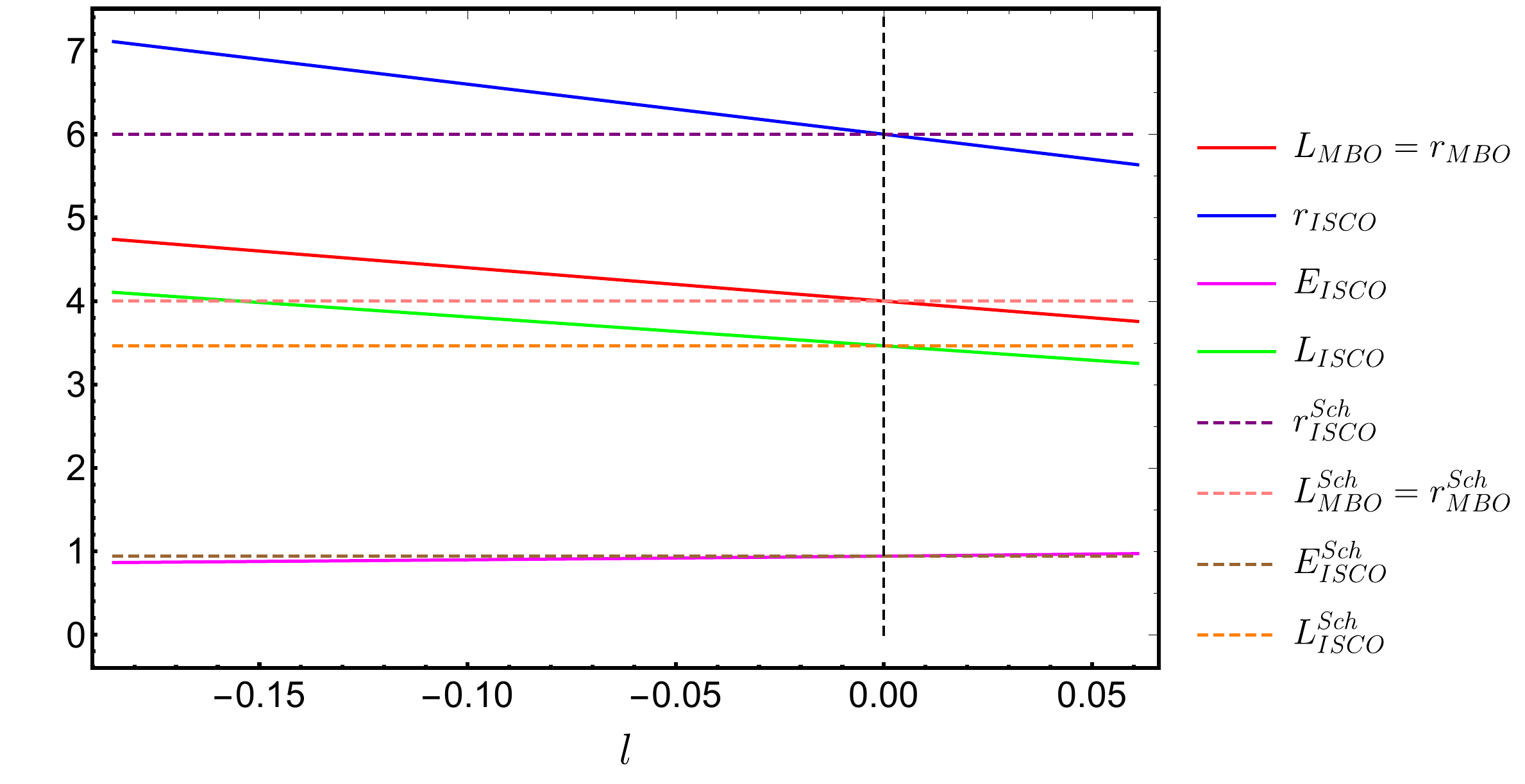} }

\caption{Graphical representation of the behaviour of the specific orbits
MBO and ISCO specific orbits as a function of the parameter $l$ and comparison with Schwarzschild marked with the dashed black line at $l=0$.}\label{fig1}
\end{figure}

In Fig.\,\ref{fig1} we depict the behaviour of these MBO and ISCO energy and angular momentum for KR gravity black holes and the Schwarzschild black hole. Note that for both Schwarzschild and KR black holes, we have $L_{\rm MBO}=r_{\rm MBO}$. When $l$ increases from its minimum value $r_{\rm MBO}$ and $L_{\rm MBO}$ decrease in the same proportion, while both $r_{\rm ISCO}$ and $L_{\rm ISCO}$ decrease proportionally but $E_{\rm ISCO}$ increases. 
Therefore, bound orbits must exist at  $  r_{\rm MBO}<r<r_{\rm ISCO}$, with $ L_{\rm ISCO}<L<L_{\rm MBO}$ and $E_{\rm ISCO}<E<E_{\rm MBO}$.

Using the above expressions we can study the general properties of bound orbits through the effective potential and the radial motion of the particles around the KR black hole.  For a given value of $l$, the angular momentum and energy of  particle in bound orbits change by
\begin{eqnarray}
L&=&L_{\rm ISCO}+\epsilon\left(L_{\rm MBO} - L_{\rm ISCO}\right)\,,\label{Lcontrol}  \\
E&=&E_{\rm ISCO}+\eta\left(E_{\rm MBO} - E_{\rm ISCO}\right)\,,\label{Econtrol}
\end{eqnarray}
where $0\leq \epsilon \leq 1 $ and $0\leq \eta \leq 1 $, respectively. 

\begin{figure*}[htb!]
\centering
\subfigure[ $l = l_{\rm min}$] 
{\label{Va}\includegraphics[width=7.75cm]{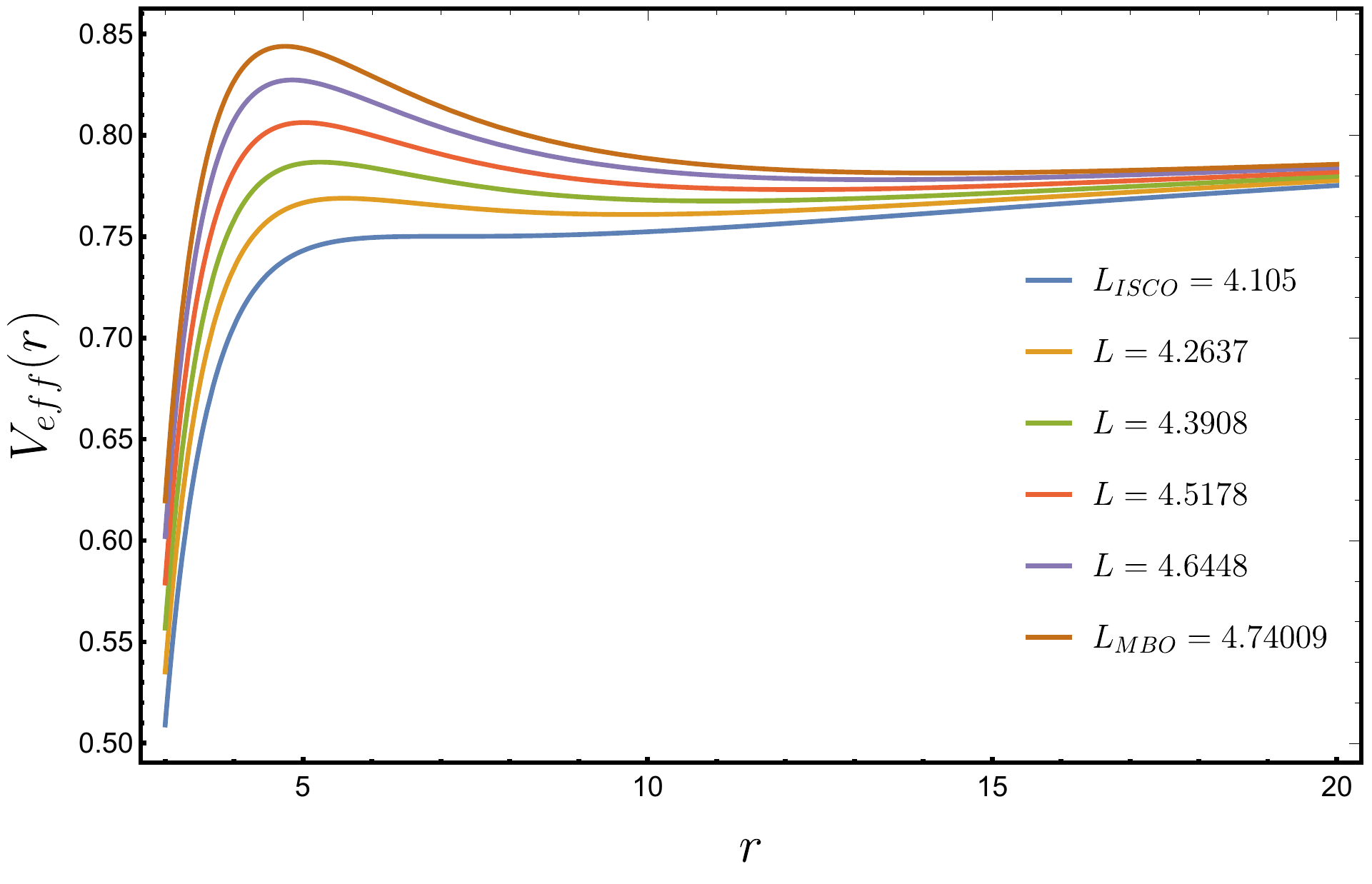} }
\hspace{0.75cm}
\subfigure[ $l = -0.01285$] 
{\label{Vb}\includegraphics[width=7.75cm]{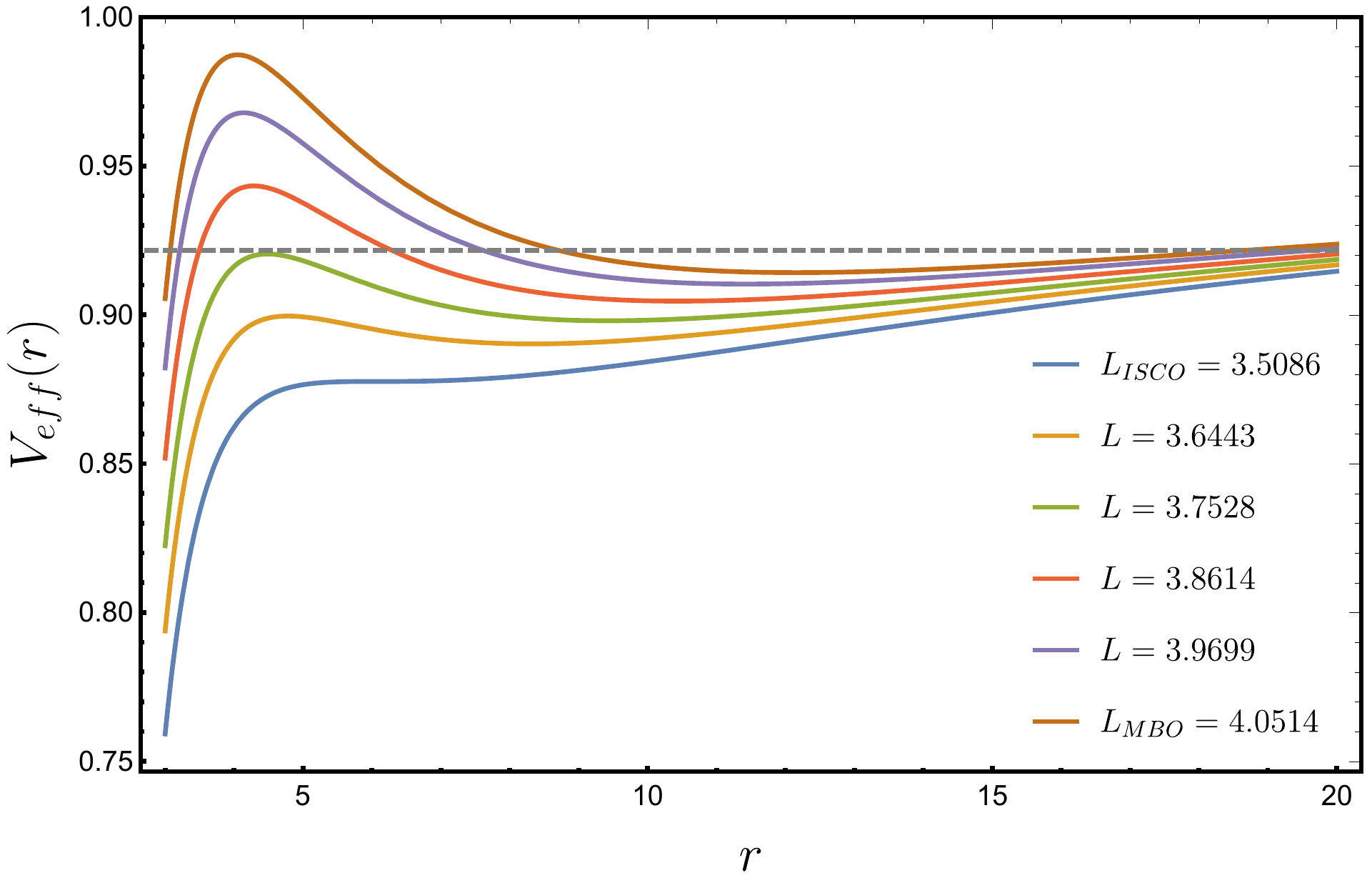} }
\subfigure[$l = 0.011746$ ] 
{\label{Vc}\includegraphics[width=7.75cm]{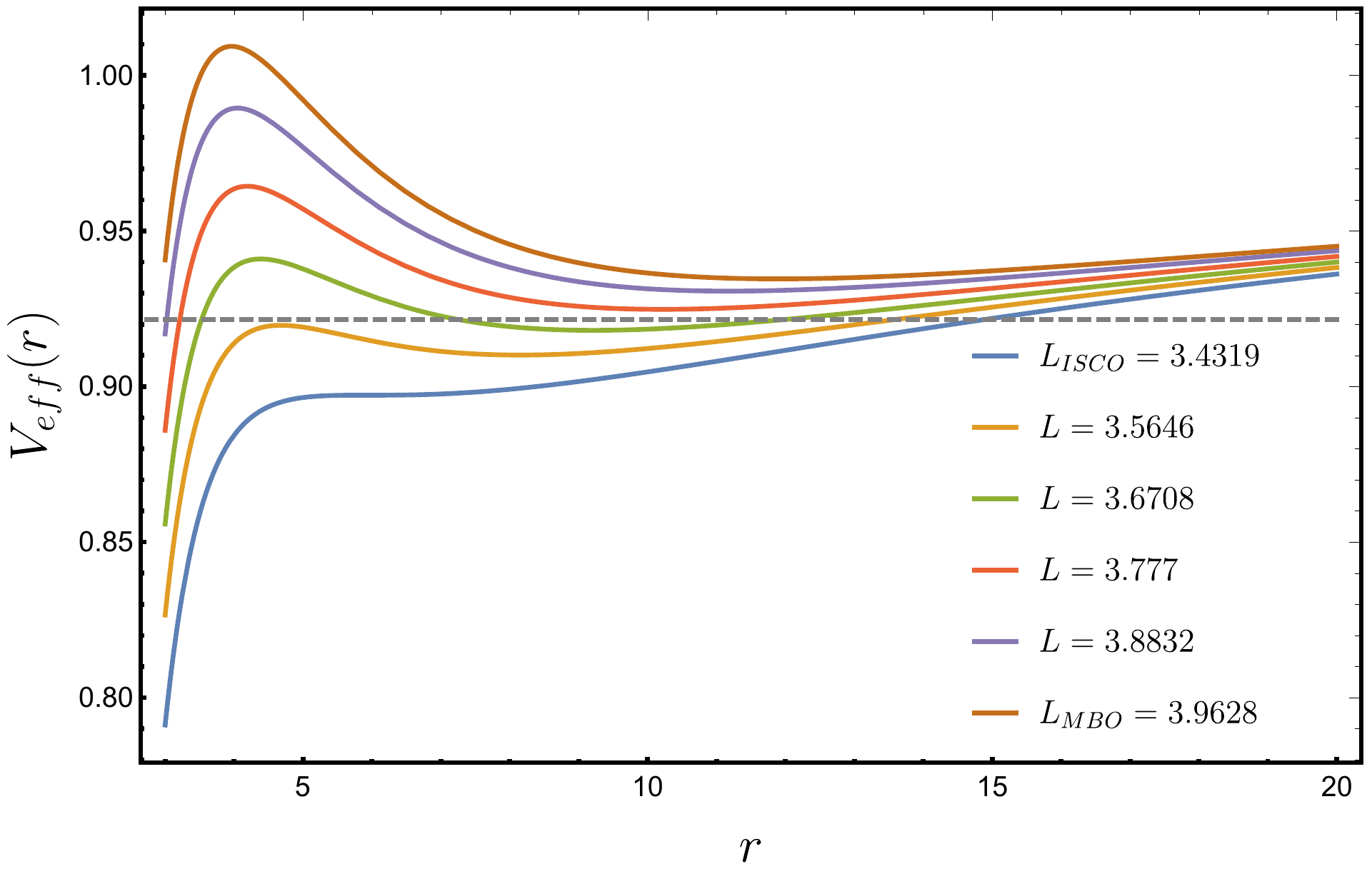}}
\hspace{0.75cm}
\subfigure[$l = l_{\rm max}$] 
{\label{Vd}\includegraphics[width=7.75cm]{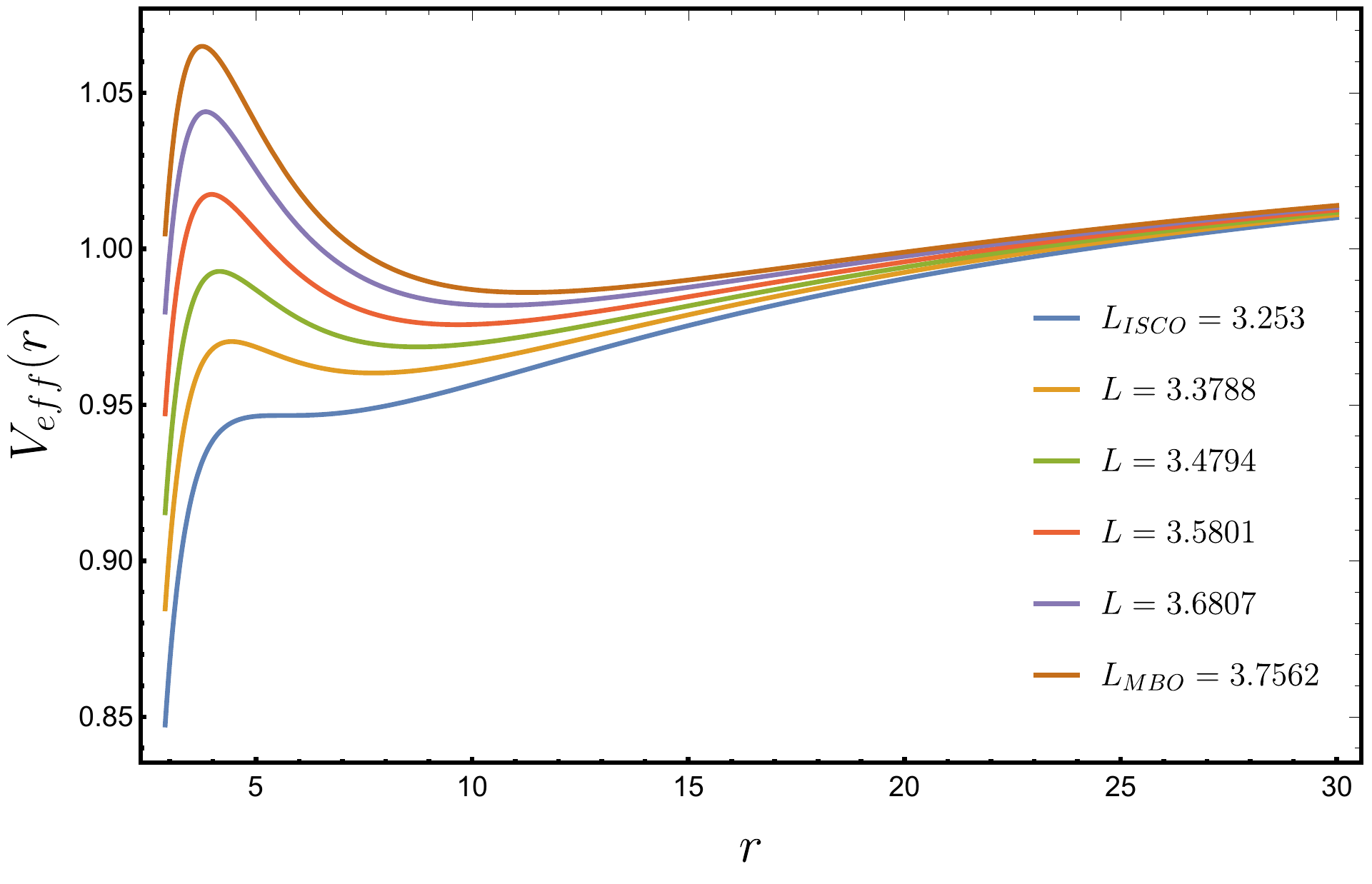}}
\caption{Representation of the effective potential $V_{eff}$ for four values of $L$ varying according to Eq.\eqref{Lcontrol}. On each plot each curve corresponds to values  of  $\epsilon=\{0,0.2,0.4,0.6,0.8,1.0 \} $.  In \ref{Vb} and \ref{Vc}, the dashed line indicates a specific reference value for particle energy  $E=0.96$.}\label{Fig3}
\end{figure*}

In the set of plots in Fig.\,\ref{Fig3}, these depict the effective potential \eqref{Veff} for four values of the KR parameter $l$ and varying the angular momentum $L$ according to the expression (\ref{Lcontrol}) taking $\epsilon=\{0,0.2,0.4,0.6,0.8,1.0 \} $ to produce six curves. This way, in Figs. \ref{Va}, \ref{Vb}, \ref{Vc} and  \ref{Vd} the uppermost curve corresponds to the effective potential for  $L_{\rm MBO}$, which has always two extrema: the maximum corresponds to the unstable circular orbit where a small perturbation will cause the particle to collapse into the black hole or escape away from it, while the minimum is the stable circular orbit where a small perturbation will cause the particle to merely oscillate around it. The shape of the potential curves shifts as $l$ increases, though. For negative $l$ (Figs. \ref{Va} and \ref{Vb}) the maximum energy is below $E=1$ (corresponding to the maximum value for the Schwarzschild metric), while for positive $l$ (Figs \ref{Vc} and \ref{Vd}) the maximum energy is above $E=1$. 
In Figs.\,\ref{Vb} and \ref{Vc} the dashed line represents a reference energy value of $E=0.96$. We note that the range of the existence of periodic orbits is modified by the value of $l$; for instance, in Fig \ref{Vb} the red curve has two turning points around the minimum potential, corresponding to the range of periodic orbits, while in Fig. \ref{Vc} the red curve has no turning points for this energy value, indicating the absence of periodic orbits at the momentum and energy values adopted. 

\begin{figure*}[htb!]
\centering
\subfigure[Values for $\alpha=0$, $\alpha=\frac{l_{\rm min}}{l_{\rm min}-l_{\rm max}}$, and $\alpha=1$.] 
{\label{ELa}\includegraphics[width=7.75cm]{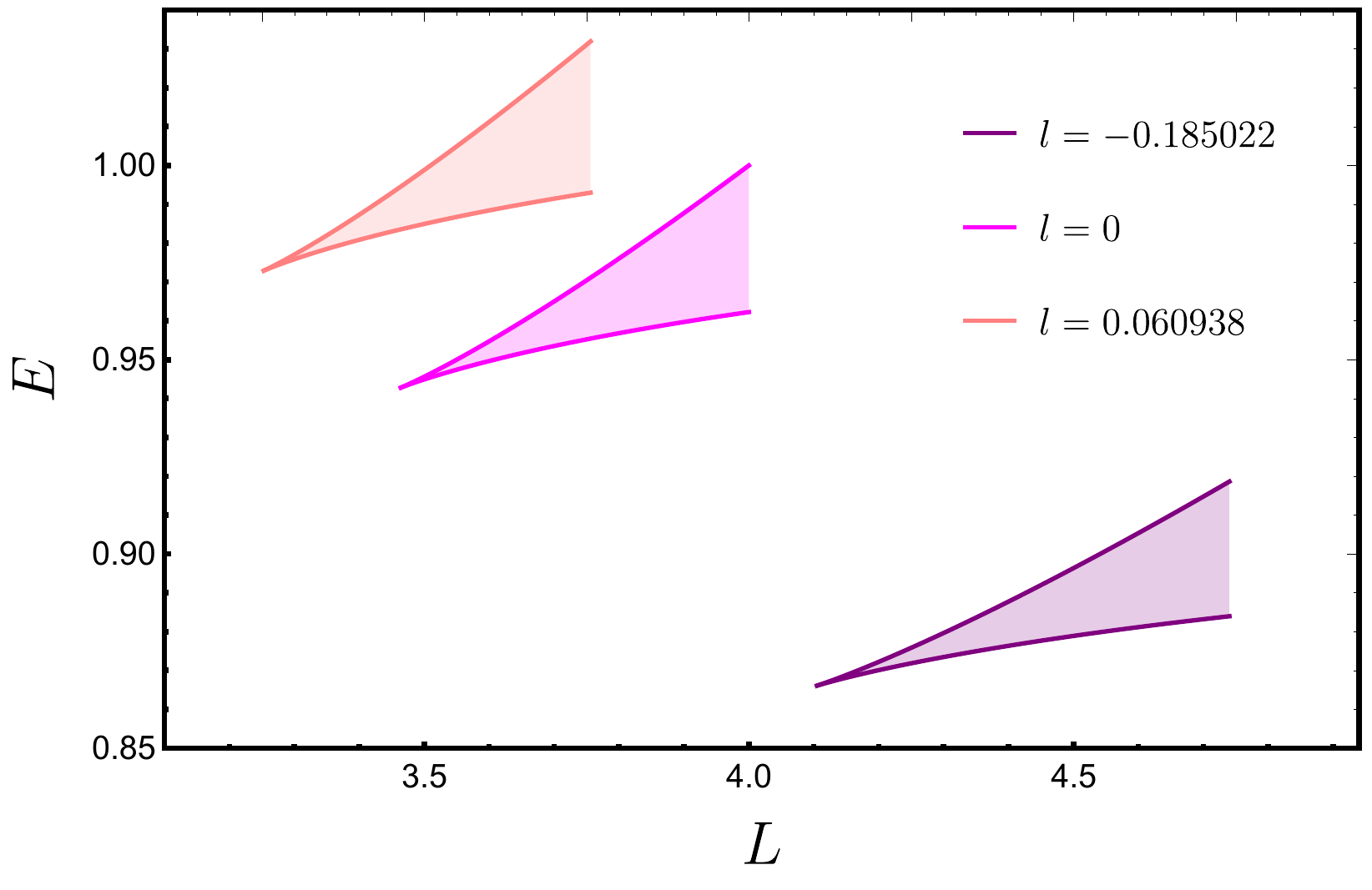} }
\hspace{0.75cm}
\subfigure[Values for $\alpha=0.7$, $\alpha=\frac{l_{\rm min}}{l_{\rm min}-l_{\rm max}}$, and $\alpha=0.8$.] 
{\label{ELb}\includegraphics[width=7.75cm]{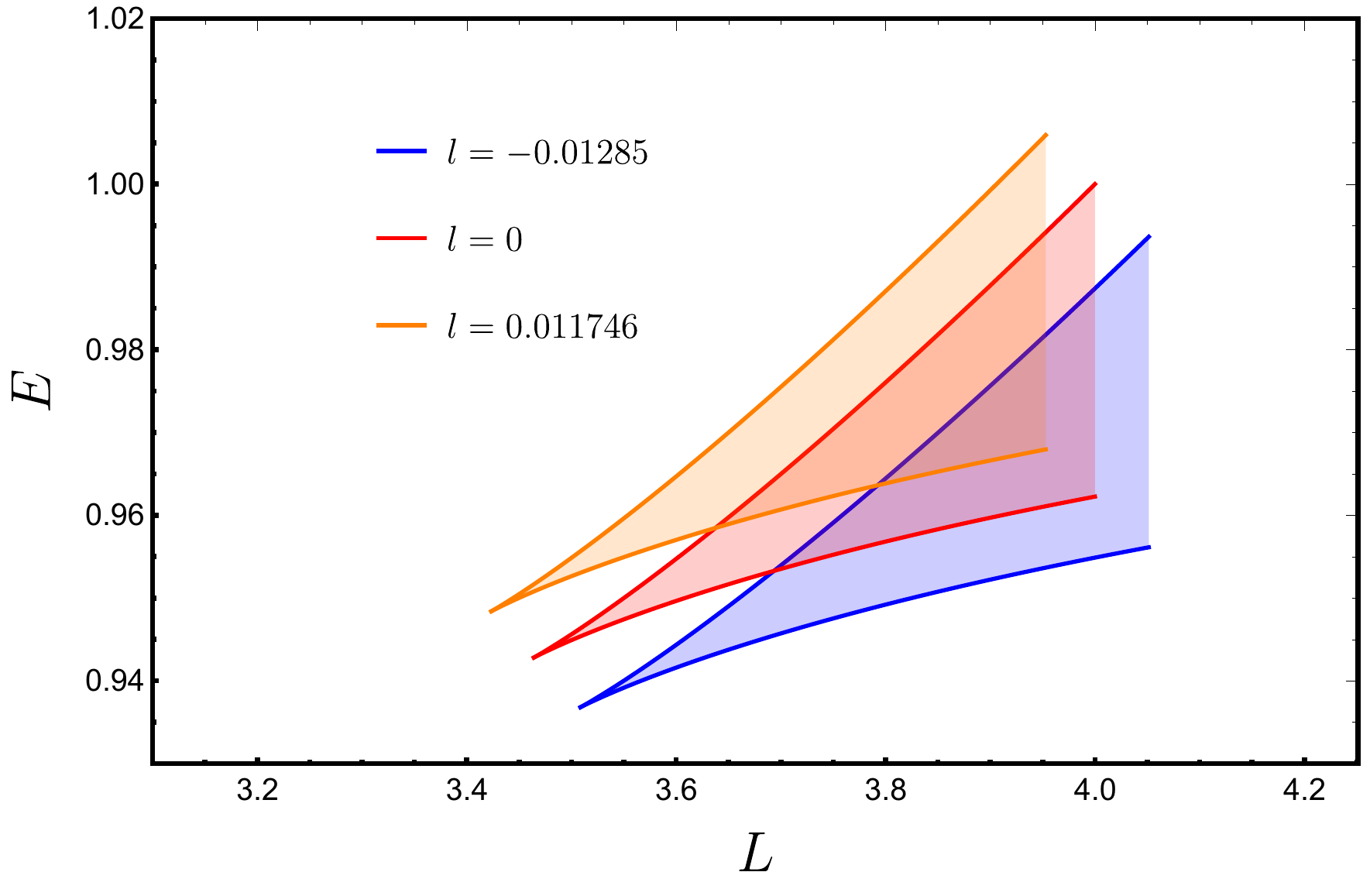} }
\caption{Representation of the allowed region for energy and momentum of the marginally bound orbits. In Fig. \ref{ELa}, the plane $(E-L)$ is depicted for $l_{\rm min}=-0.185022$ and $l_{\rm max}=0.060938$ and for Schwarzschild with $l=0$. In Fig. \ref{ELb} we use values for $l$ modified according to Eq.\eqref{lcontrol} by changing values of $\alpha$. }\label{Fig4}
\end{figure*}

The energy and momentum regions in the plane $(E-L)$ are determined by assuming $\frac{dV_{\rm eff}}{dr}=0$, whose roots are found and substituted into $V_{\rm eff}=E^2$ from which we get 
\begin{equation}
E^2_{\pm}(L)=\pm\frac{2 \left(L \left(\sqrt{L^2-12 (l-1)^2}\mp L\right)\pm4 (l-1)^2\right)^2}{ L (l-1) \left(\sqrt{ L^2-12 (l-1)^2}\mp L \right)^3}\,,\label{E+-}
\end{equation} 
Therefore, for a given $L$, the allowed energy for any bound orbit is $E^2_{-}(L)\leq E^2\leq E^2_{+}(L)$. Figure\,\ref{Fig4} shows the region $(E-L)$ allowing for bound orbits around the black hole with spontaneous symmetry-breaking compared to Schwarzschild case.  For $E=0.96$, which we took as a reference in the analysis of the effective potential above, it is not contained in the maximum and minimum $l$ values, and therefore there are no bound orbits in this case, as shown in Fig.\,\ref{ELa}. However, for this fixed energy value, in Fig.\,\ref{ELb} the corresponding value of $l$ does allow for the existence of bound orbits.
  
\begin{figure*}[htb!]
\centering
\subfigure[$l=l_{\rm min}$] 
{\label{rLa}\includegraphics[width=7.75cm]{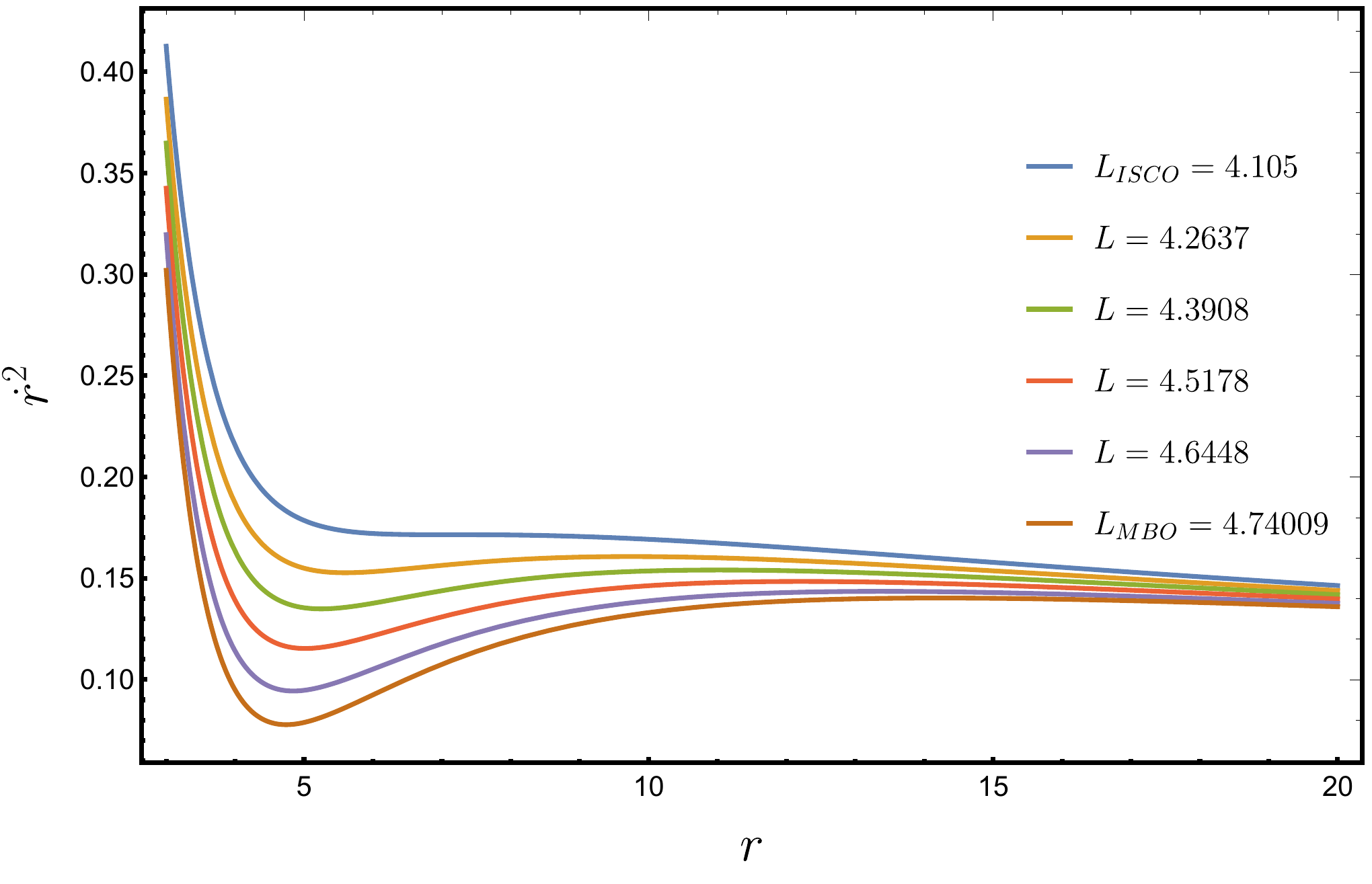} }
\hspace{0.75cm}
\subfigure[  $l=-0.01285$] 
{\label{rLb}\includegraphics[width=7.75cm]{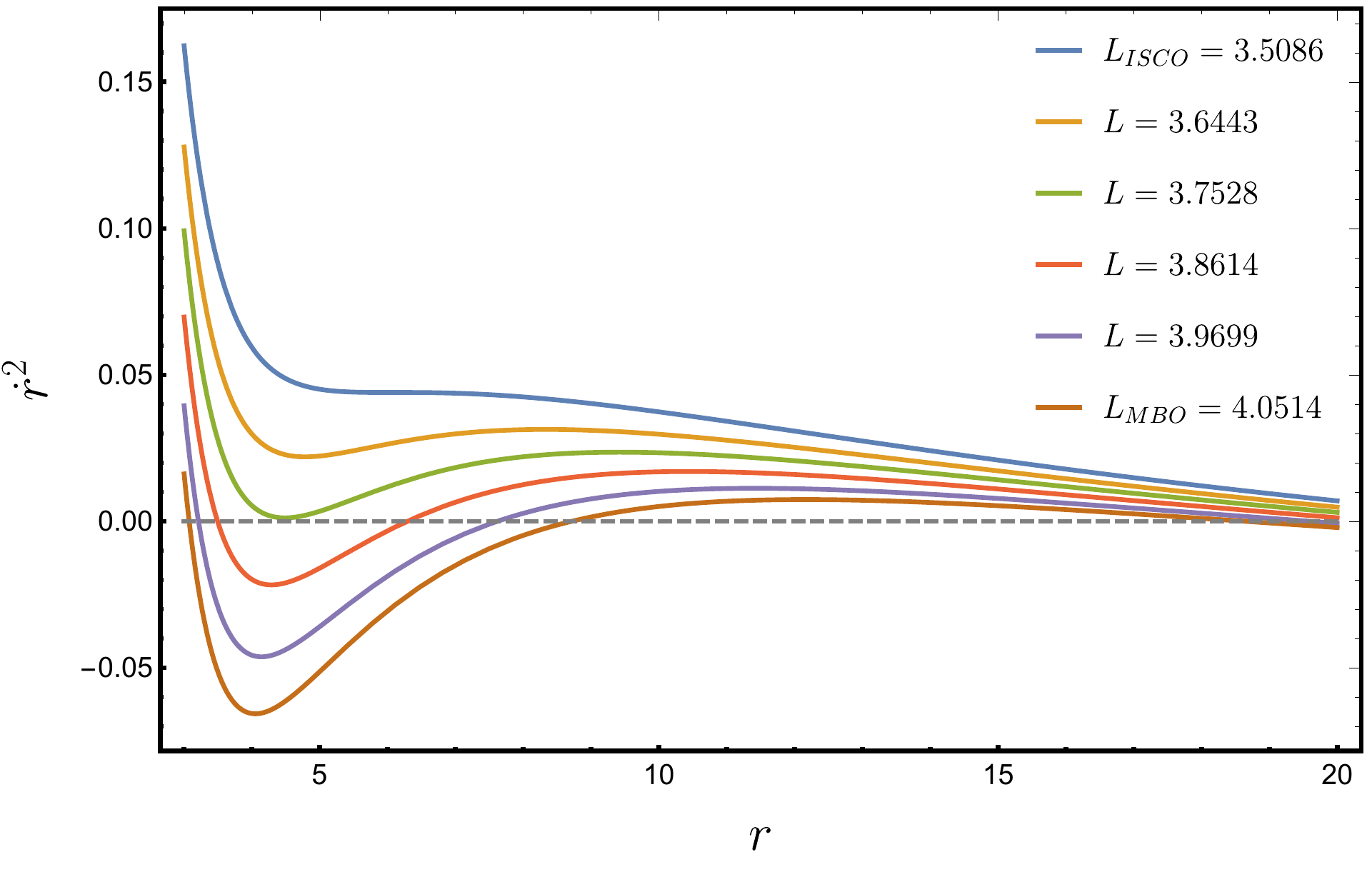} }
\subfigure[ $l=0.011746$] 
{\label{rLc}\includegraphics[width=7.75cm]{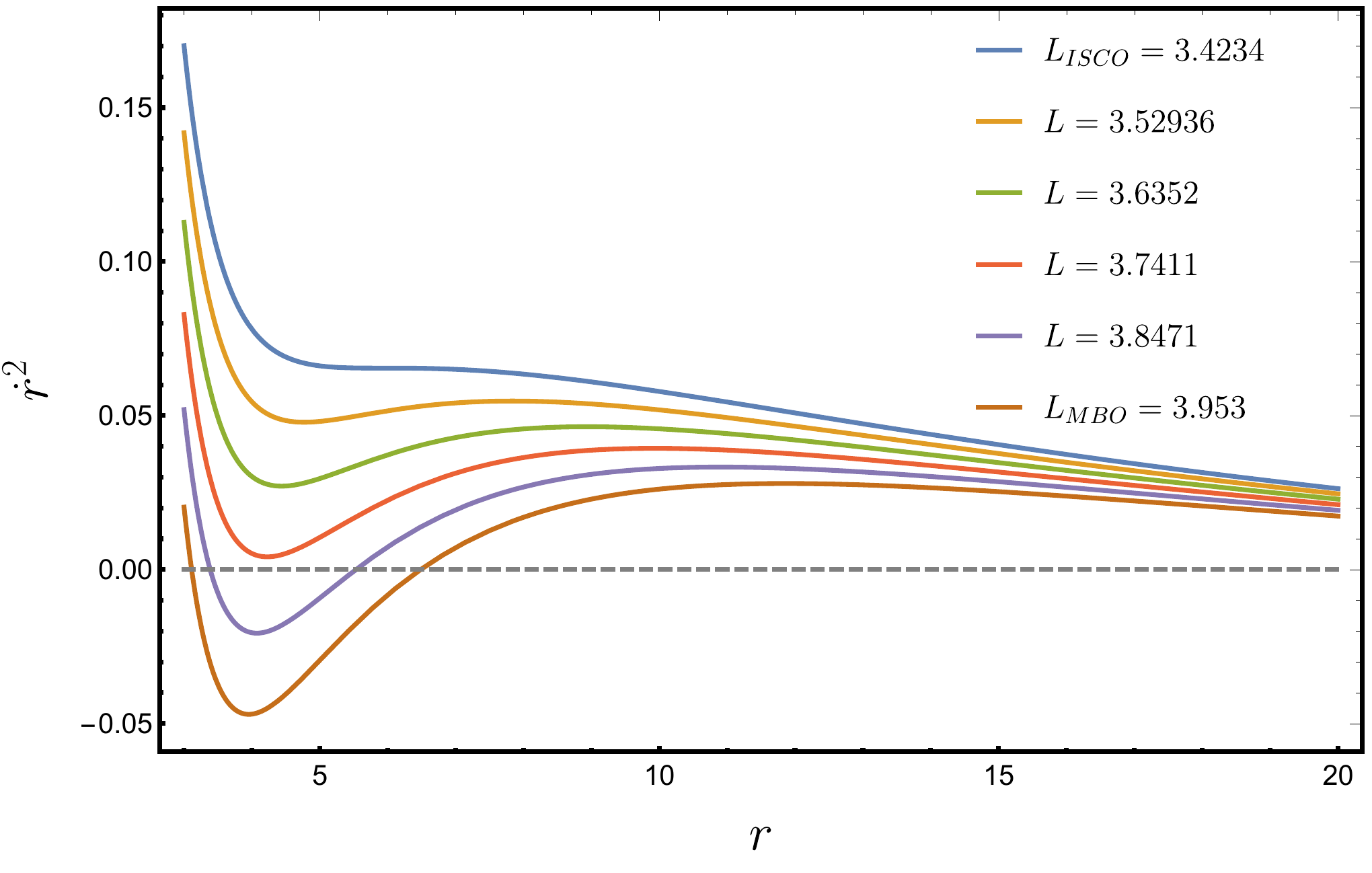}}
\hspace{0.75cm}
\subfigure[$l = l_{\rm max}$] 
{\label{rLd}\includegraphics[width=7.75cm]{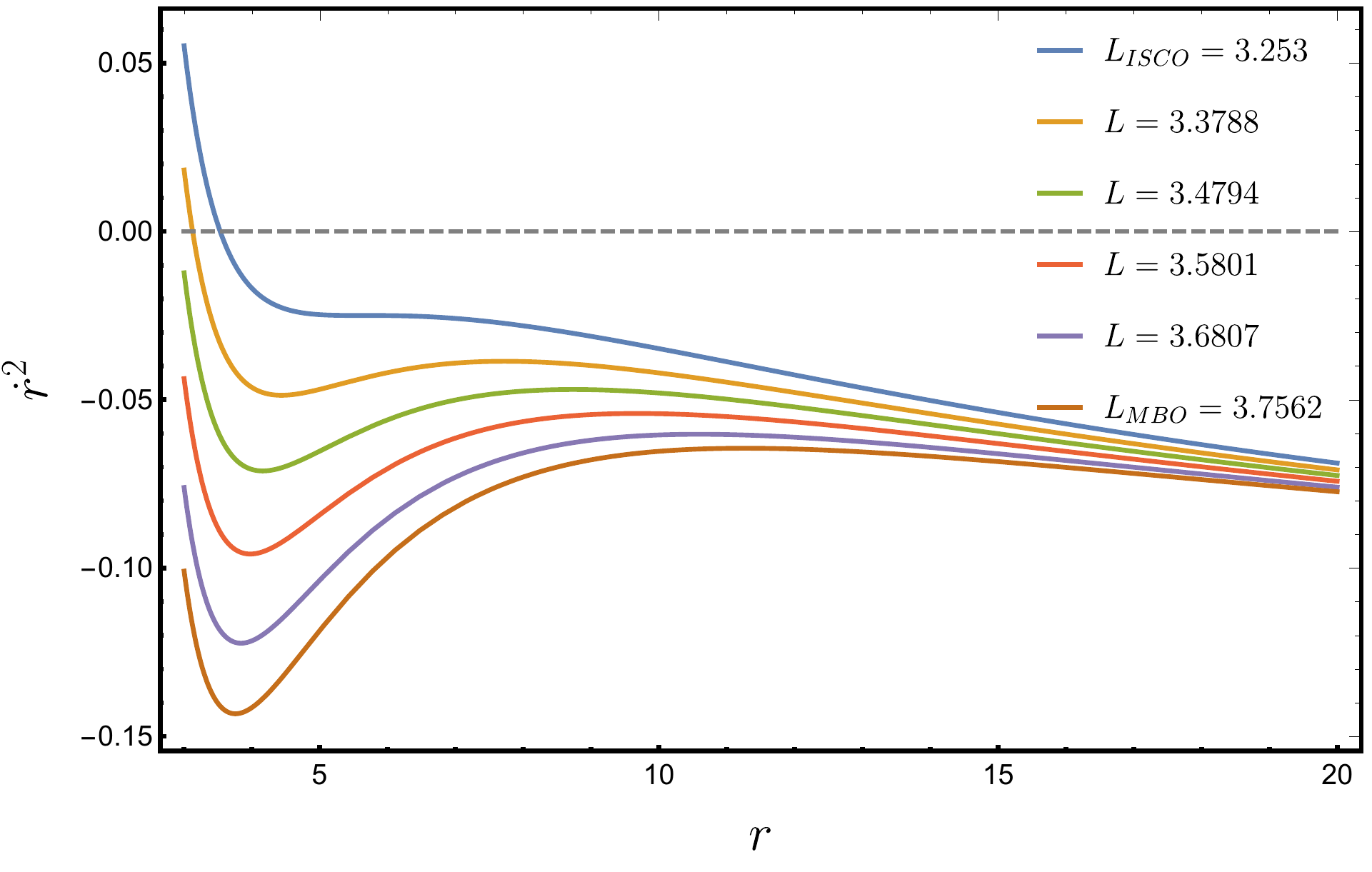}}
\caption{Graphical representation of $\dot{r}^2$ depending on $r$ with $E=0.96$  fixed and different values of $l$, with $L$ varying for $\epsilon=0$, $\epsilon=0.25$, $\epsilon=0.45$, $\epsilon=0.65$, $\epsilon=0.85$, $\epsilon=1$. Bound orbits only exist when $\dot{r}^2=0$ has at least two roots.}\label{Fig5}
\end{figure*}

We now turn our attention to the behaviour of $\dot{r}^2$ within the allowed regions of energy and momentum for different values of the KR parameter $l$ given by Eq.\eqref{lcontrol}. In Fig.\,\ref{Fig5} we fix $E=0.96$ and vary $L$ according to Eq.(\ref{Lcontrol}) by using regular intervals of $\epsilon$. In Figs.\,\ref{rLa} and \ref{rLd}, we observe that there are no bound orbits for $l_{\rm min}$ and $l_{\rm max}$ between $L_{\rm ISCO}$ and $L_{\rm MBO}$ since $\dot{r}^2=0$ has no roots. On the other hand, in Fig.\,\ref{rLb} (red, purple and orange curves) and Fig.\,\ref{rLc} (purple and orange curves), as $L$ increases from $L_{\rm ISCO}$ there are three roots and the bound orbits are in the regime between the outermost two roots with $\dot{r}^2 > 0$.

\begin{figure*}[htb!]
\centering
\subfigure[$l=l_{\rm min}$] 
{\label{rEaa}\includegraphics[width=7.75cm]{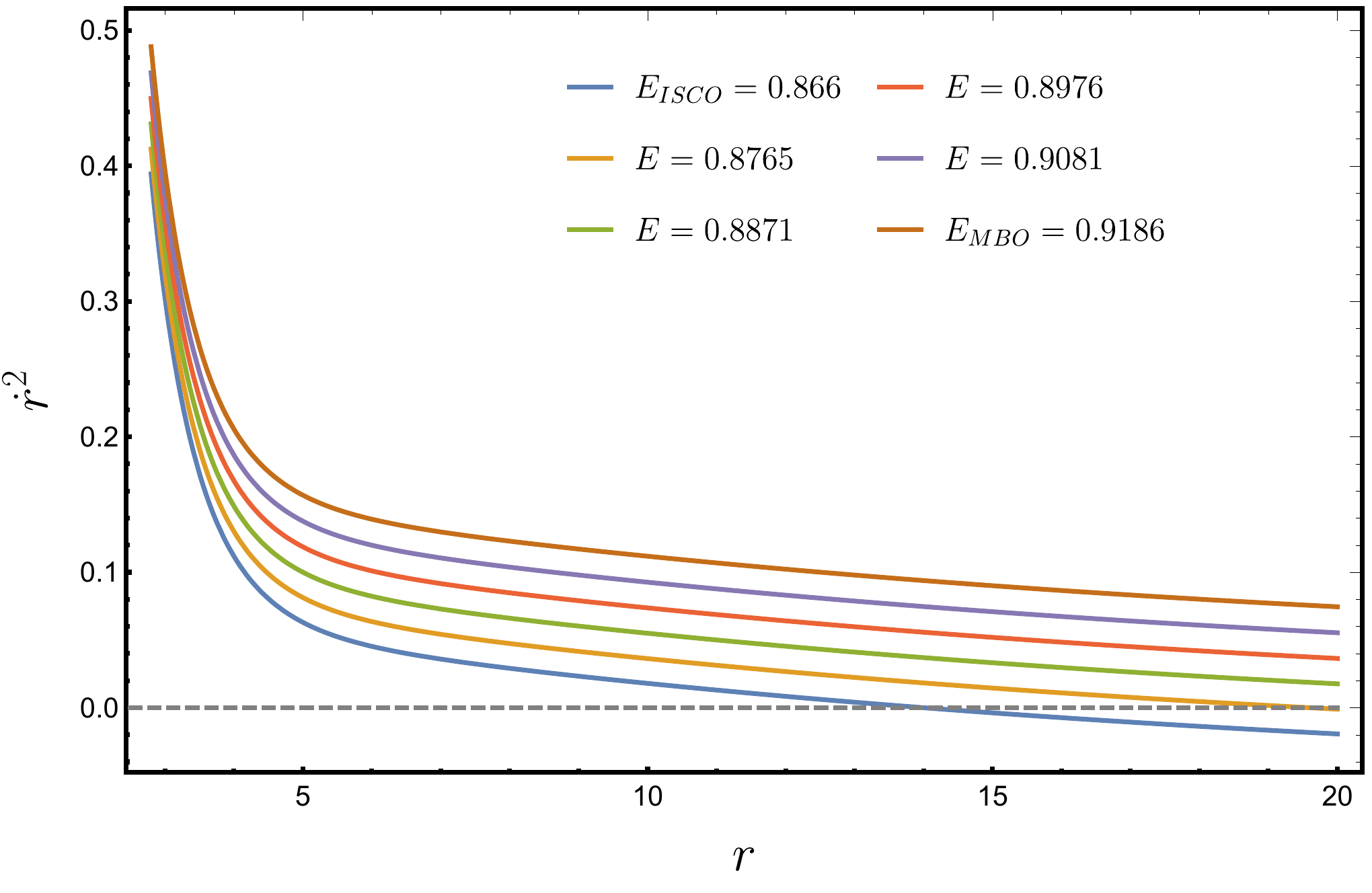} }
\hspace{0.75cm}
\subfigure[  $l=-0.01285$] 
{\label{rEbb}\includegraphics[width=7.75cm]{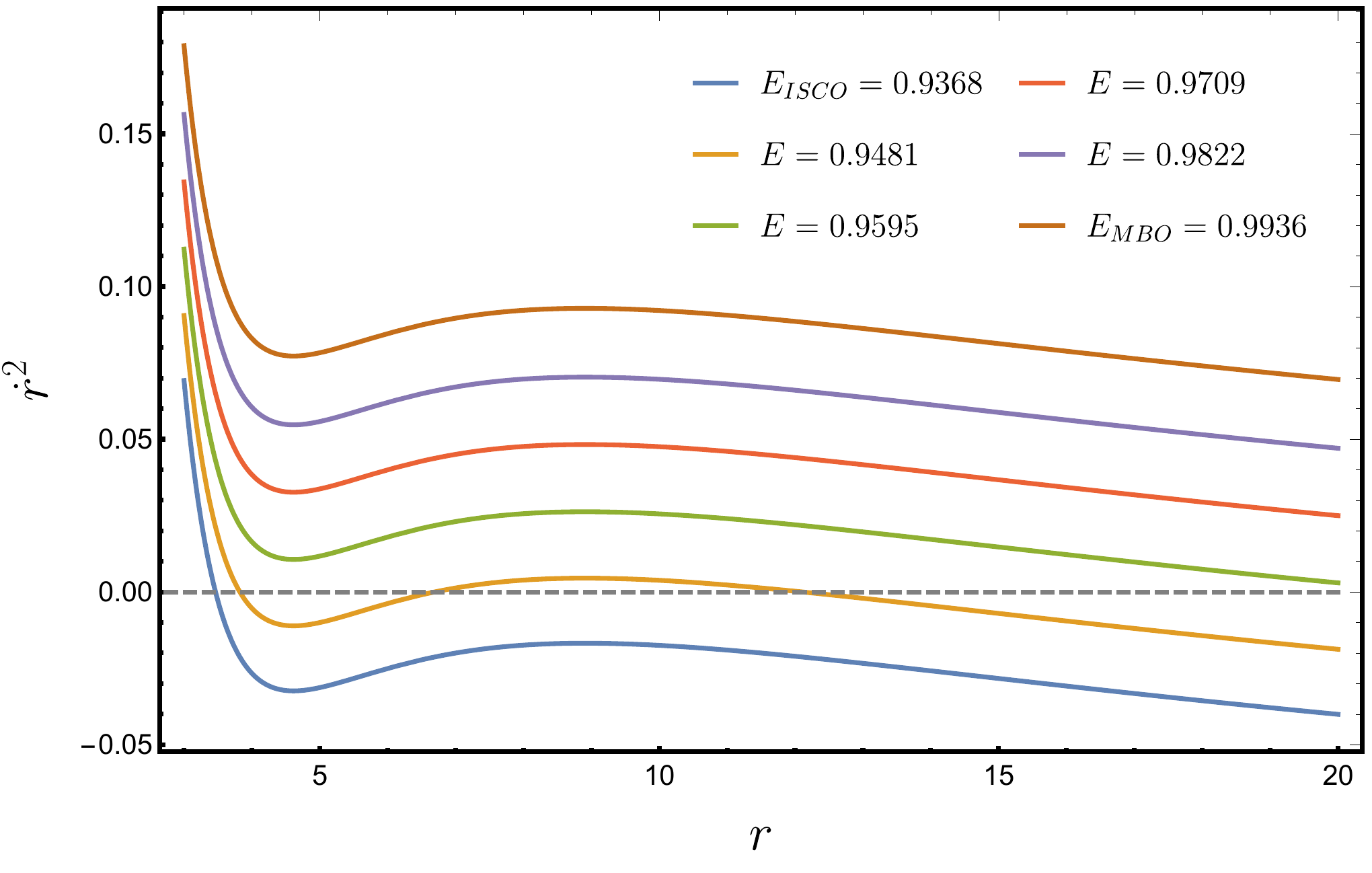} }
\subfigure[ $l=0.011746$] 
{\label{rEc}\includegraphics[width=7.75cm]{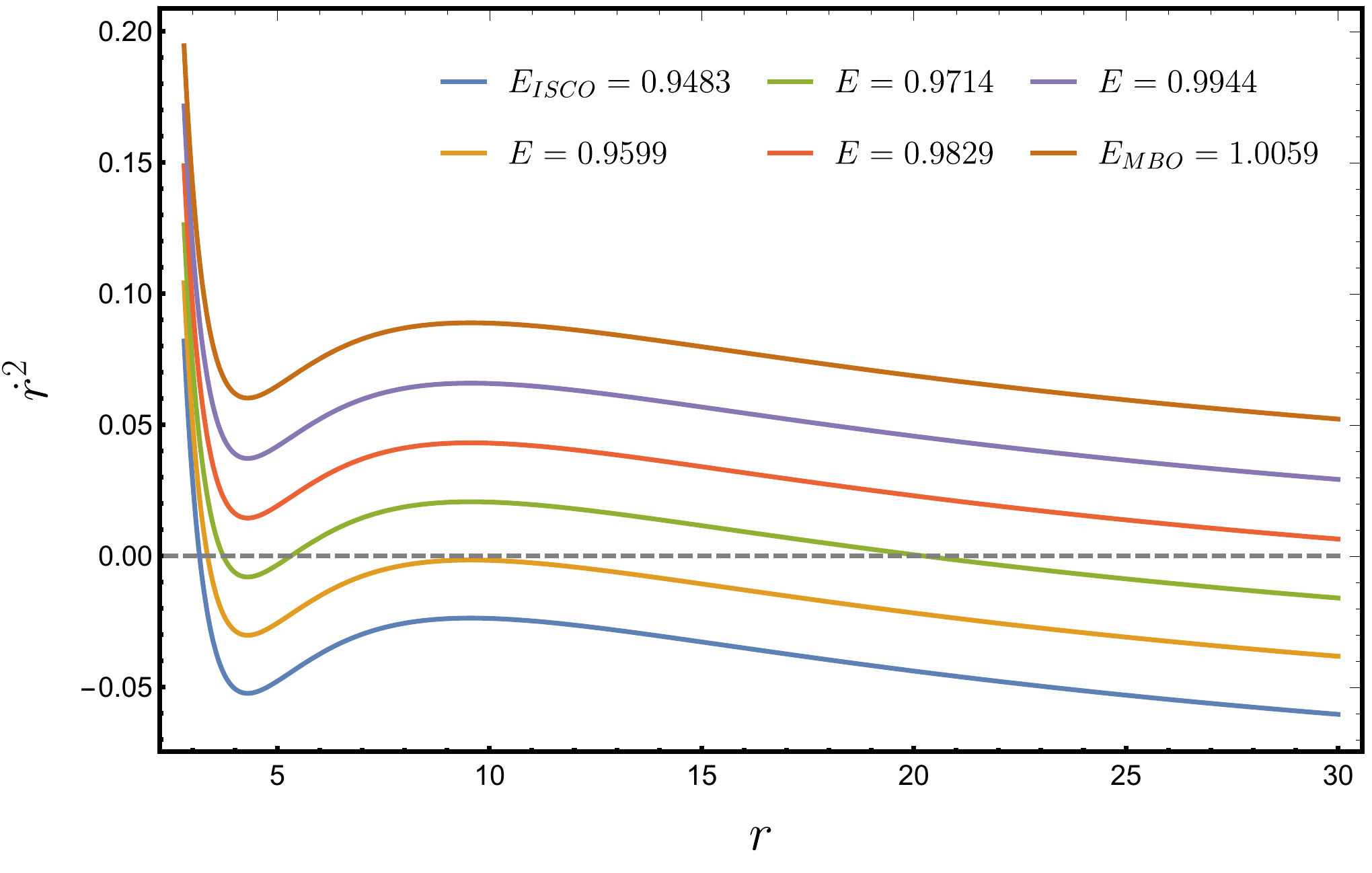}}
\hspace{0.75cm}
\subfigure[$l = l_{\rm max}$] 
{\label{rEd}\includegraphics[width=7.75cm]{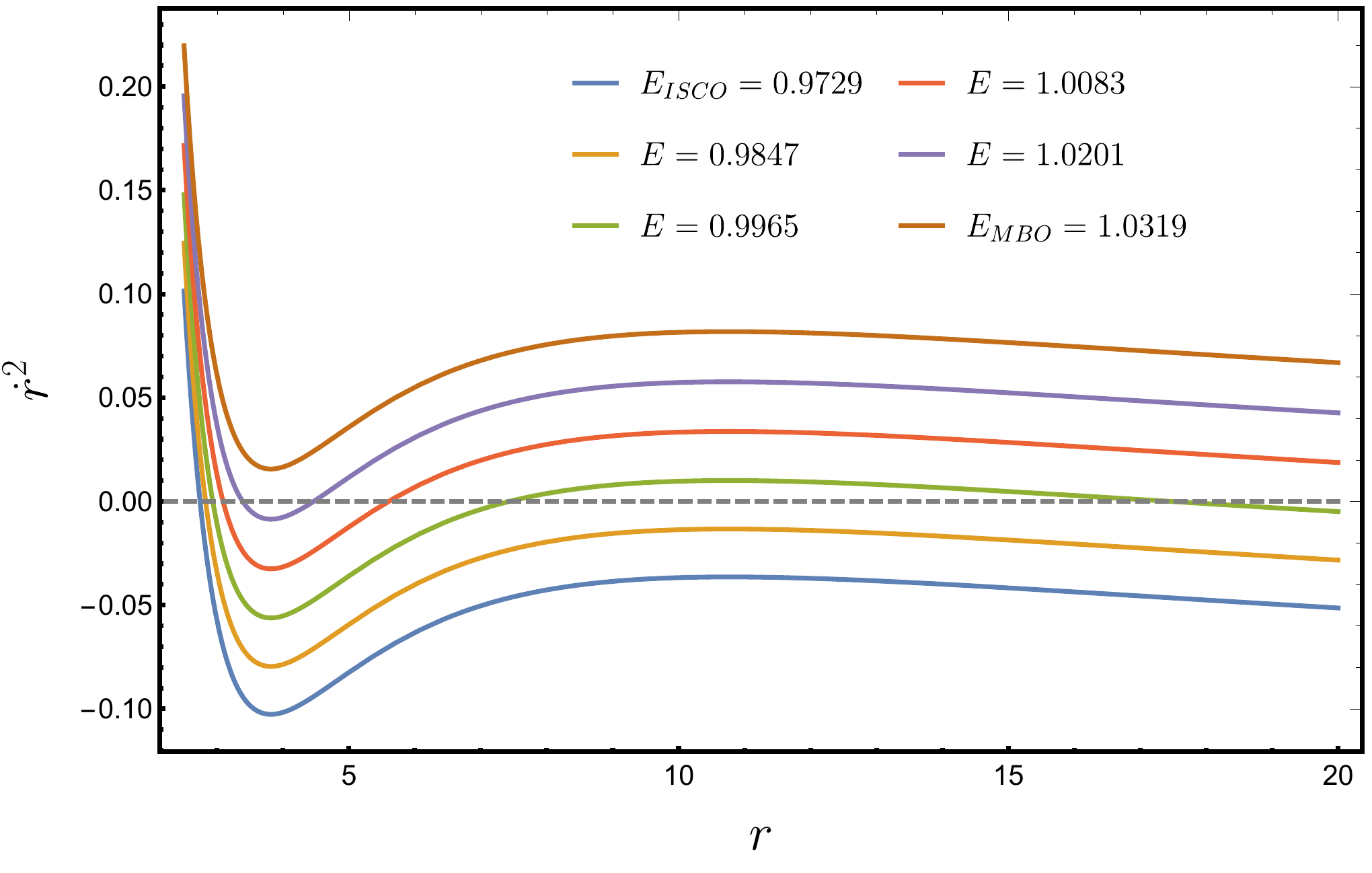}}
\caption{Graphical representation of $\dot{r}^2$ depending on $r$ with a fixed values of $L=3.7$ and different values of $l$, with $E$ varying for $\eta=0$, $\eta=0.2$, $\eta=0.4$, $\eta=0.6$, $\eta=0.8$, $\eta=1$ according to Eq.(\ref{Econtrol}). Bound orbits only exist when $\dot{r}^2=0$ has at least two roots.}\label{Fig6}
\end{figure*}

Fig.\,\ref{Fig6} shows the behavior of $\dot{r}^2$ as a function of $r$ for $L=3.7$ and $E$ varying according to \eqref{Econtrol} for different values of the $l$ parameter. In Fig.\,\ref{rEaa} there are no bound orbits for $L=3.7$, which is expected as seen in Fig.\,\ref{ELa}.  In Fig.\,\ref{rEbb} there are bound orbits with $E_{\rm ISCO}=0.9368 < E < 0.9595$ with $\dot{r}^2>0$ whose $\dot{r}^2=0$ has at least two roots.
In Figs.\,\ref{rEc} and \ref{rEd}, increasing the value of $l$ leads to a shift in the curves, thereby increasing the allowed energy range for the occurrence of bound orbits. Our results consistently recover those obtained for the Schwarzschild solution when $l\rightarrow 0$  \cite{TAX}.

\section{PERIODIC ORBITS: numerical results }\label{PO}

In this section we will study a particular class of bound orbits for a time-like particle dubbed as a {\it periodic orbit}, located between the MBO and the ISCO, around the KR black hole. We hereafter adopt the taxonomy proposed in \cite{TAX} to identify different types of orbits around the black hole, called zoom-whirl. These orbits are unstable because they occur between the MBO and the ISCO. They are associated with the unstable circular orbits of the effective potential and occur close to the maximum of the potential, that is $\frac{d^2V_{\rm eff}}{dr^2}<0$. Therefore, small deviations cause the particle to be ejected or fall into the black hole. A rational number $q$ is defined as a triplet of the integers $(z, w, v)$ where $z$ is the zoom number (number of leaves in the orbit), $w$ is the whirl number, and $v$ is the vertex number that relates successive apastron, and which are connected via the equation
\begin{equation}
q=w+\frac{v}{z}\,.\label{q1}
\end{equation}
Periodic orbits are defined as those that return to the same initial position after a finite time. Orbital precession takes place when a small  perturbation occur in periodic orbits, i.e,  $q=w+v/z\pm\delta$, with $0<\delta\ll 1$.  

Since any irrational number can be approximated by a rational one, a generic orbit can be approximated by a periodic orbit around the black hole such that the ratio between the oscillation frequencies $\frac{\omega_r}{\omega_{\phi}}=\Delta \phi/2\pi $ is a rational number and, consequently
\begin{eqnarray}
q\equiv \frac{\Delta \phi}{2\pi}-1\,,\label{curvaq1}
\end{eqnarray}
where 
\begin{eqnarray}
\Delta \phi = 2 \int^{r_a}_{r_p} \frac{\dot{\phi}}{\dot{r}}dr =  2 \int^{r_a}_{r_p}\frac{L}{r^2\sqrt{E^2-V_{\rm eff}(r)}}dr\,,\label{curvaq2}
\end{eqnarray}
where $r_a$ and $r_p$ are the turning points between the ISCO and the MBO called apastron and periastron, respectively, and are obtained from the roots of $\dot{r}^2$. Therefore, according to Eqs.\eqref{Veff} and \eqref{curvaq2} the rational number $q$ depends on $E$, $L$ and, in the KR space-time, on the parameter $l$. Studying the properties of the periodic orbits is fundamental to understanding any generic orbit \cite{TAX}. Using Eqs.\,\eqref{Lcontrol} and \eqref{Econtrol} we can obtain the angular momentum and energy for different values of $\epsilon$ and $\eta$. Subsequently, we solve Eq.\,\eqref{curvaq1} numerically and analyze the properties of periodic orbits for different values of $l$ varying $\alpha$ according to Eq.\,\eqref{lcontrol}. Next, the polar coordinates, $\left\lbrace r(\phi)\cos\phi, r(\phi)\sin\phi\right\rbrace$, are used for the final plot of each orbit.

\begin{figure*}[htb!]
\centering
\subfigure[$L = 3.7911$, \,$(1, 1, 0)$] 
{\label{Lb1}\includegraphics[width=4.75cm]{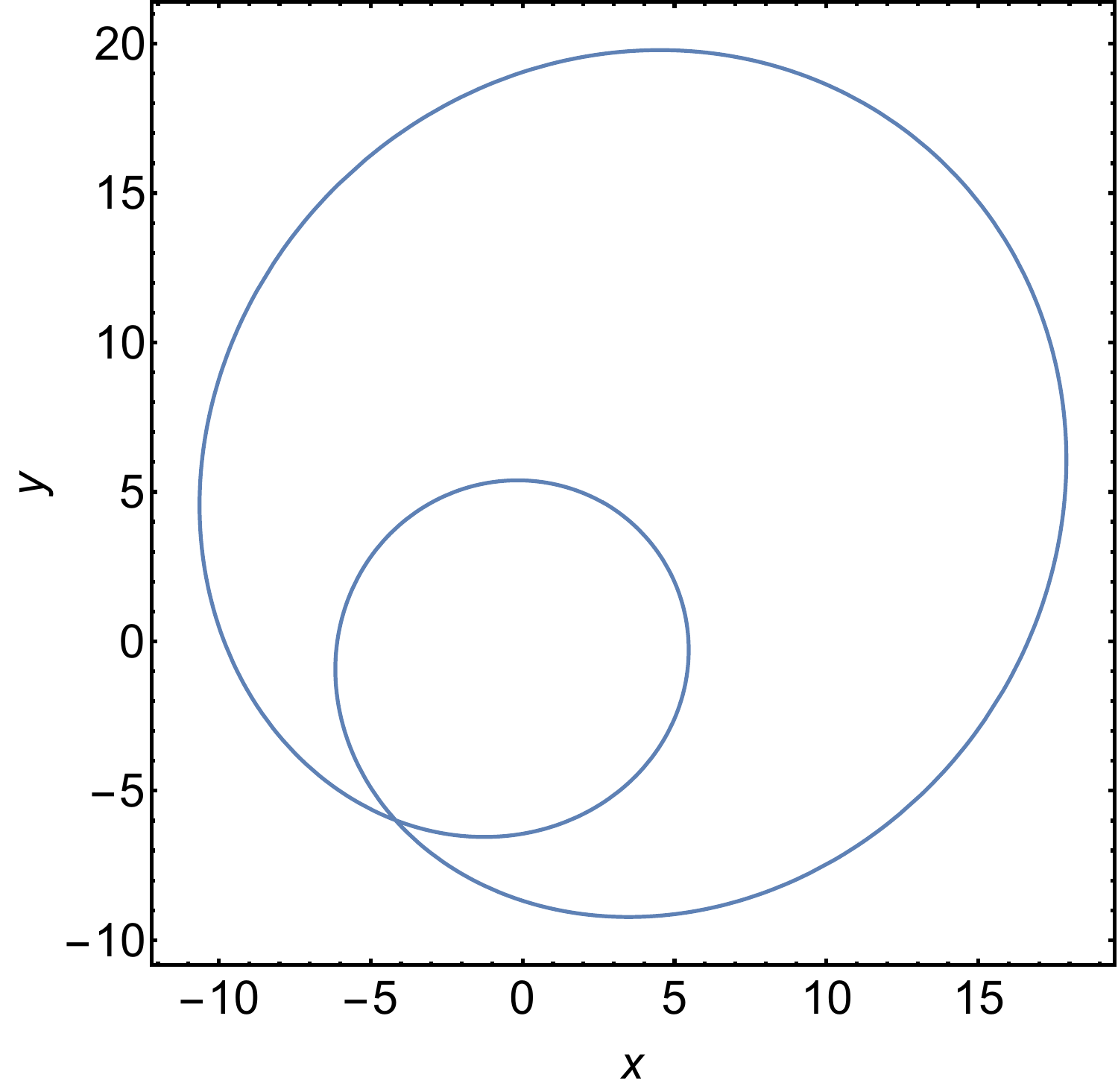} }
\hspace{0.75cm}
\subfigure[  $L = 3.7591$, \,$(1, 2, 0)$] 
{\label{Lb2}\includegraphics[width=4.75cm]{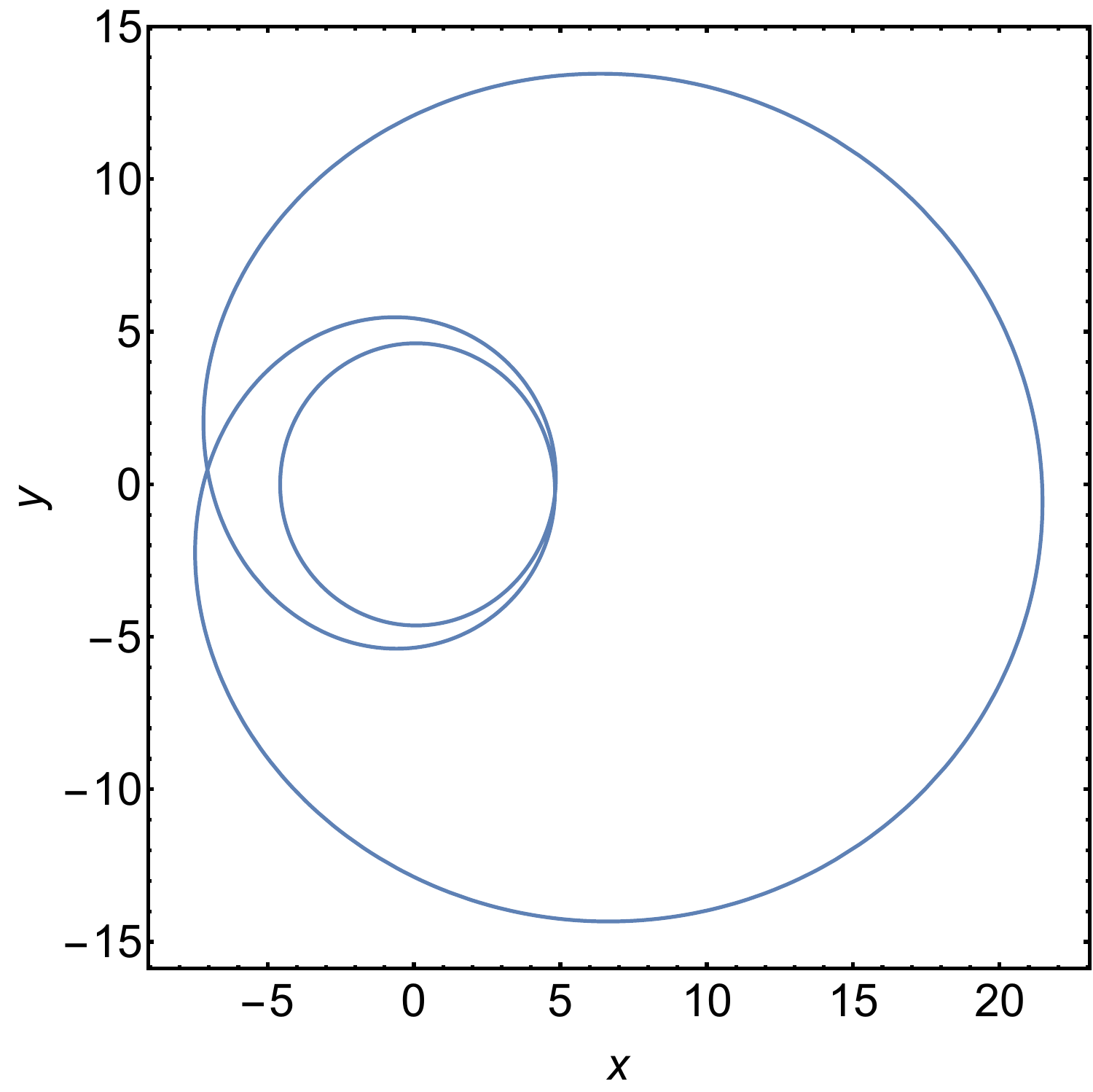} }
\hspace{0.75cm}
\subfigure[ $L = 3.7628$, \,$(2, 1, 1)$] 
{\label{Lb3}\includegraphics[width=4.75cm]{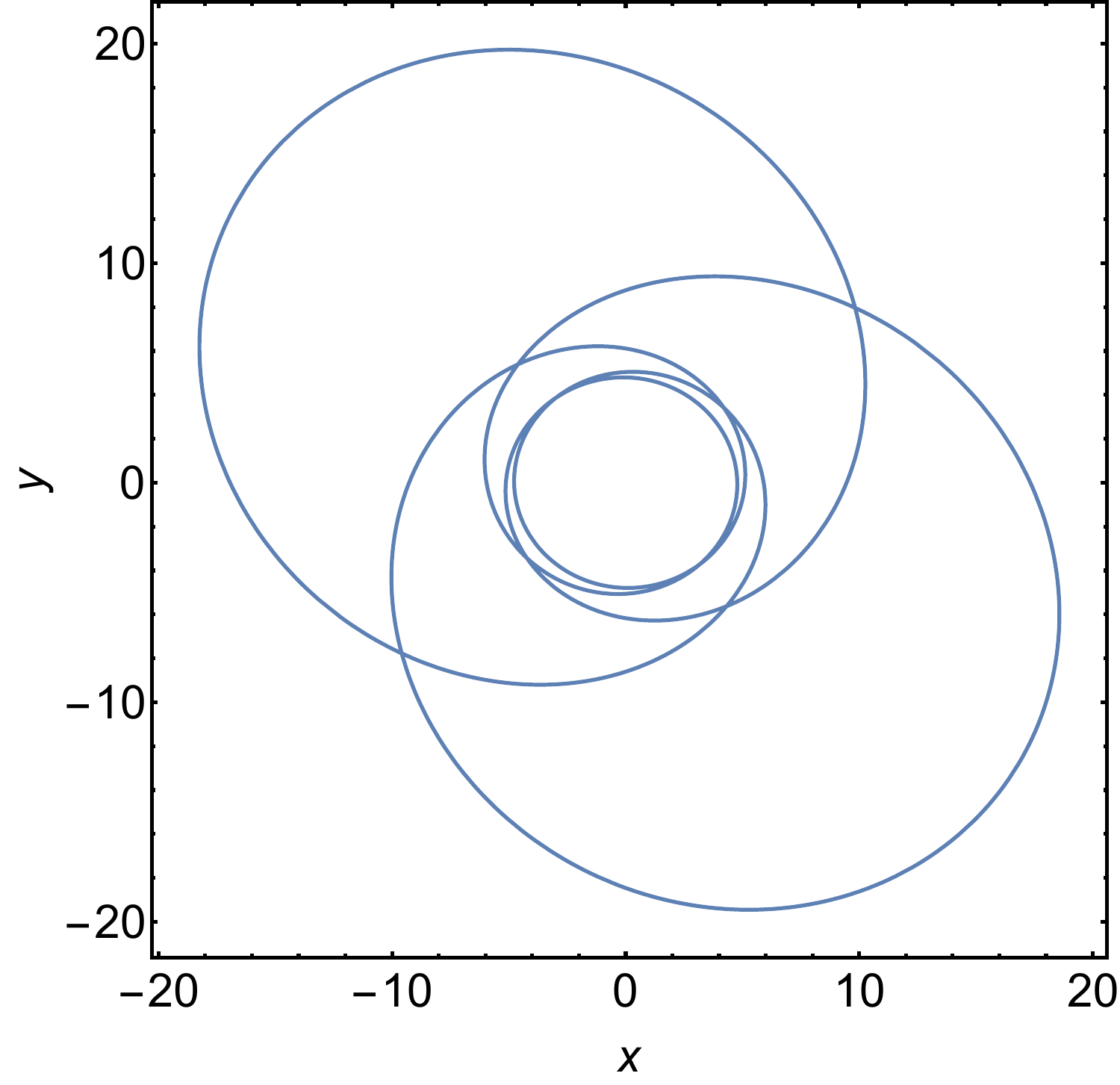}}
\hspace{0.75cm}
\subfigure[$L = 3.7587$, \, $(2, 2, 1)$] 
{\label{Lb4}\includegraphics[width=4.75cm]{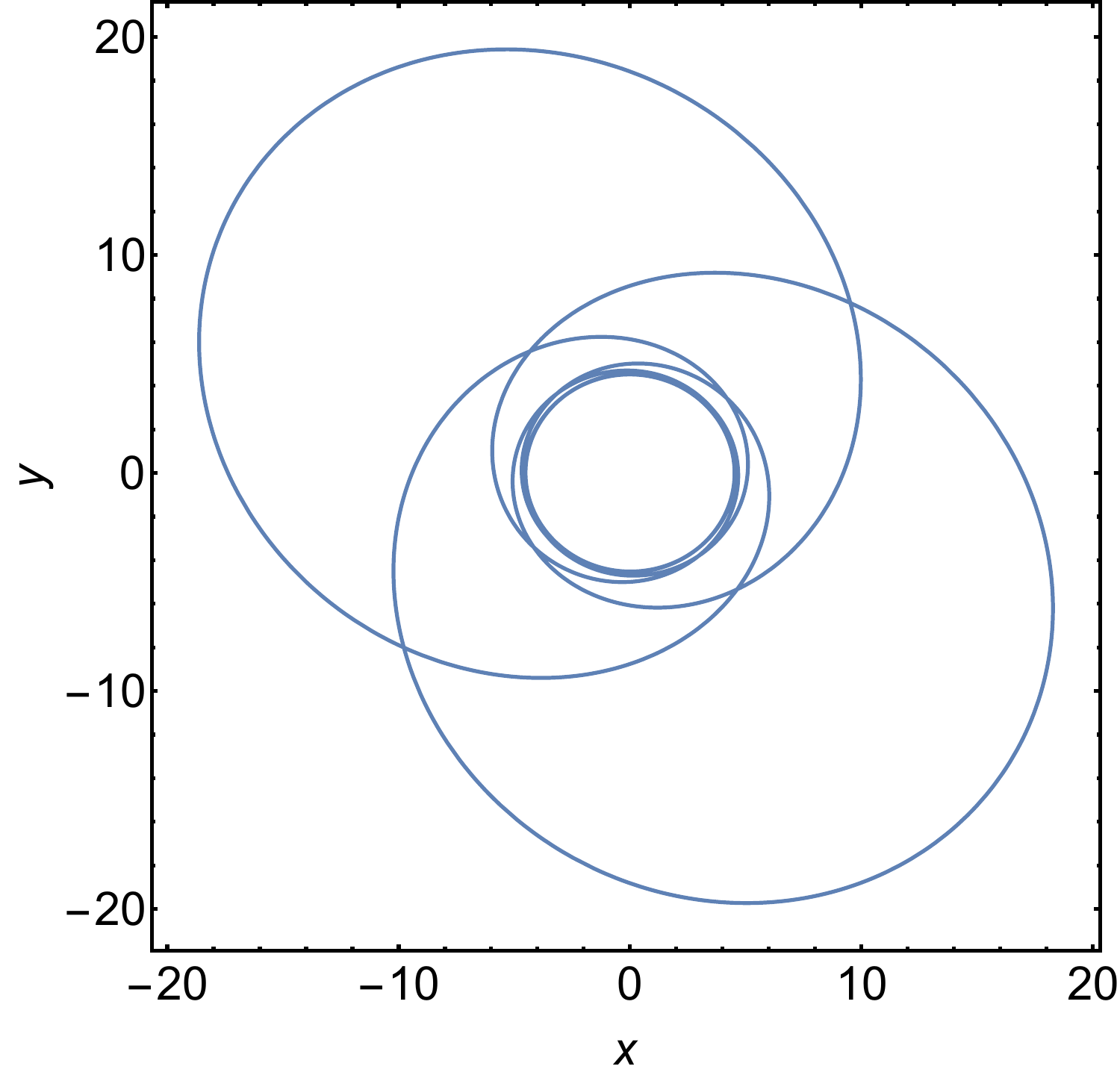}}
\hspace{0.75cm}
\subfigure[$L = 3.7609$, \, $(3, 1, 2)$] 
{\label{Lb5}\includegraphics[width=4.75cm]{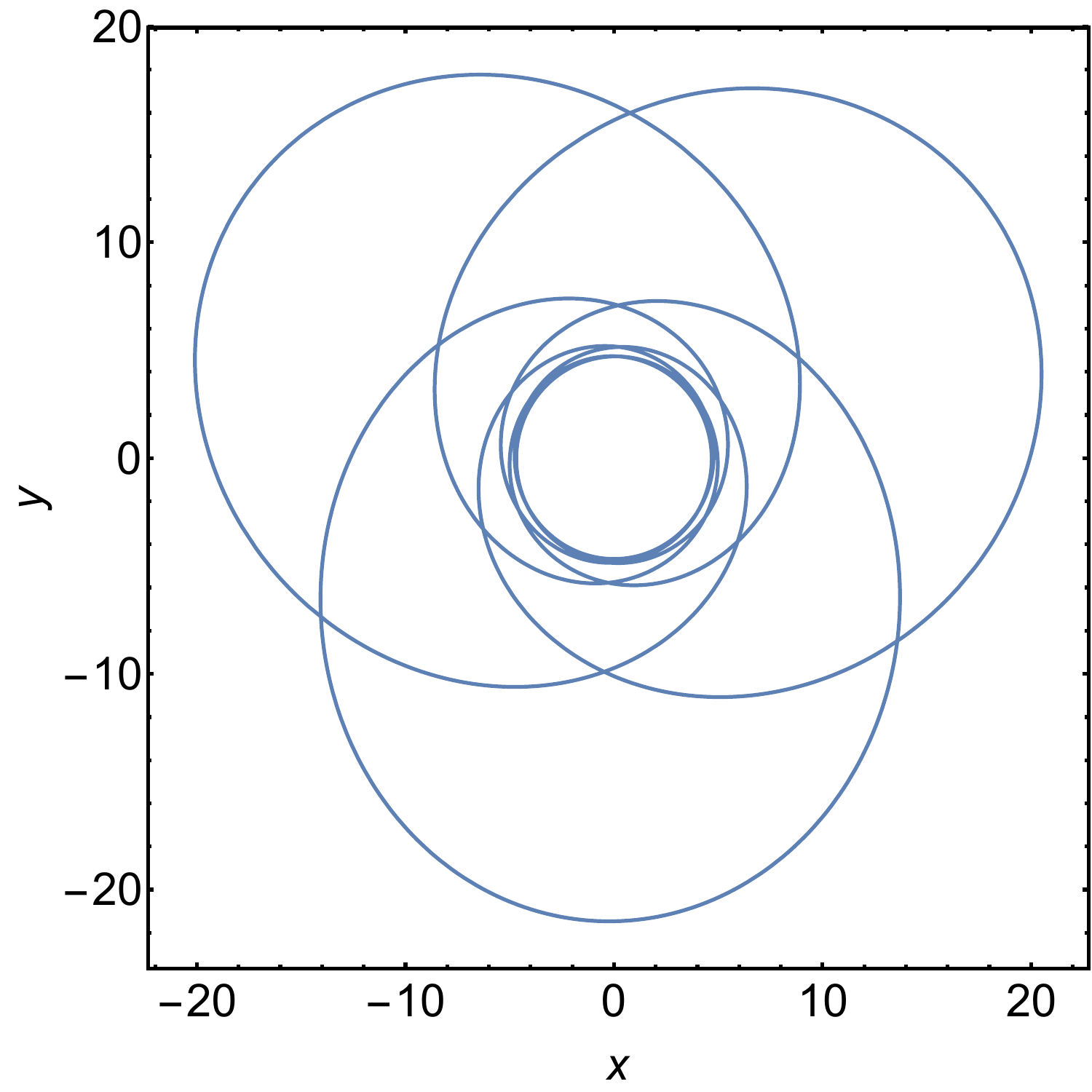}}
\hspace{0.75cm}
\subfigure[$L = 3.75865$, \, $(3, 2, 2)$] 
{\label{Lb6}\includegraphics[width=4.75cm]{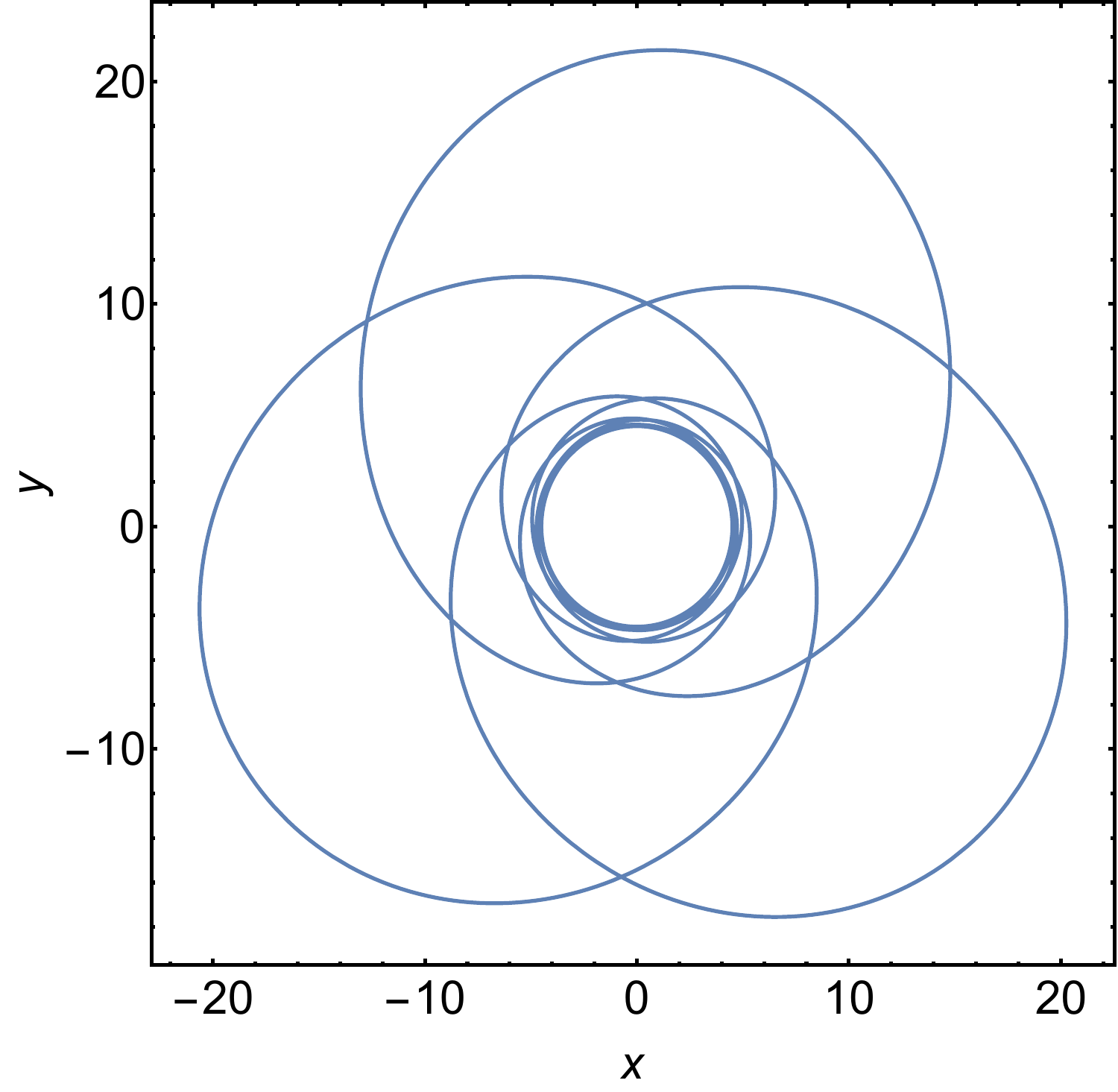}}
\hspace{0.75cm}
\subfigure[$L = 3.7602$, \, $(4, 1, 3)$] 
{\label{Lb7}\includegraphics[width=4.75cm]{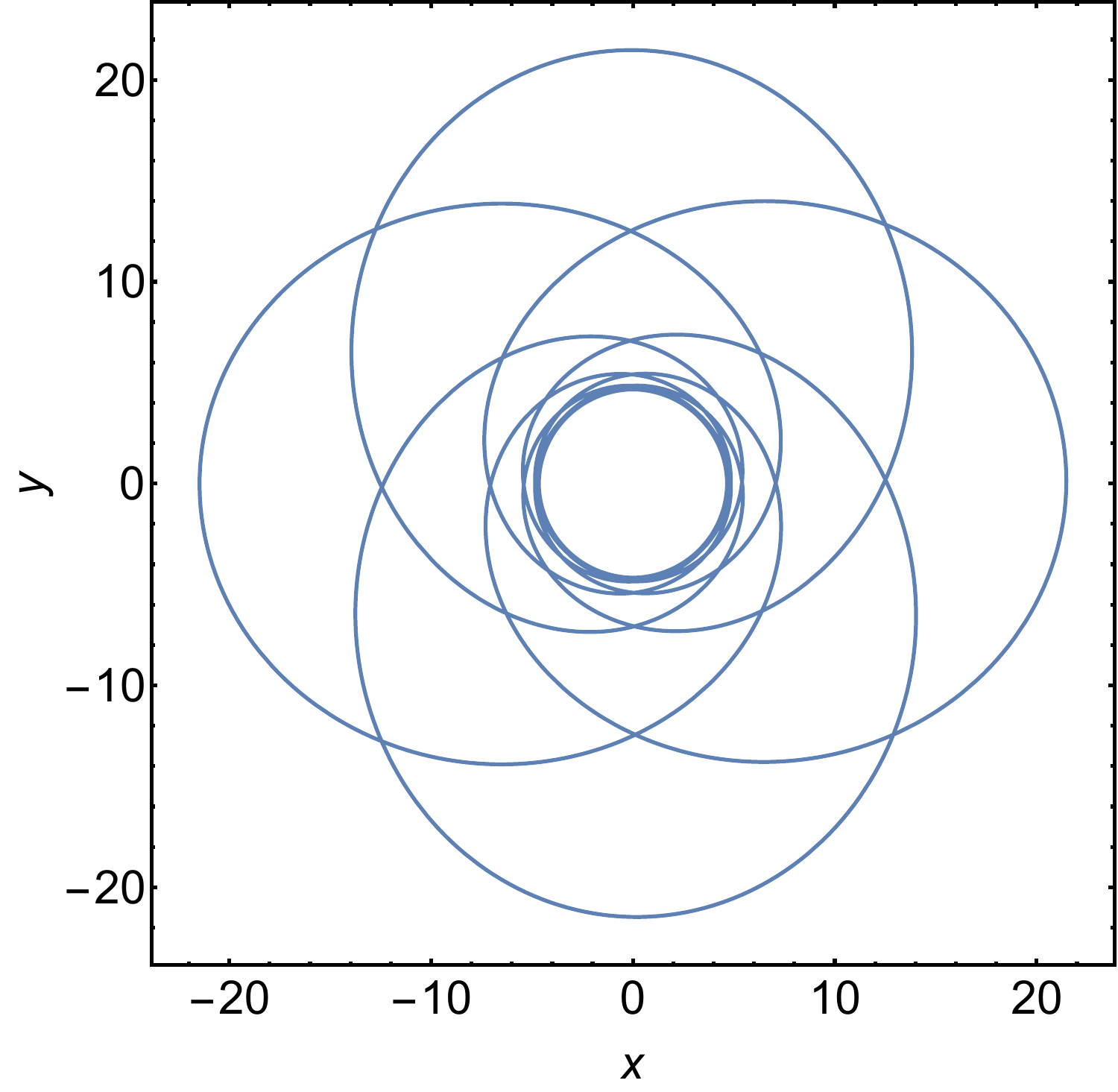}}
\hspace{0.75cm}
\subfigure[$L = 3.75864$, \, $(4, 2, 3)$] 
{\label{Lb8}\includegraphics[width=4.75cm]{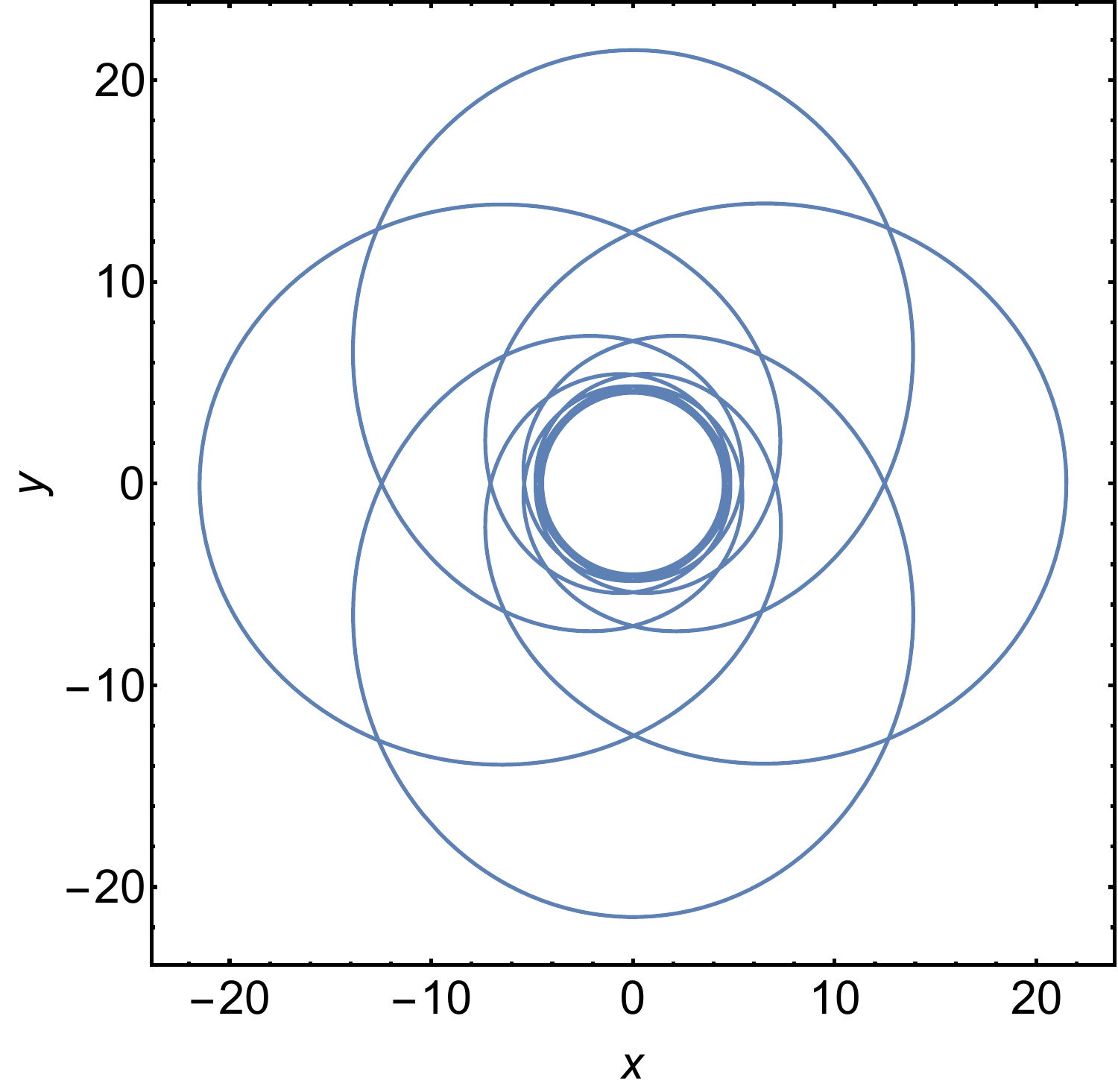}}
\hspace{0.75cm}
\subfigure[$L = 3.7589 $, \, $(5, 2, 1)$] 
{\label{Lb9}\includegraphics[width=4.75cm]{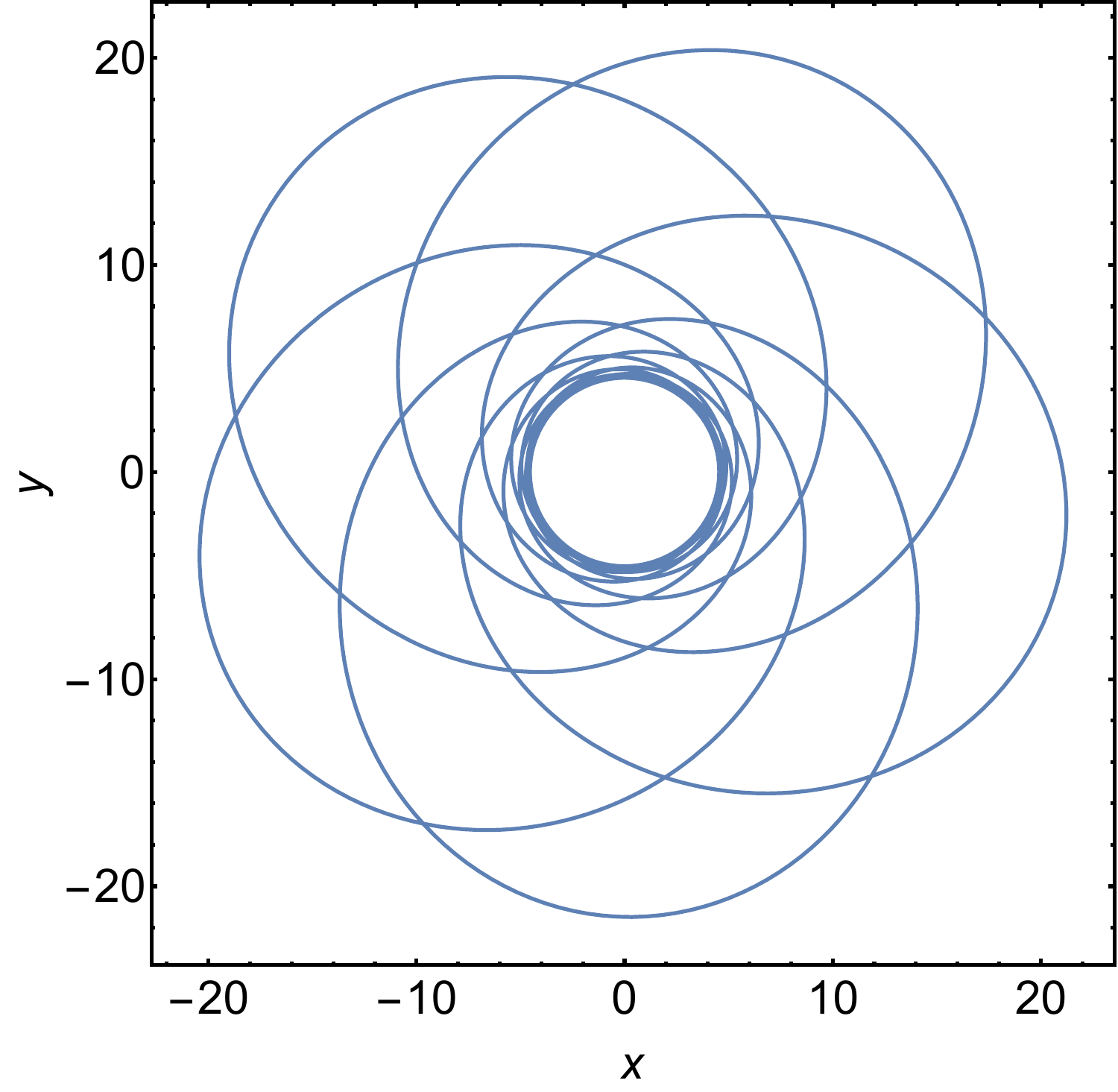}}
\caption{Periodic orbits for different values of $L$ and $(z, w, v)$ with $E=0.96$ and $\alpha=0.7$ $(l=-0.01285)$ in \eqref{lcontrol}. Here ${\rm x}$ and ${\rm y}$ have units of meter $\left[m\right]$. }\label{Fig7}
\end{figure*}

\begin{figure*}[htb!]
\centering
\subfigure[$L = 3.5881$, \,$(1, 1, 0)$] 
{\label{La1}\includegraphics[width=4.75cm]{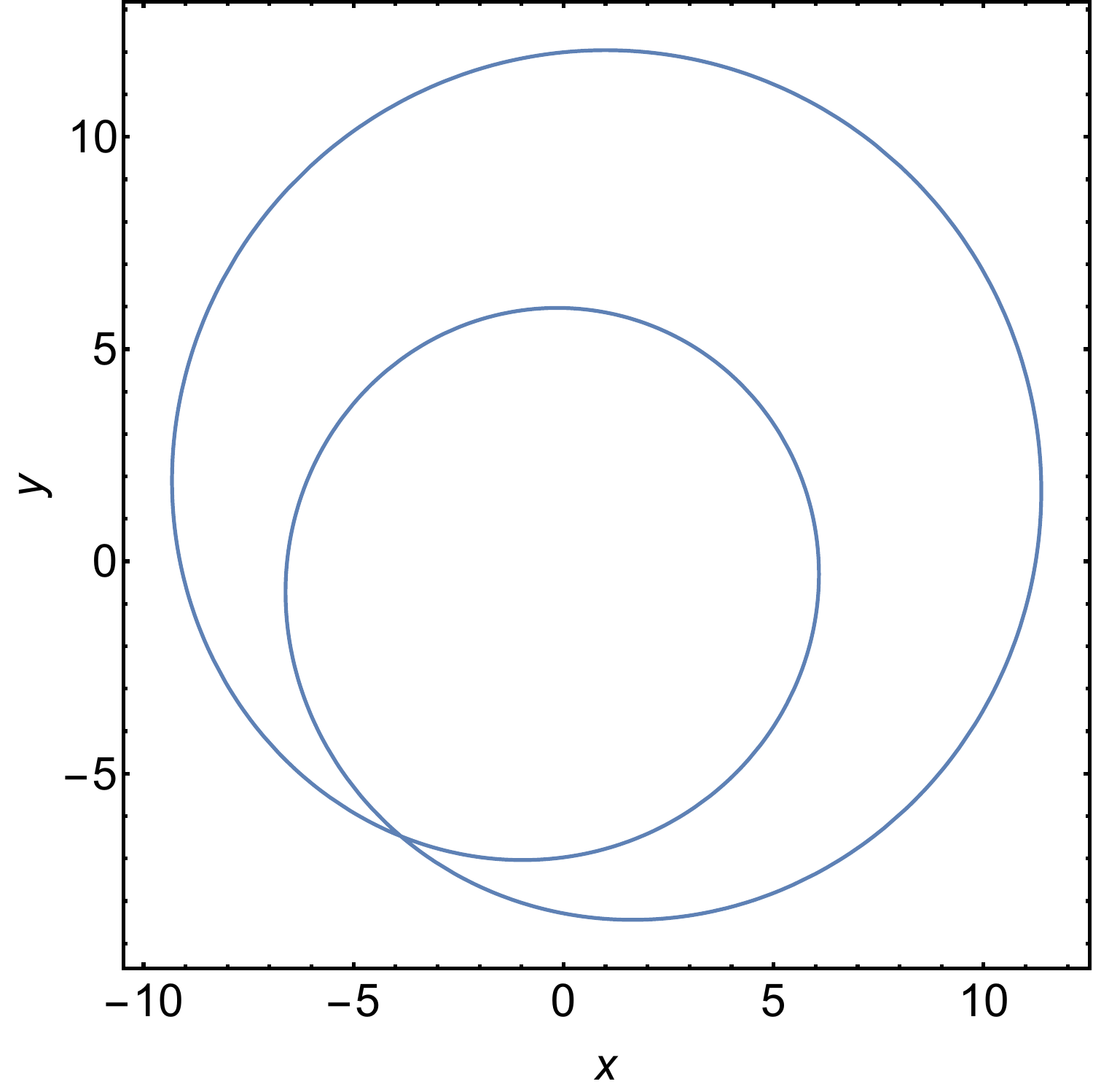} }
\hspace{0.75cm}
\subfigure[  $L = 3.5547$, \,$(1, 2, 0)$] 
{\label{La2}\includegraphics[width=4.75cm]{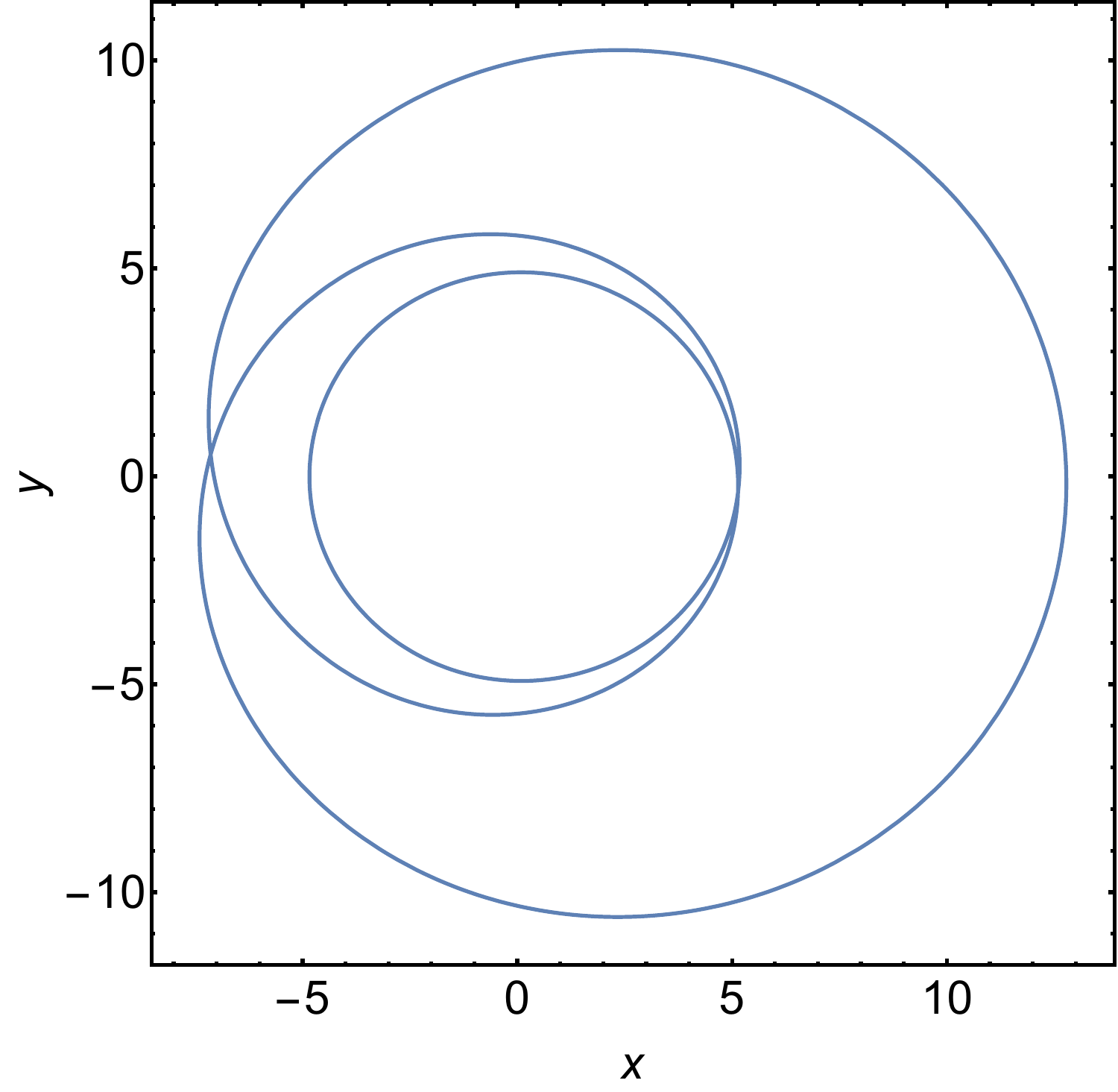} }
\hspace{0.75cm}
\subfigure[ $L = 3.5599$, \,$(2, 1, 1)$] 
{\label{La3}\includegraphics[width=4.75cm]{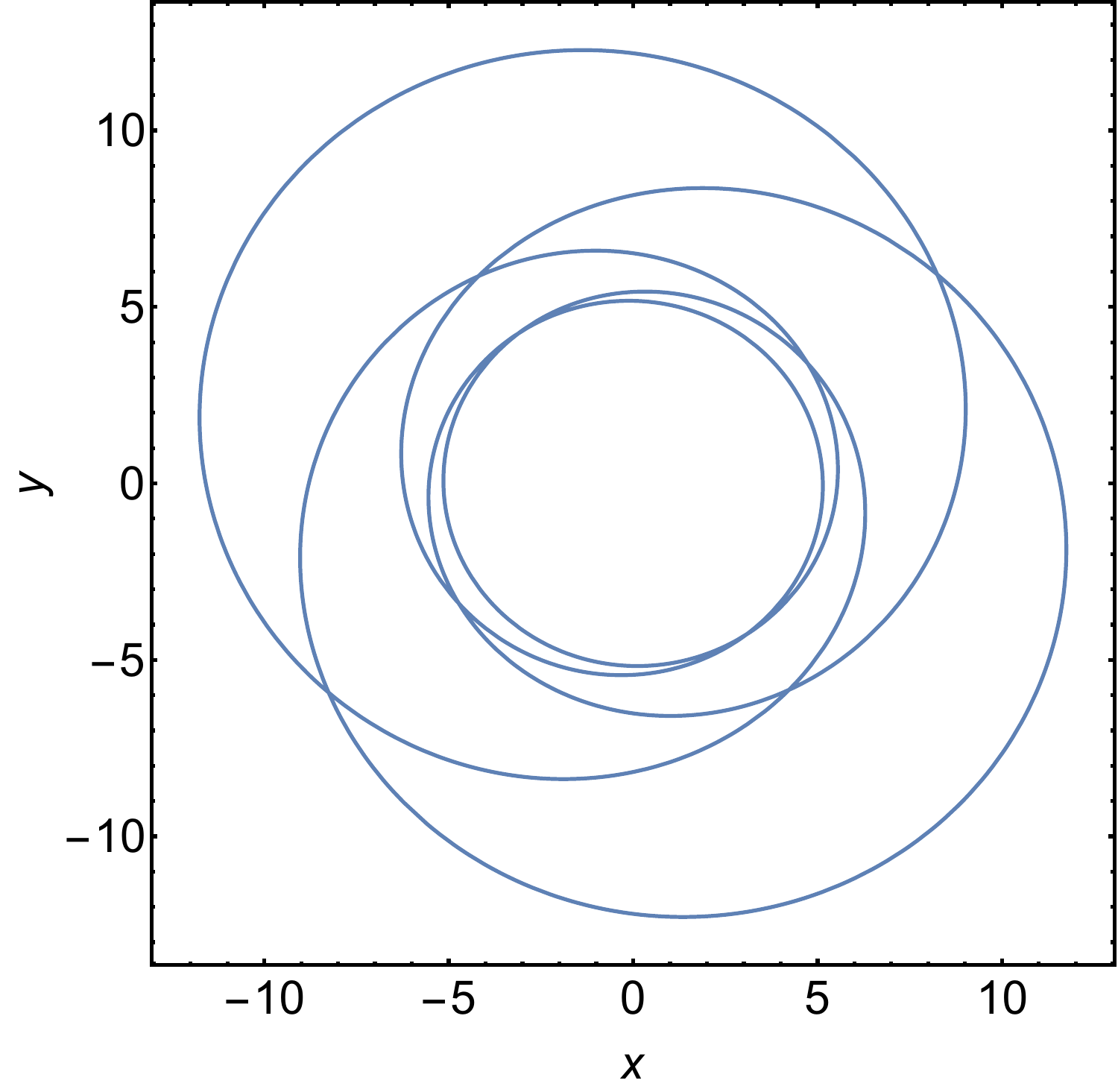}}
\hspace{0.75cm}
\subfigure[$L = 3.5540$, \, $(2, 2, 1)$] 
{\label{La4}\includegraphics[width=4.75cm]{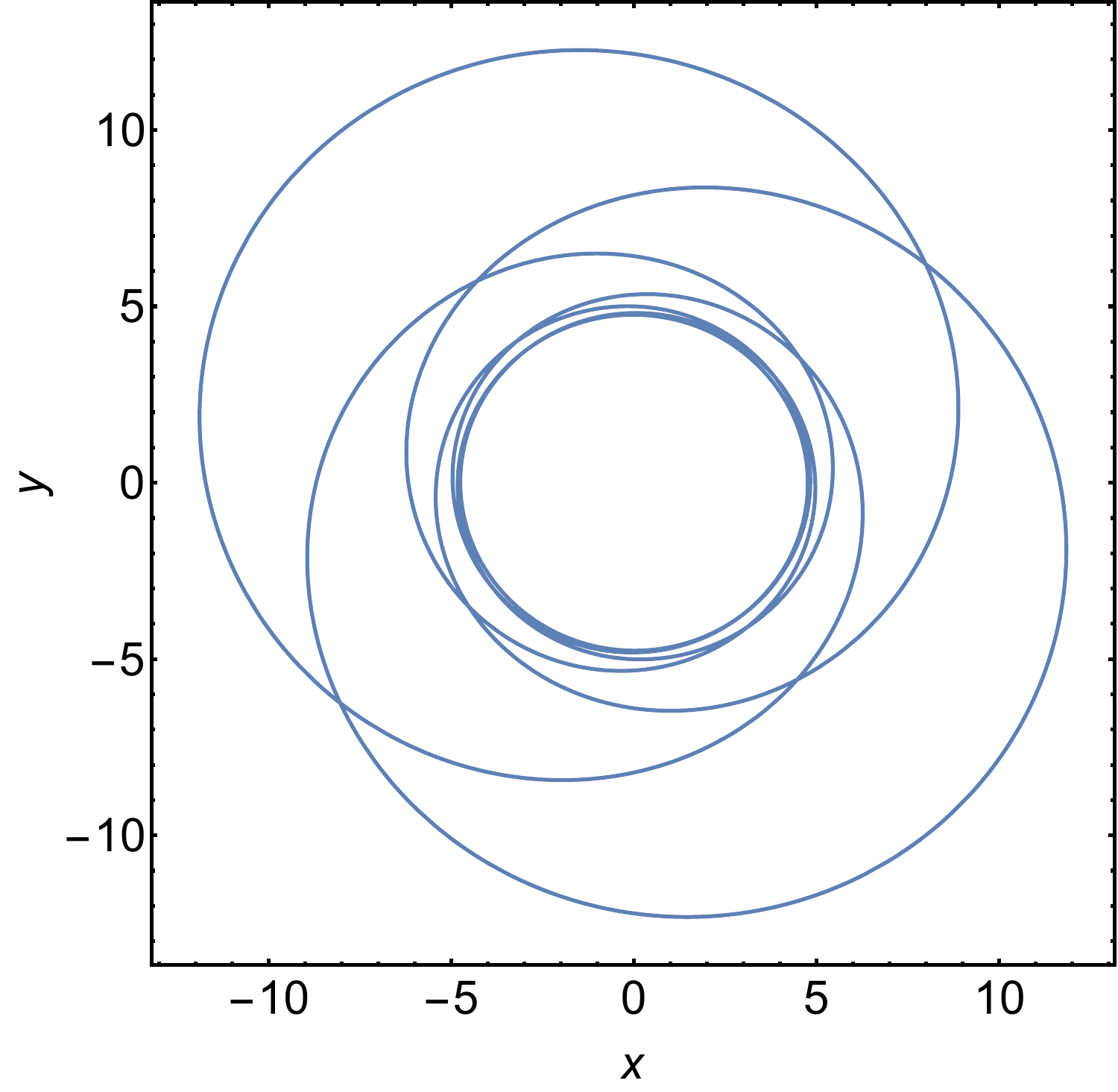}}
\hspace{0.75cm}
\subfigure[$L = 3.5572$, \, $(3, 1, 2)$] 
{\label{La5}\includegraphics[width=4.75cm]{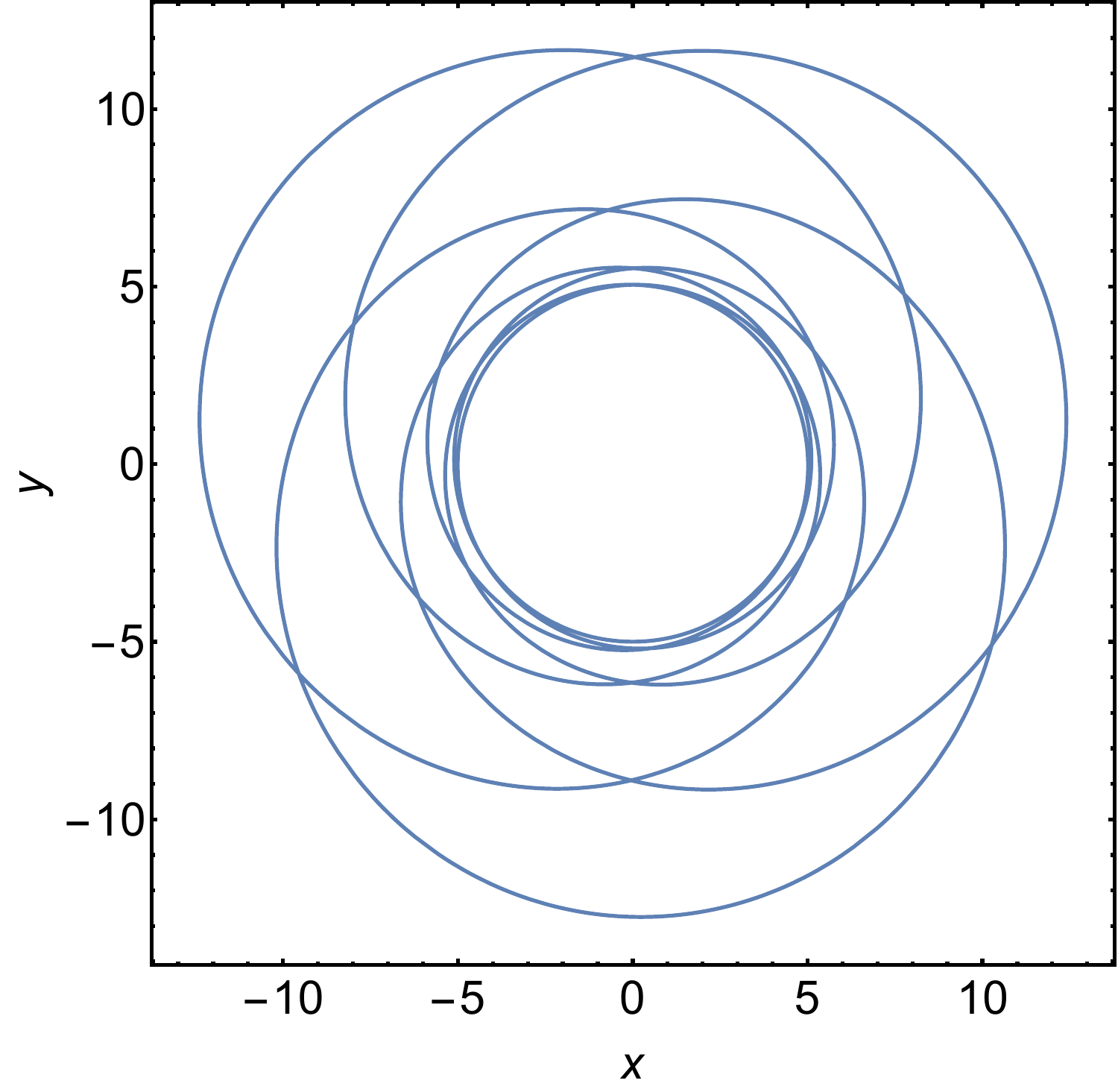}}
\hspace{0.75cm}
\subfigure[$L = 3.5539$, \, $(3, 2, 2)$] 
{\label{La6}\includegraphics[width=4.75cm]{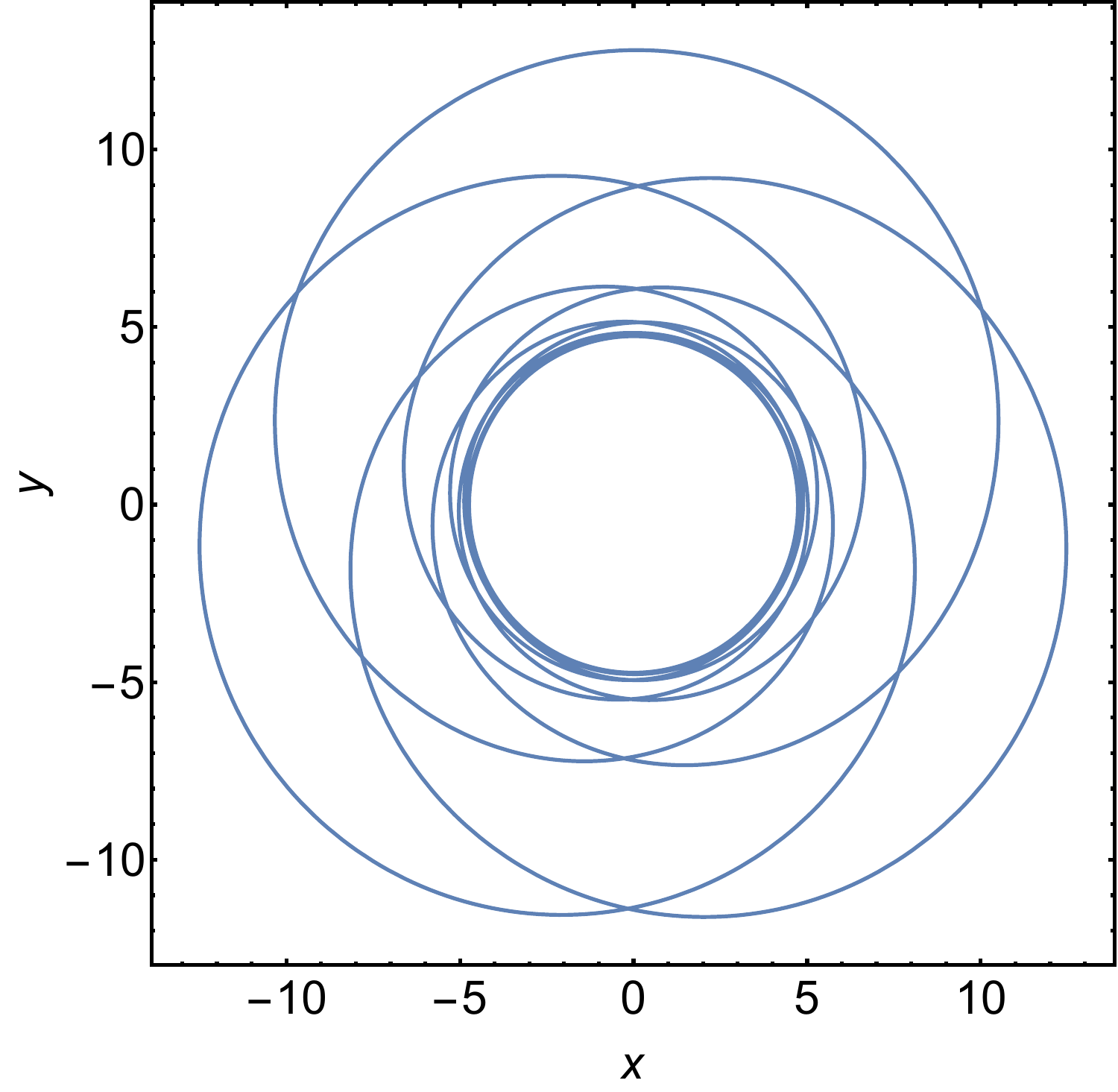}}
\hspace{0.75cm}
\subfigure[$L = 3.5564$, \, $(4, 1, 3)$] 
{\label{La7}\includegraphics[width=4.75cm]{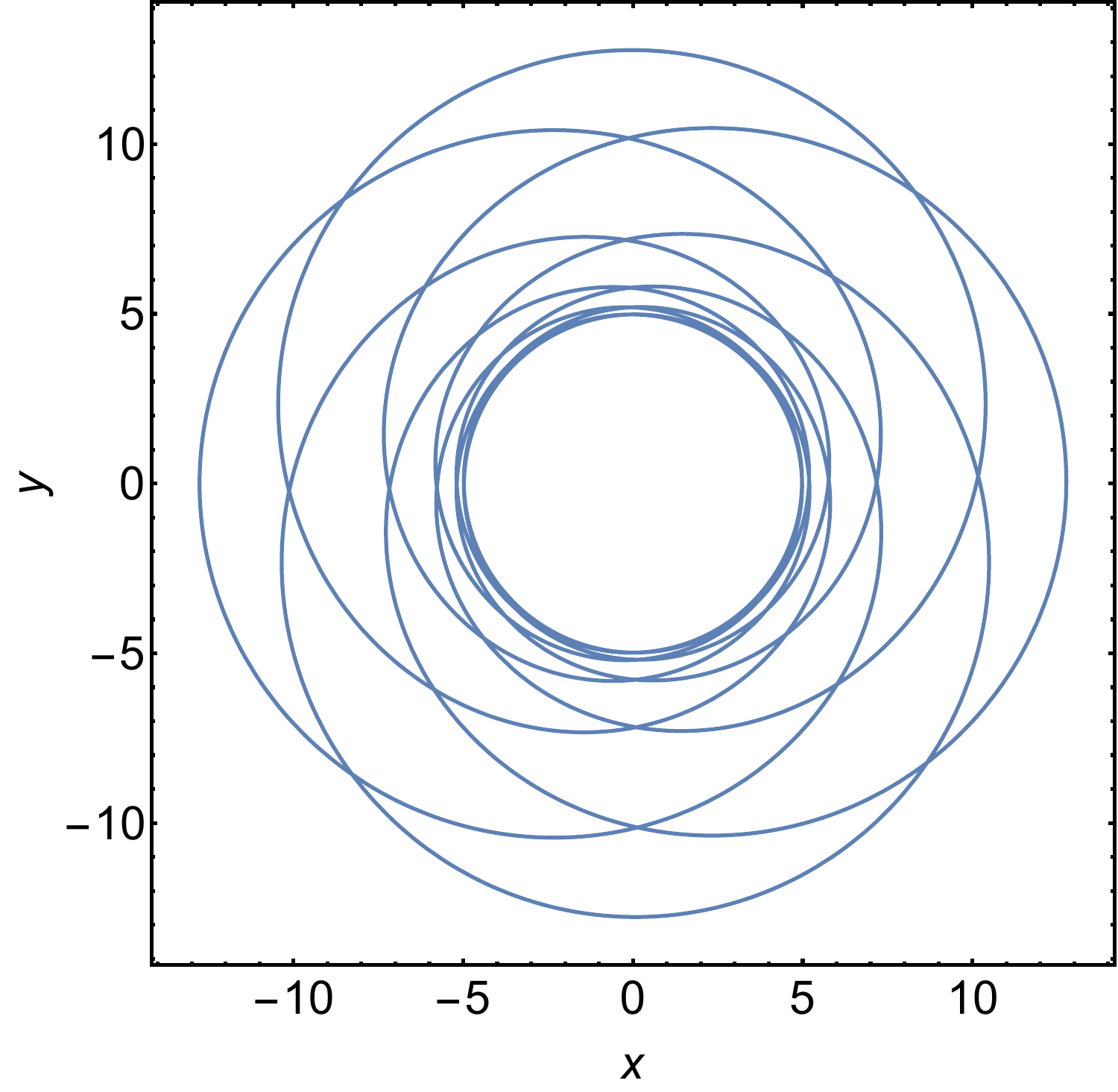}}
\hspace{0.75cm}
\subfigure[$L = 3.5539$, \, $(4, 2, 3)$] 
{\label{La8}\includegraphics[width=4.75cm]{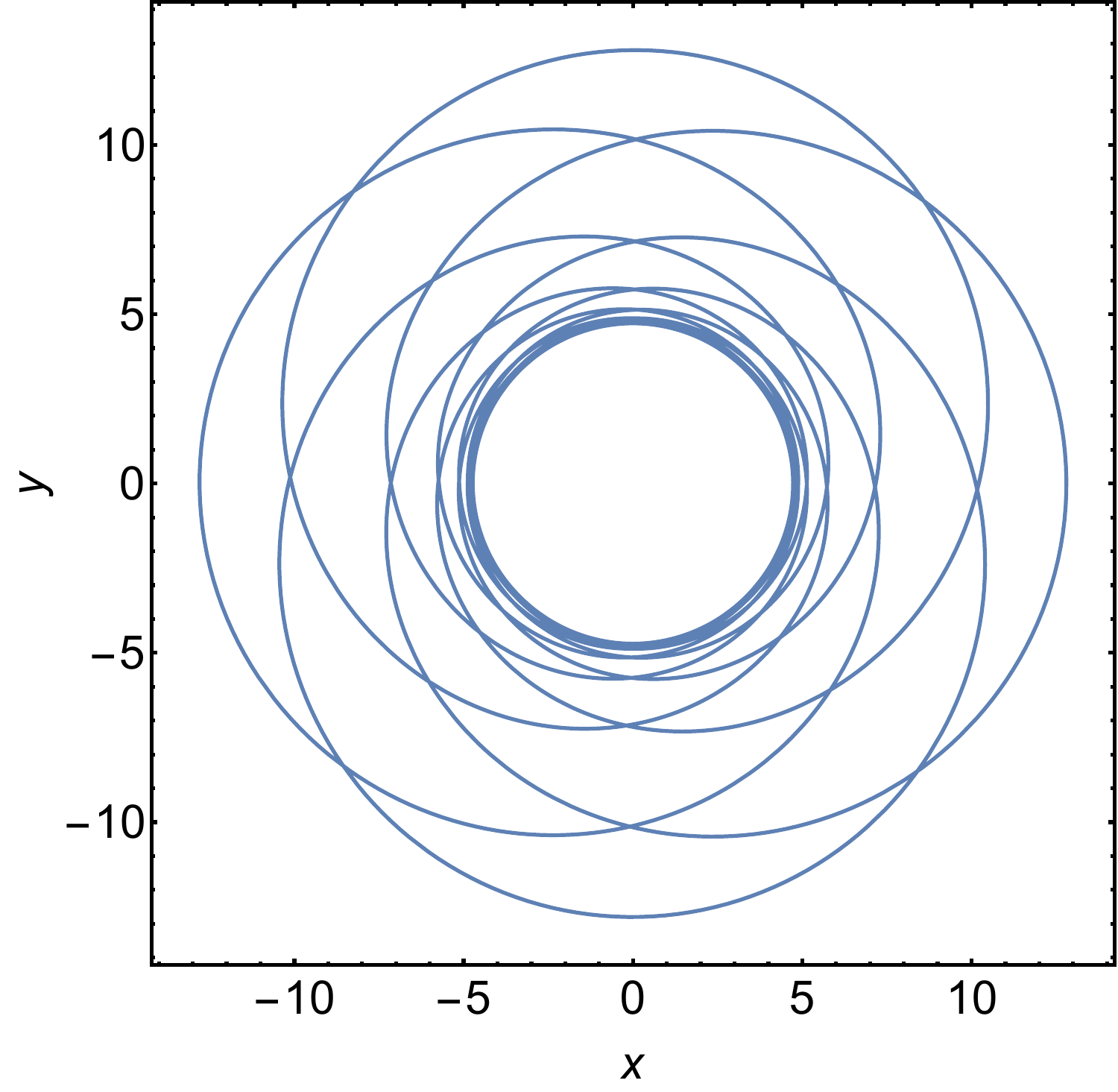}}
\hspace{0.75cm}
\subfigure[$L = 3.5544 $, \, $(5, 2, 1)$] 
{\label{La9}\includegraphics[width=4.75cm]{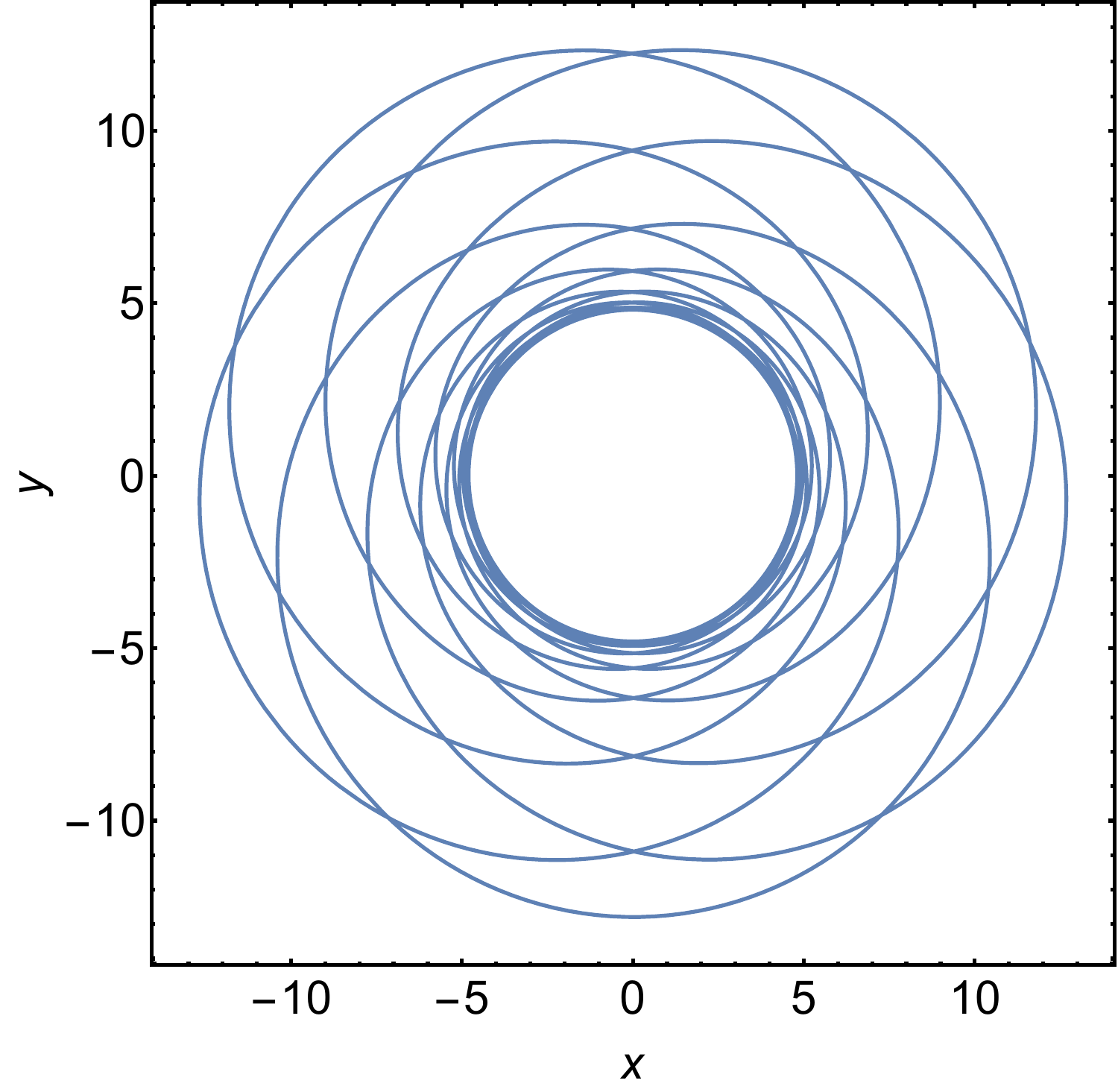}}
\caption{Periodic orbits for different values of $L$ and $(z, w, v)$ with $E=0.96$ and $\alpha=0.8$ $(l=0.011746)$ in \eqref{lcontrol}. Here ${\rm x}$ and ${\rm y}$ have units of meter $\left[m\right]$.}\label{Fig8}
\end{figure*}

In Fig.\,\ref{Fig7} and Fig.\,\ref{Fig8} we show the behaviour of the periodic orbits for a fixed energy $E=0.96$, which according to Fig.\,\ref{ELb} is contained within a certain interval $L$, taking values for $\alpha=0.7$ and $\alpha=0.8$ in Eq.\,\eqref{lcontrol}. Comparing the result, we see that for $l<0$ the orbits have higher momentum values compared to the orbits of the same taxonomy $(z, w, v)$ for $l>0$ and, therefore, a higher eccentricity.

\begin{figure*}[htb!]
\centering
\subfigure[$E = 0.9510$, \,$(1, 1, 0)$] 
{\label{Eb1}\includegraphics[width=4.75cm]{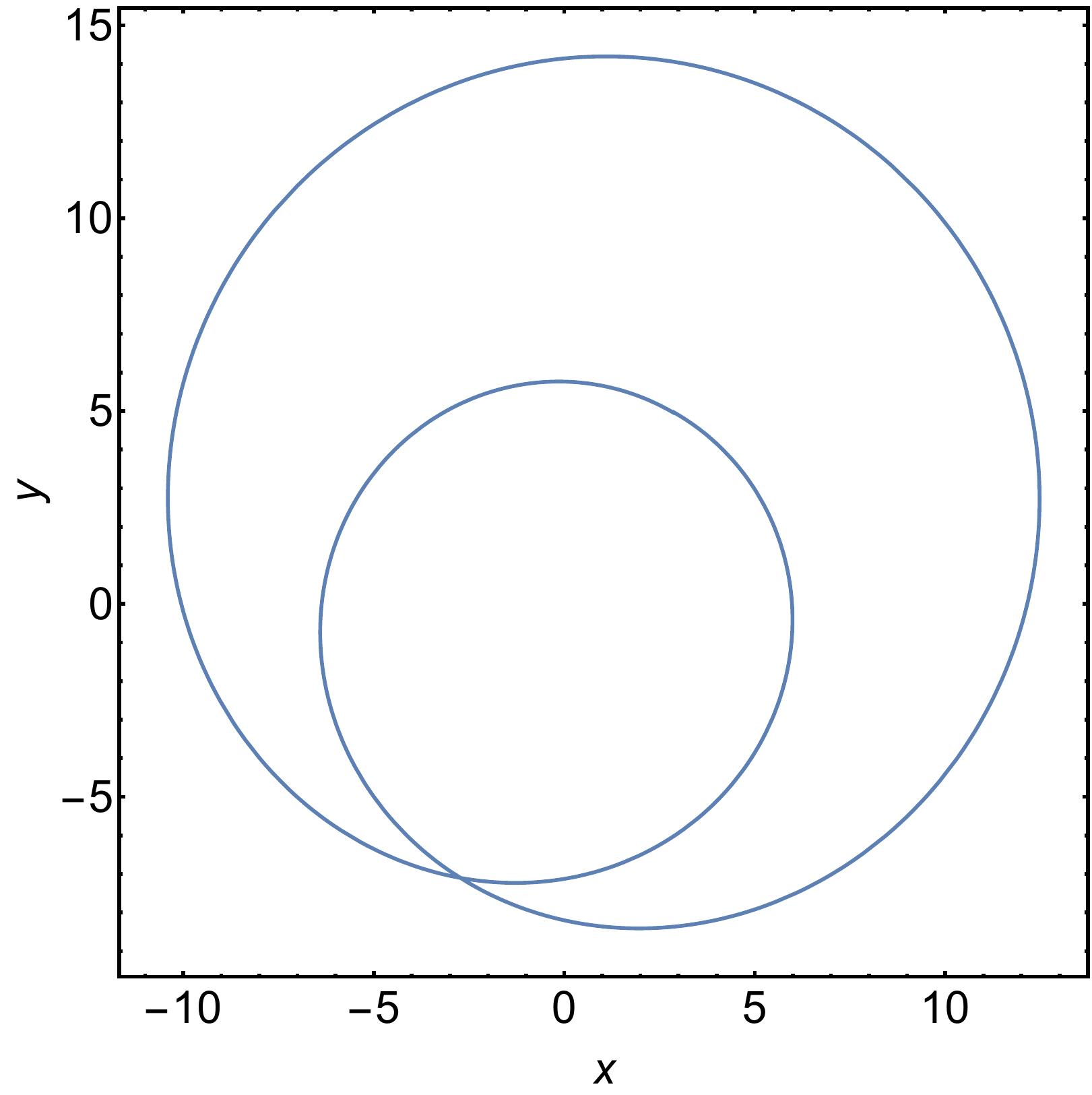} }
\hspace{0.75cm}
\subfigure[  $E = 0.9538$, \,$(1, 2, 0)$] 
{\label{Eb2}\includegraphics[width=4.75cm]{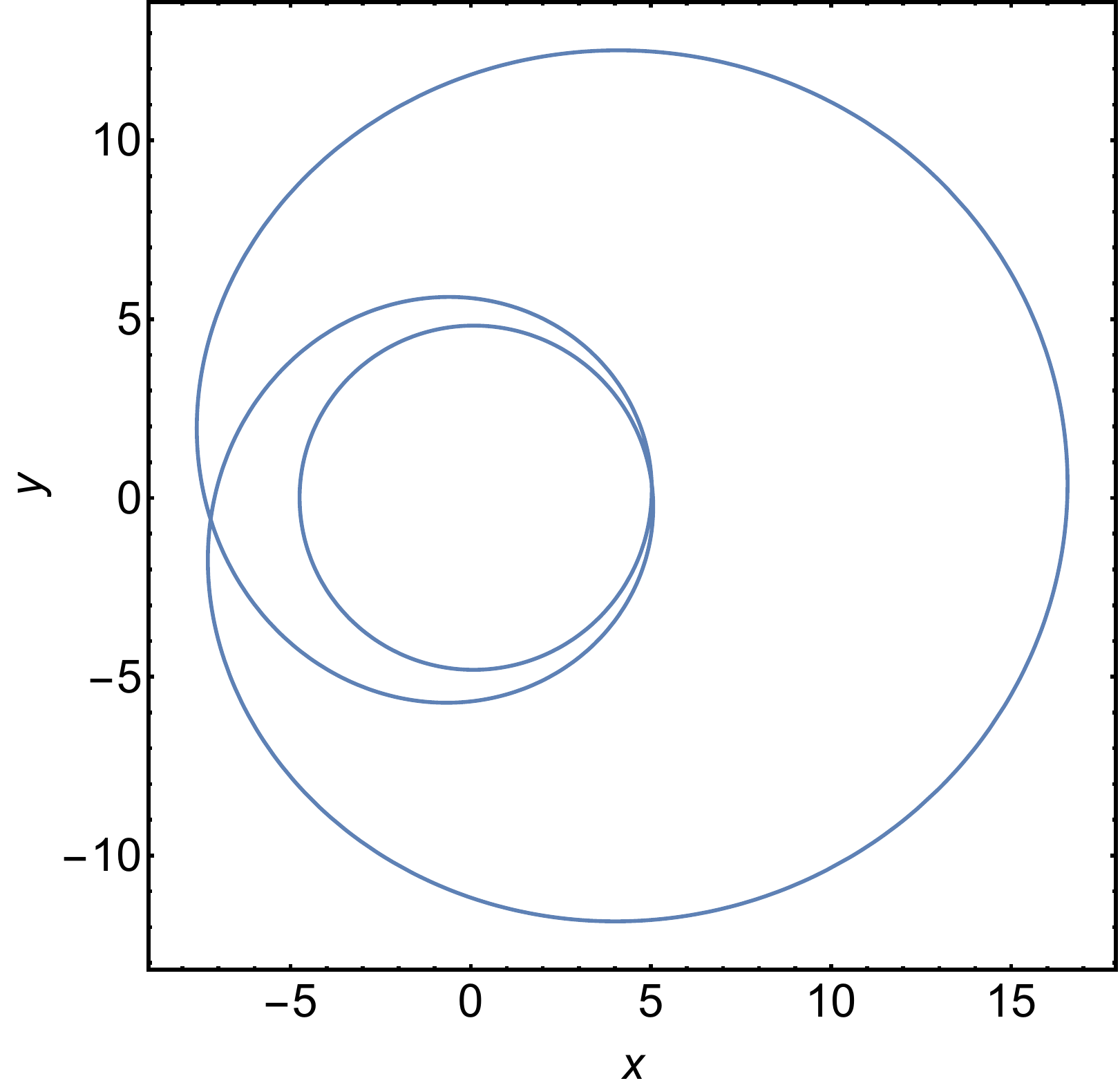} }
\hspace{0.75cm}
\subfigure[ $E = 0.9534$, \,$(2, 1, 1)$] 
{\label{Eb3}\includegraphics[width=4.75cm]{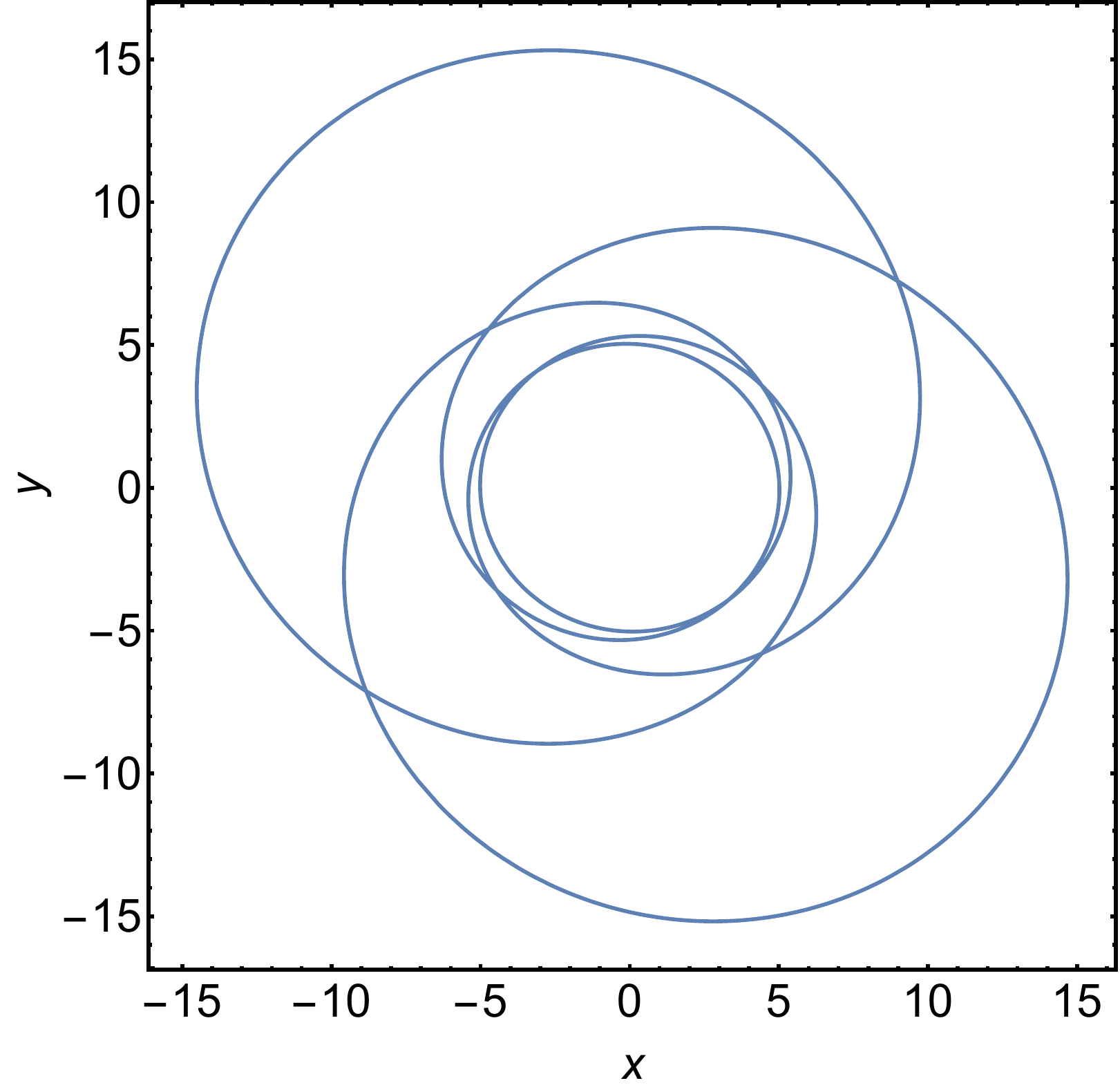}}
\hspace{0.75cm}
\subfigure[$E = 0.9539$, \, $(2, 2, 1)$] 
{\label{Eb4}\includegraphics[width=4.75cm]{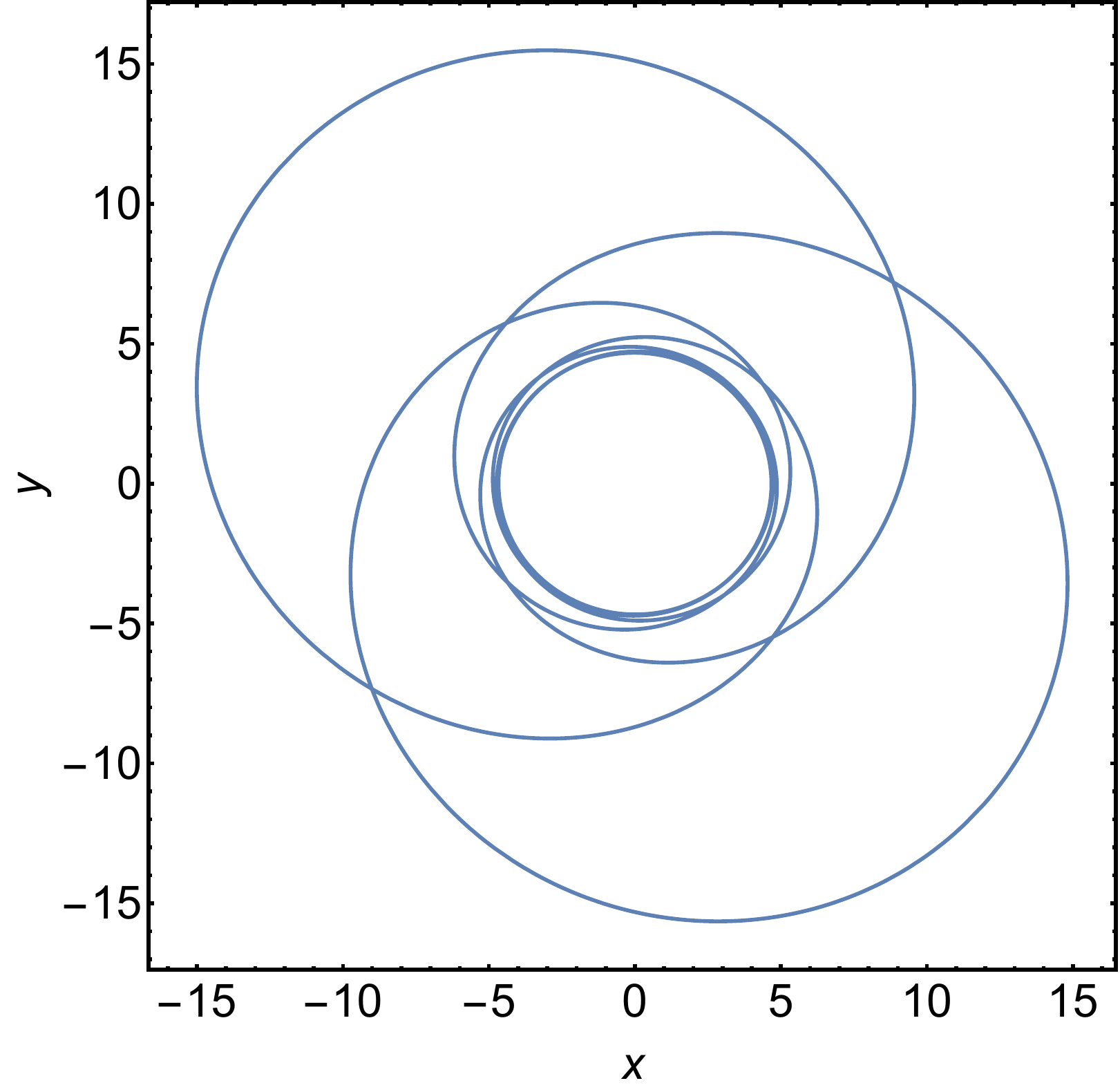}}
\hspace{0.75cm}
\subfigure[$E = 0.9536$, \, $(3, 1, 2)$] 
{\label{Eb5}\includegraphics[width=4.75cm]{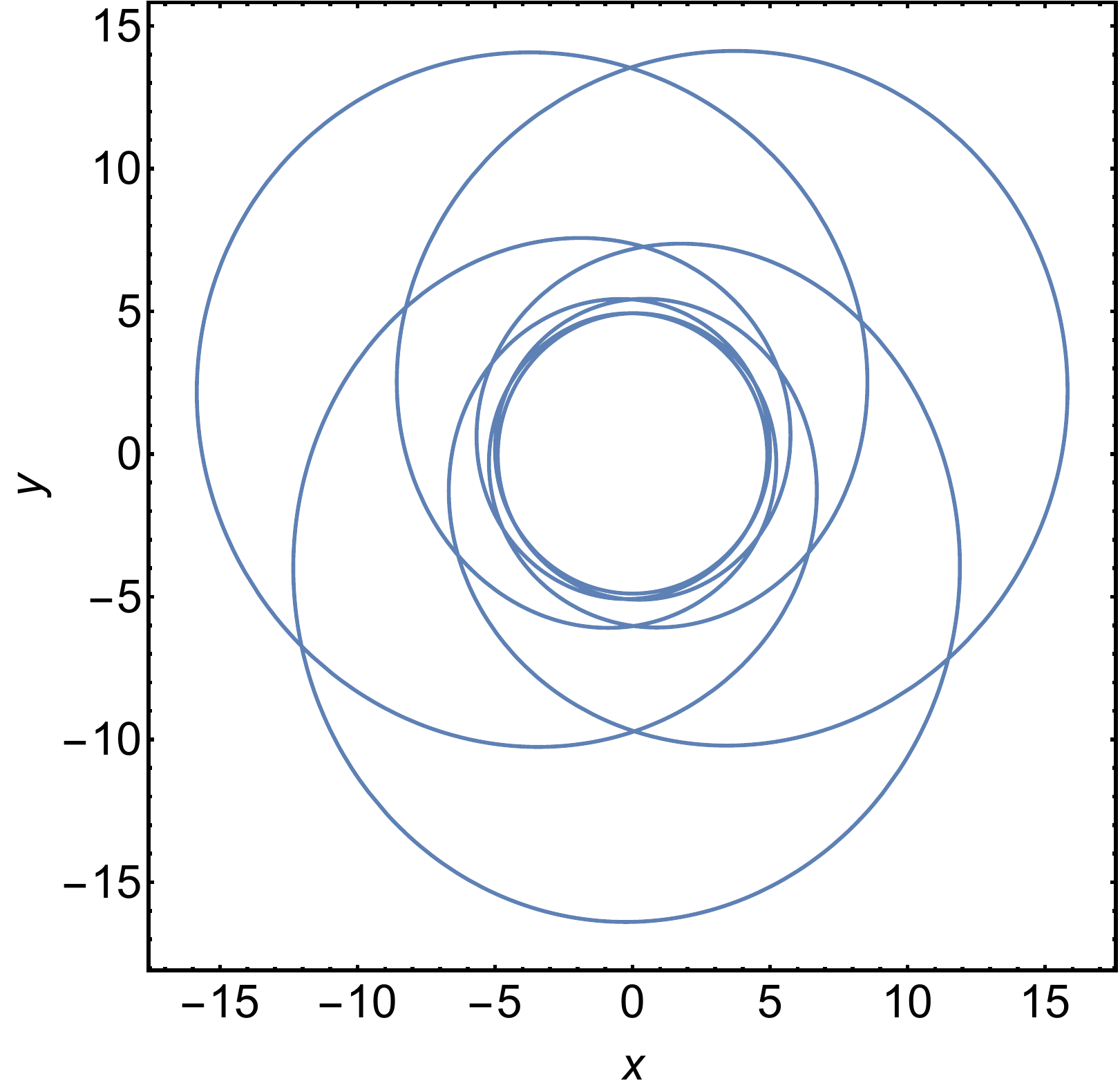}}
\hspace{0.75cm}
\subfigure[$E = 0.95393$, \, $(3, 2, 2)$] 
{\label{Eb6}\includegraphics[width=4.75cm]{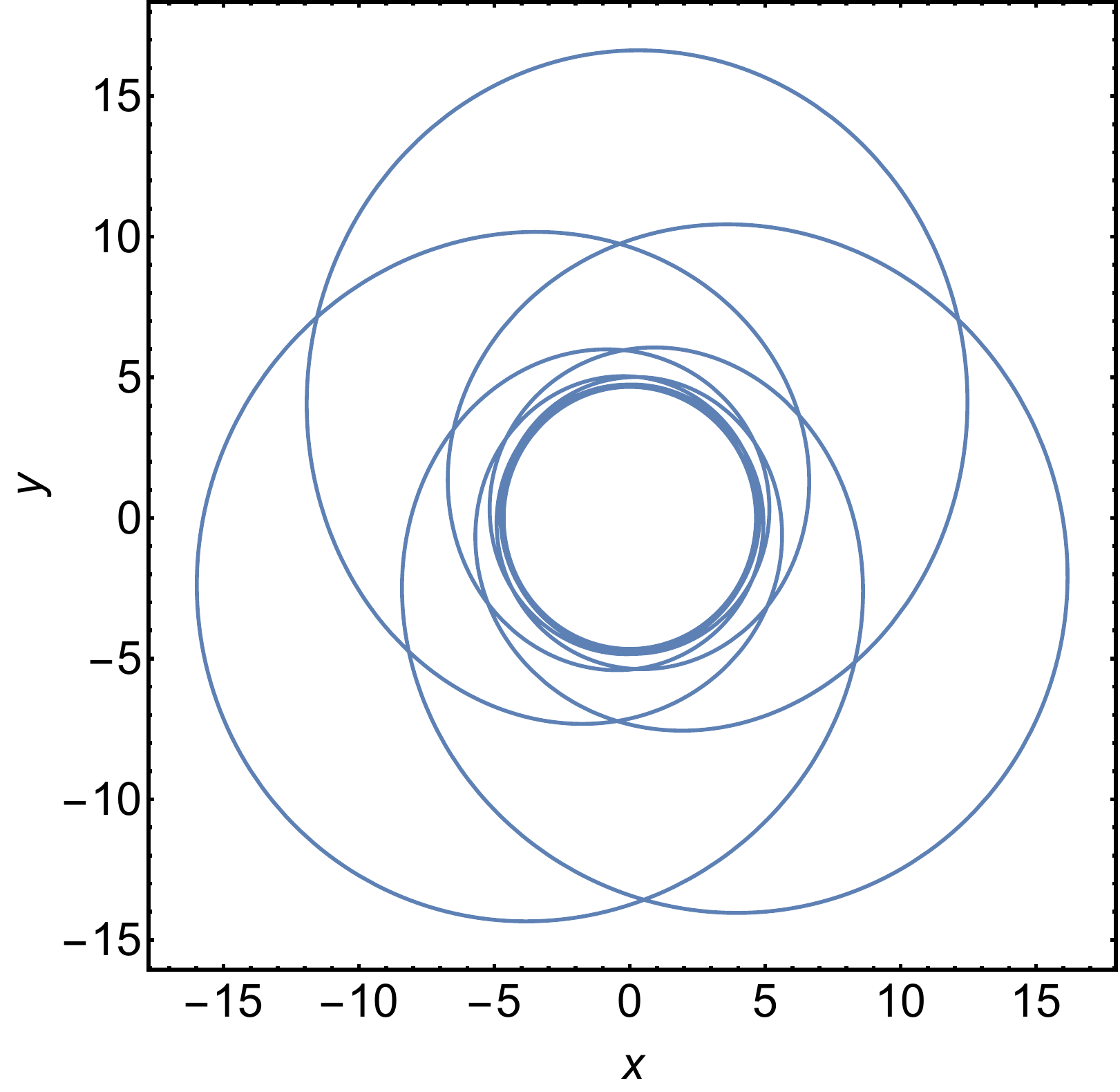}}
\hspace{0.75cm}
\subfigure[$E = 0.9537$, \, $(4, 1, 3)$] 
{\label{Eb7}\includegraphics[width=4.75cm]{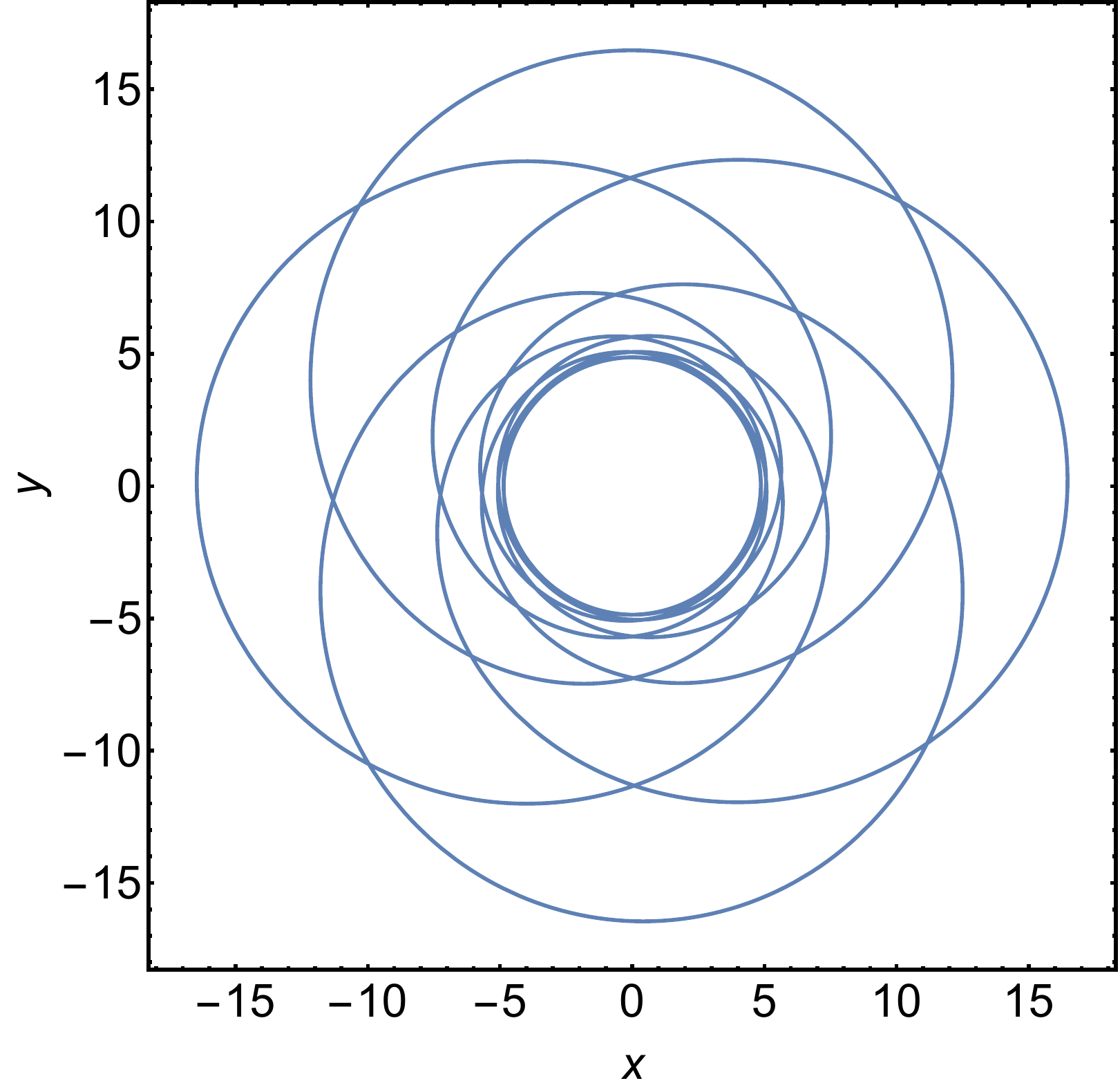}}
\hspace{0.75cm}
\subfigure[$E = 0.953936$, \, $(4, 2, 3)$] 
{\label{Eb8}\includegraphics[width=4.75cm]{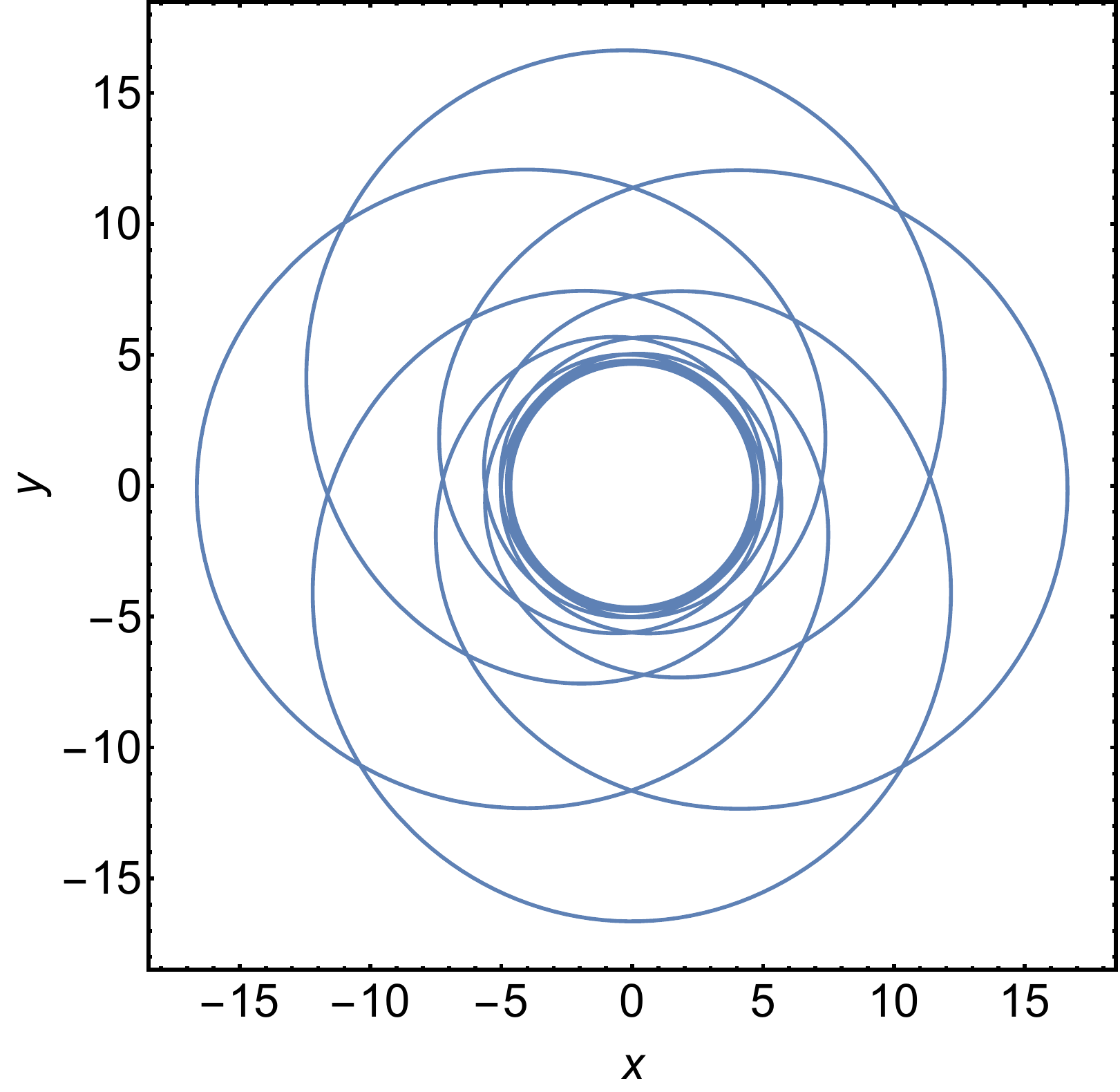}}
\hspace{0.75cm}
\subfigure[$E = 0.95389 $, \, $(5, 2, 1)$] 
{\label{Eb9}\includegraphics[width=4.75cm]{Ea8.pdf}}
\caption{Periodic orbits for different values of $E$ and $(z, w, v)$ with $L=3.7$ and $\alpha=0.7$ $(l=-0.01285)$ in Eq.\eqref{lcontrol}. Here ${\rm x}$ and ${\rm y}$ have units of meter $\left[m\right]$. }\label{Fig9}
\end{figure*}

\begin{figure*}[htb!]
\centering
\subfigure[$E = 0.9725$, \,$(1, 1, 0)$] 
{\label{Ea1}\includegraphics[width=4.75cm]{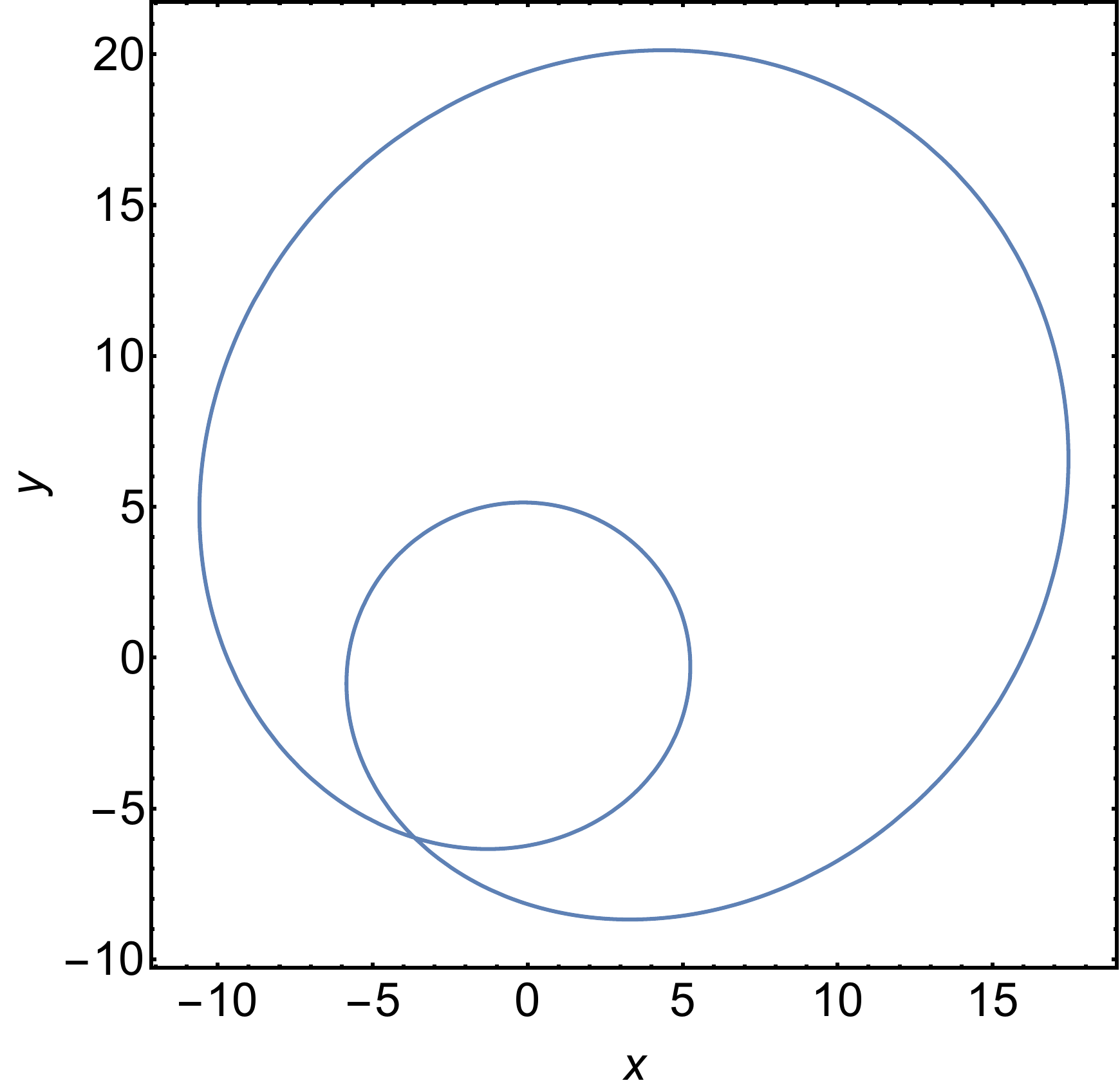} }
\hspace{0.75cm}
\subfigure[  $E = 0.97544$, \,$(1, 2, 0)$] 
{\label{Ea2}\includegraphics[width=4.75cm]{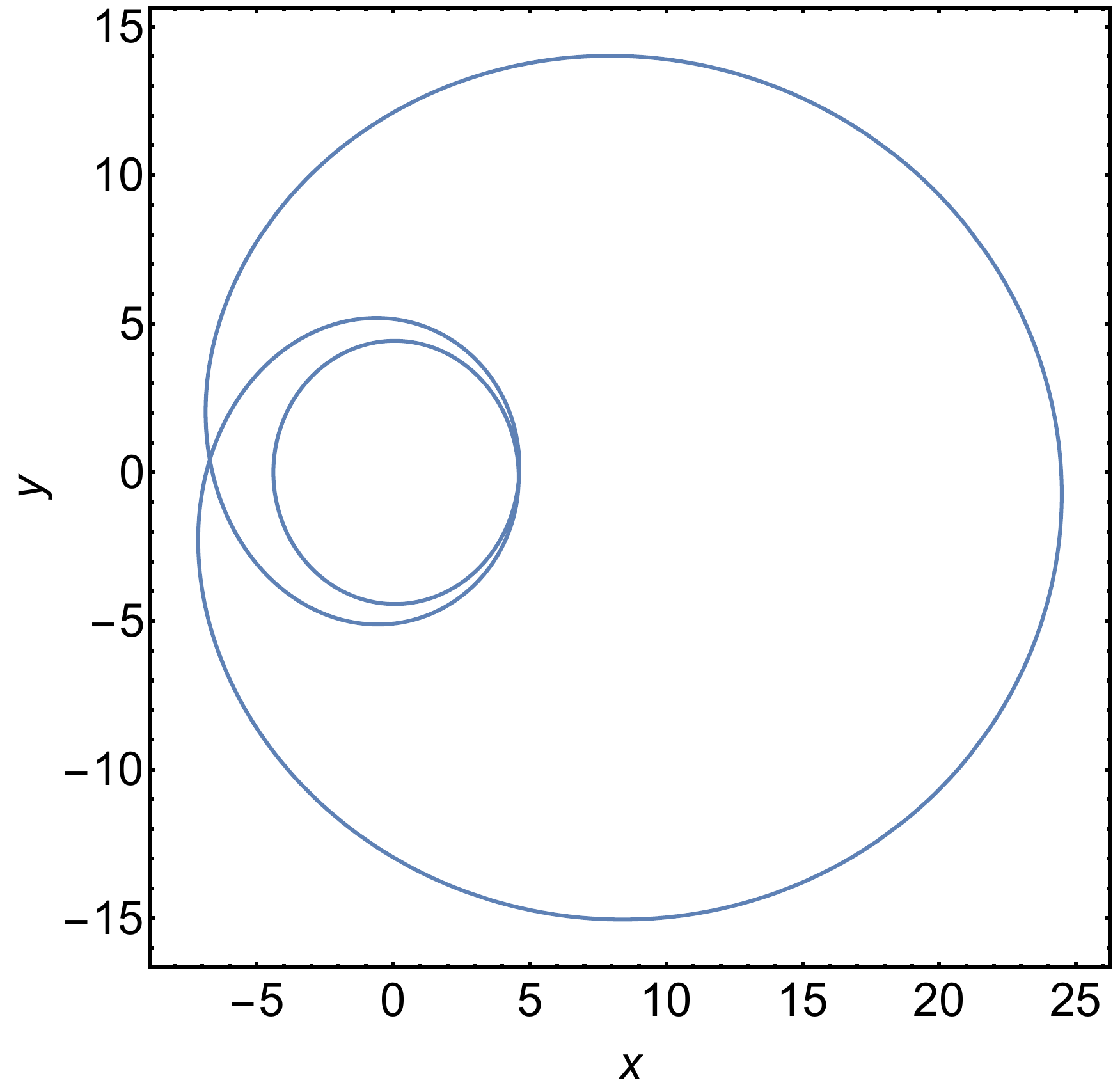} }
\hspace{0.75cm}
\subfigure[ $E = 0.9751$, \,$(2, 1, 1)$] 
{\label{Ea3}\includegraphics[width=4.75cm]{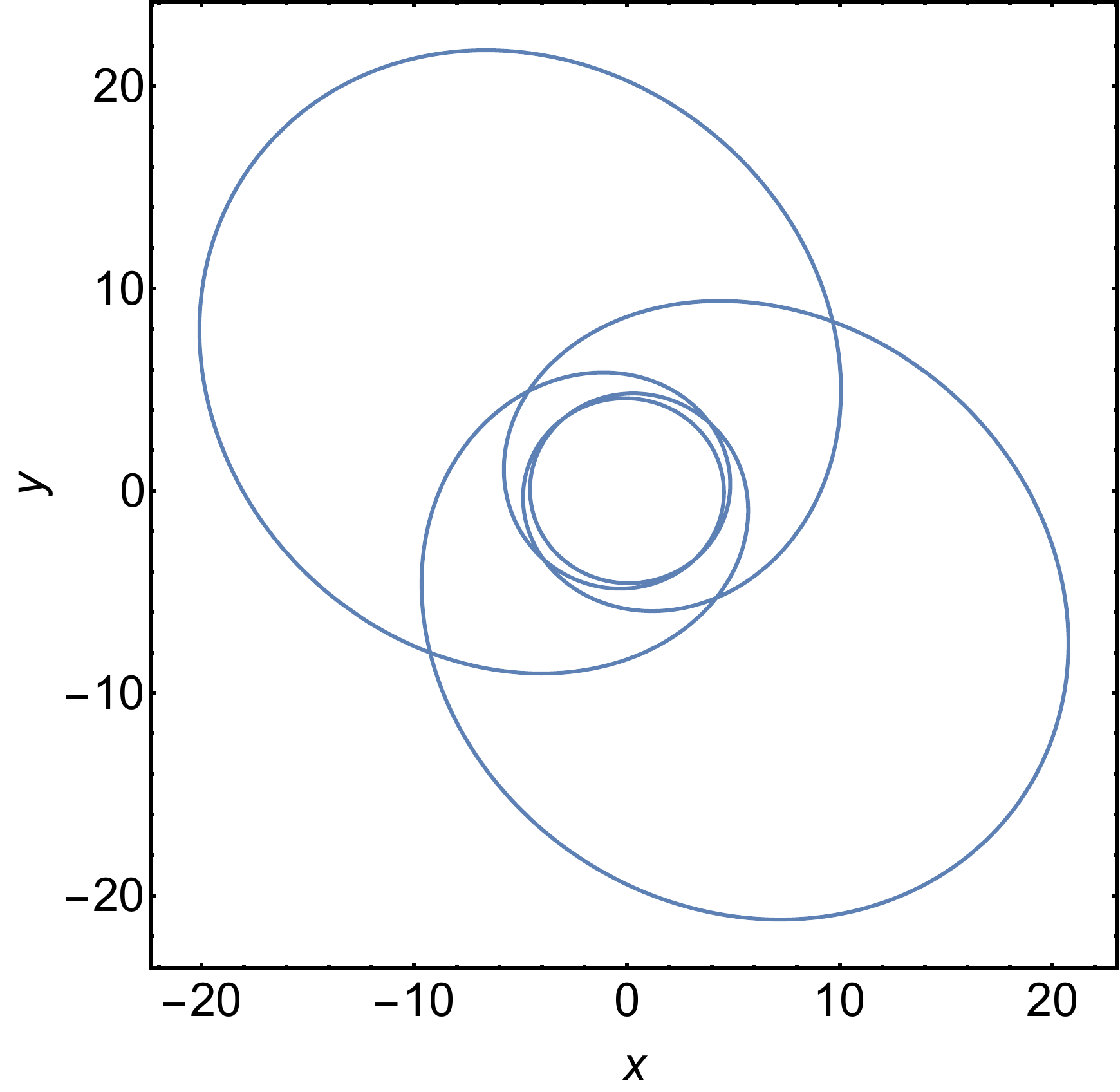}}
\hspace{0.75cm}
\subfigure[$E = 0.97549$, \, $(2, 2, 1)$] 
{\label{Ea4}\includegraphics[width=4.75cm]{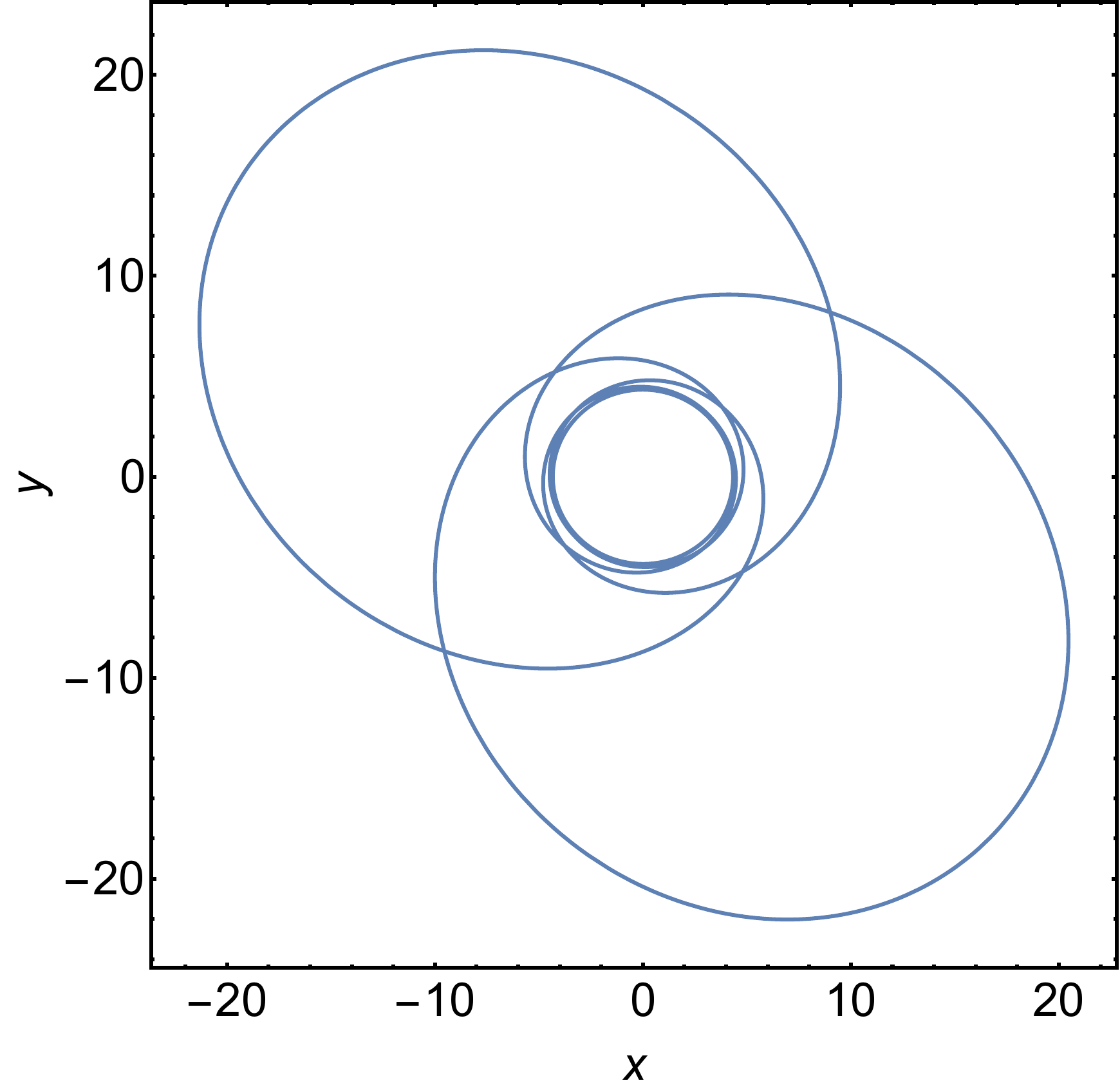}}
\hspace{0.75cm}
\subfigure[$E = 0.9753$, \, $(3, 1, 2)$] 
{\label{Ea5}\includegraphics[width=4.75cm]{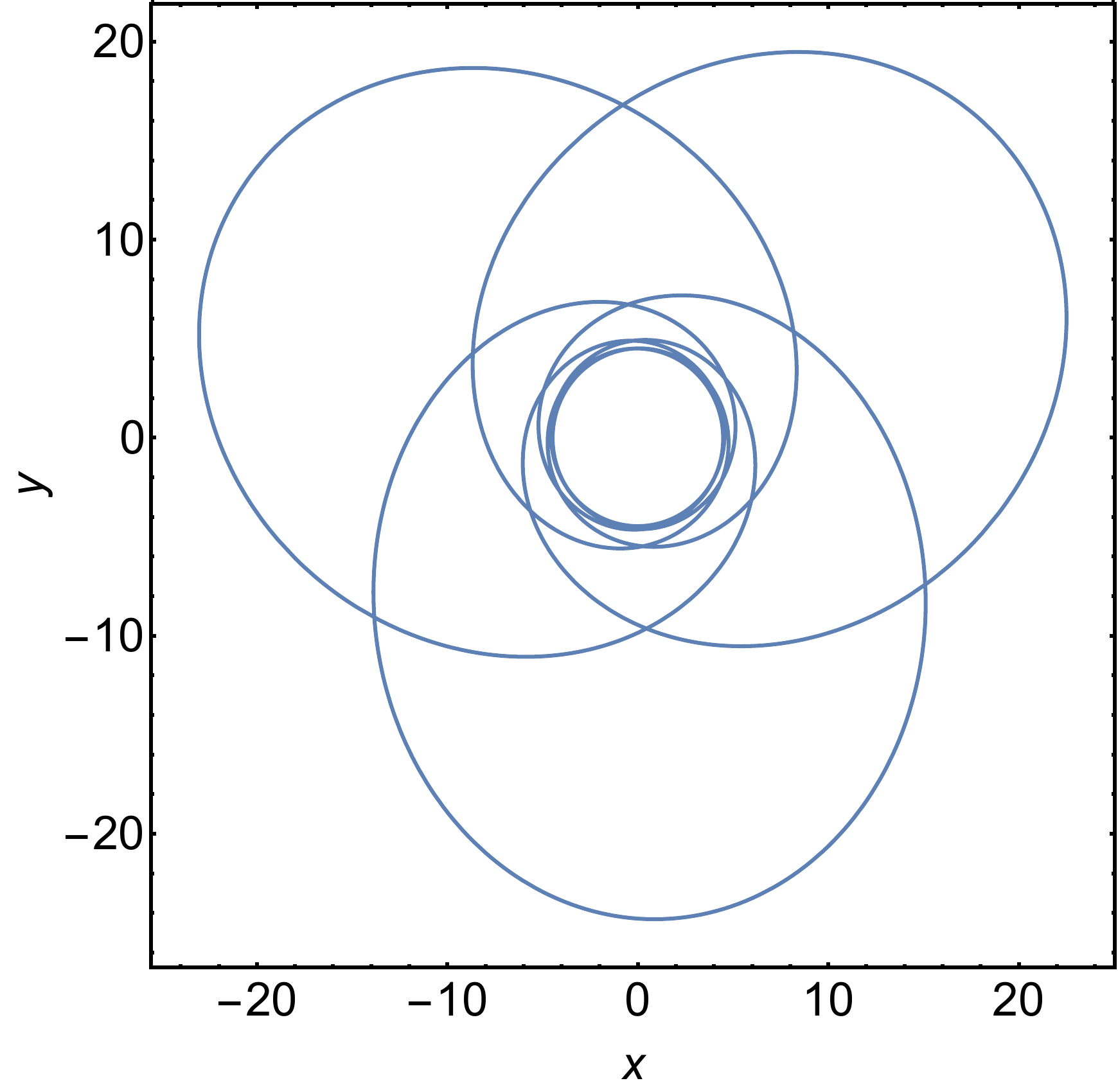}}
\hspace{0.75cm}
\subfigure[$E = 0.975495$, \, $(3, 2, 2)$] 
{\label{Ea6}\includegraphics[width=4.75cm]{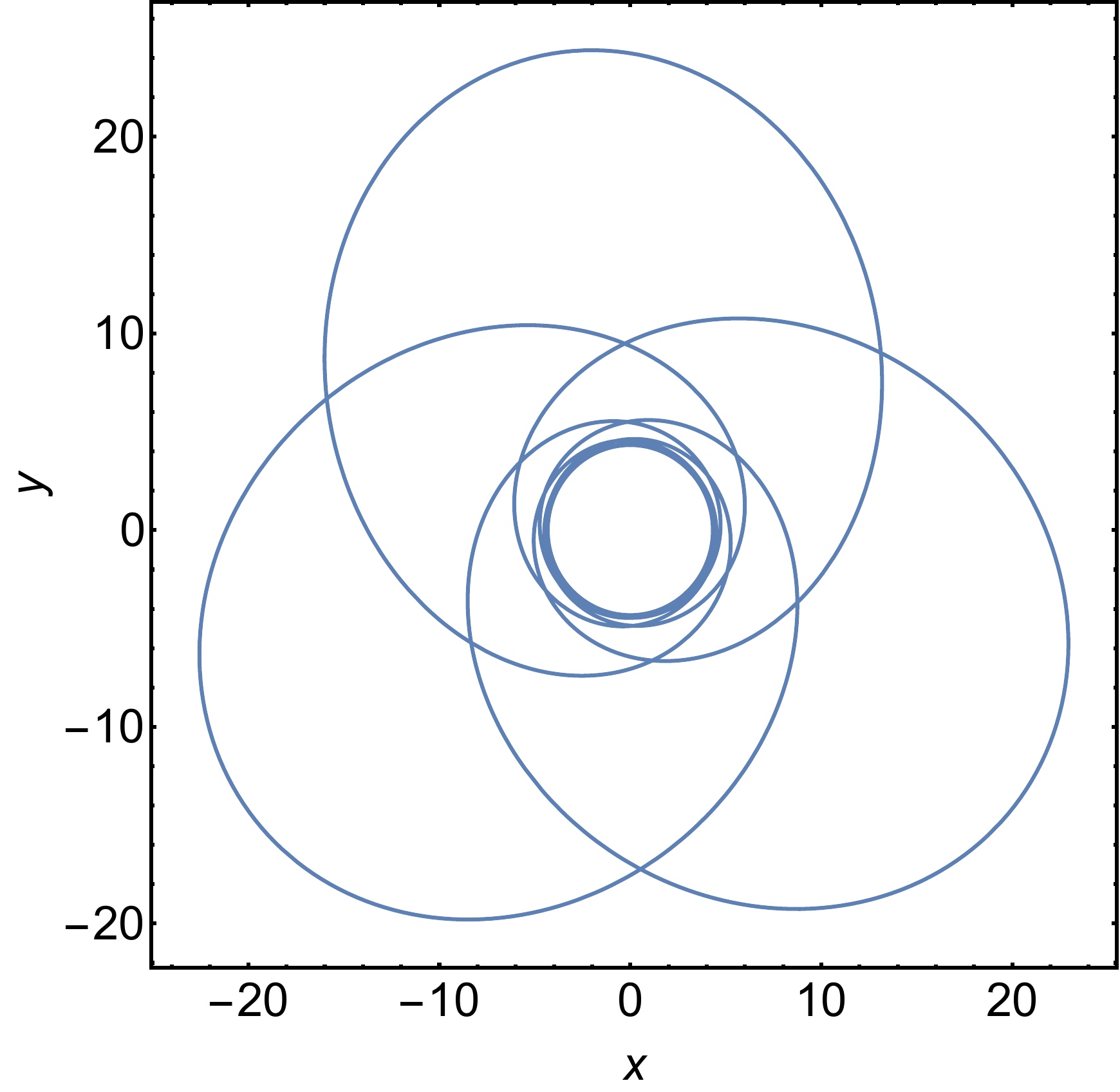}}
\hspace{0.75cm}
\subfigure[$E = 0.97535$, \, $(4, 1, 3)$] 
{\label{Ea7}\includegraphics[width=4.75cm]{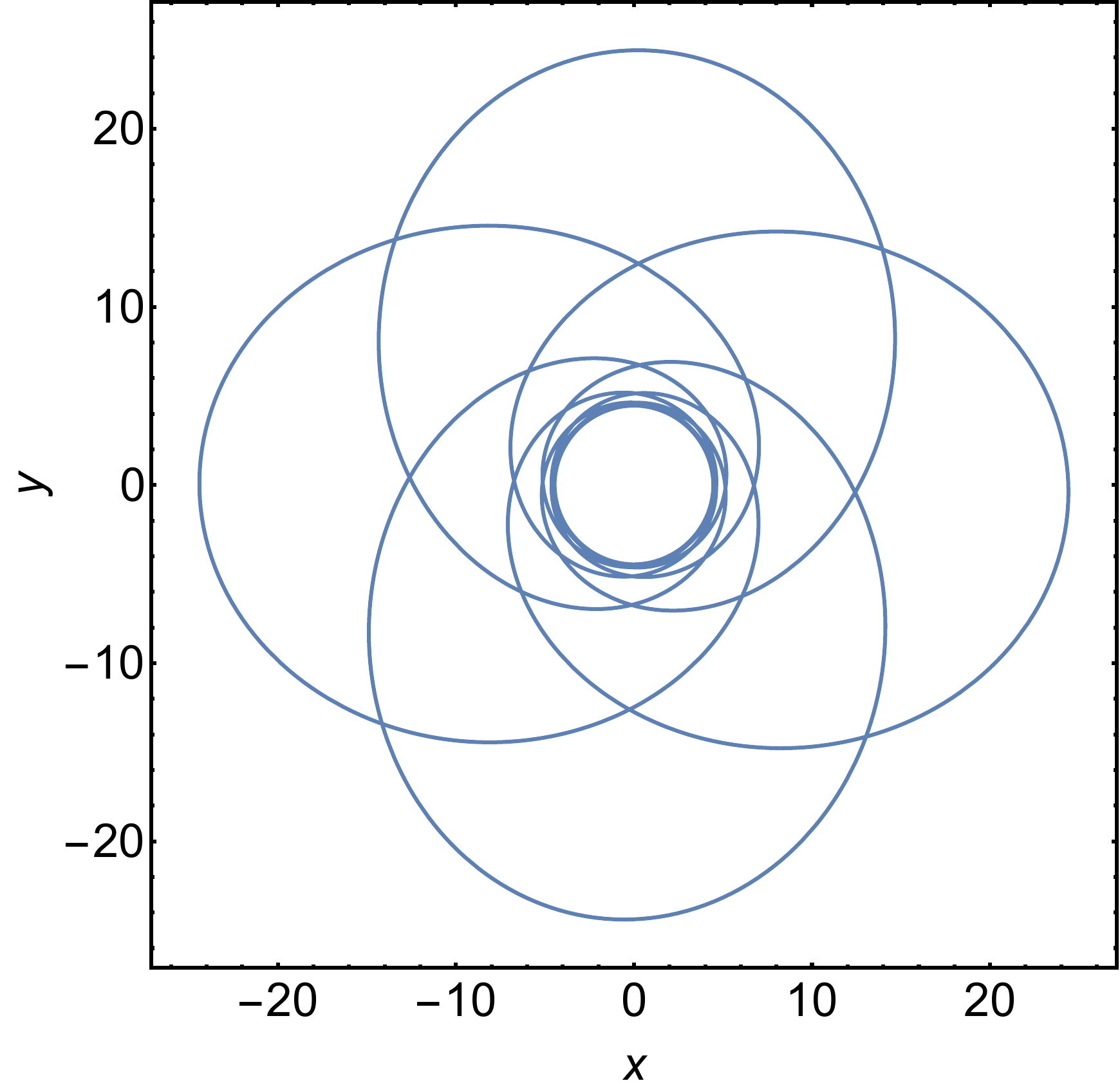}}
\hspace{0.75cm}
\subfigure[$E = 0.975496$, \, $(4, 2, 3)$] 
{\label{Ea8}\includegraphics[width=4.75cm]{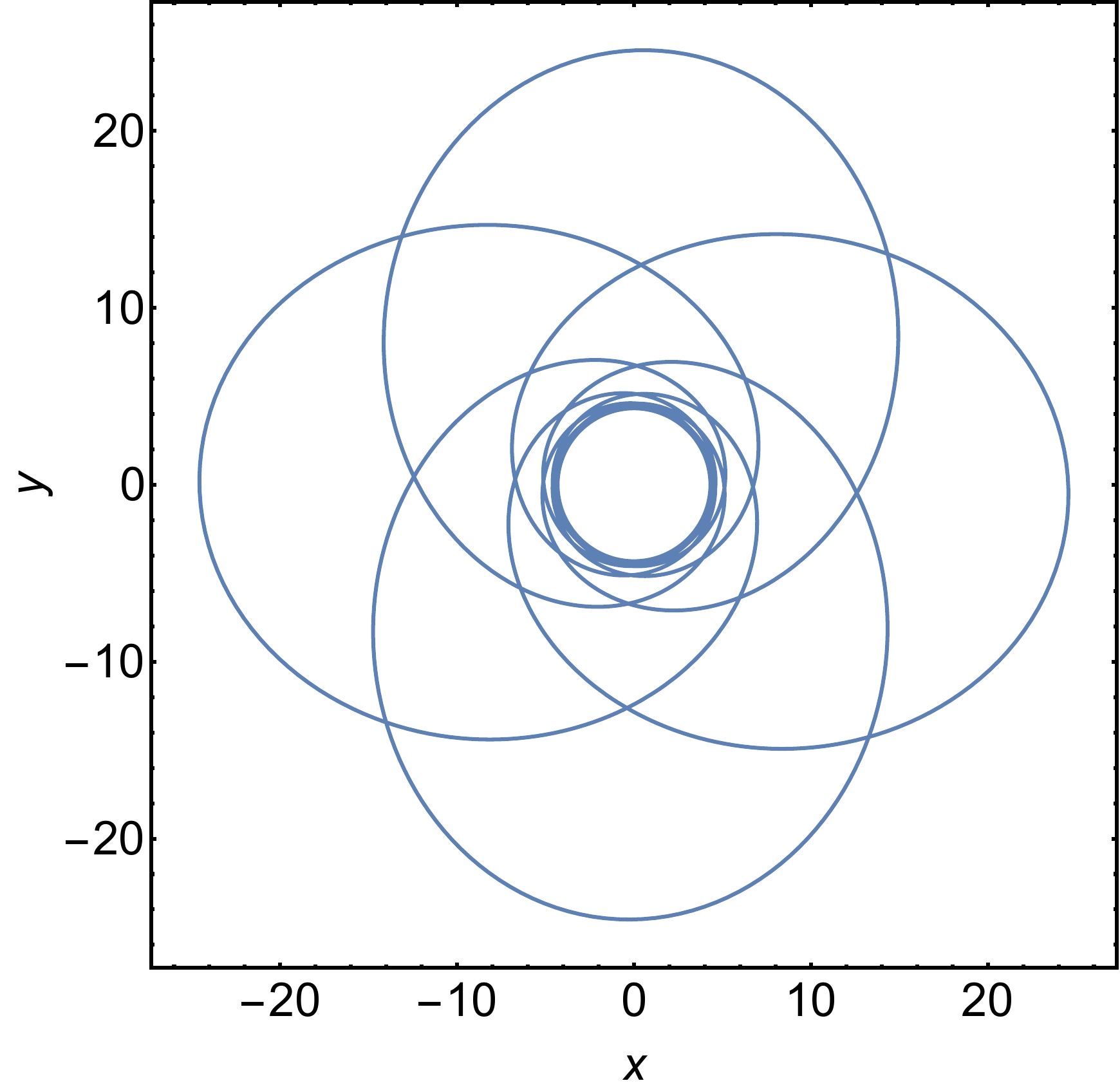}}
\hspace{0.75cm}
\subfigure[$E = 0.97547 $, \, $(5, 2, 1)$] 
{\label{Ea9}\includegraphics[width=4.75cm]{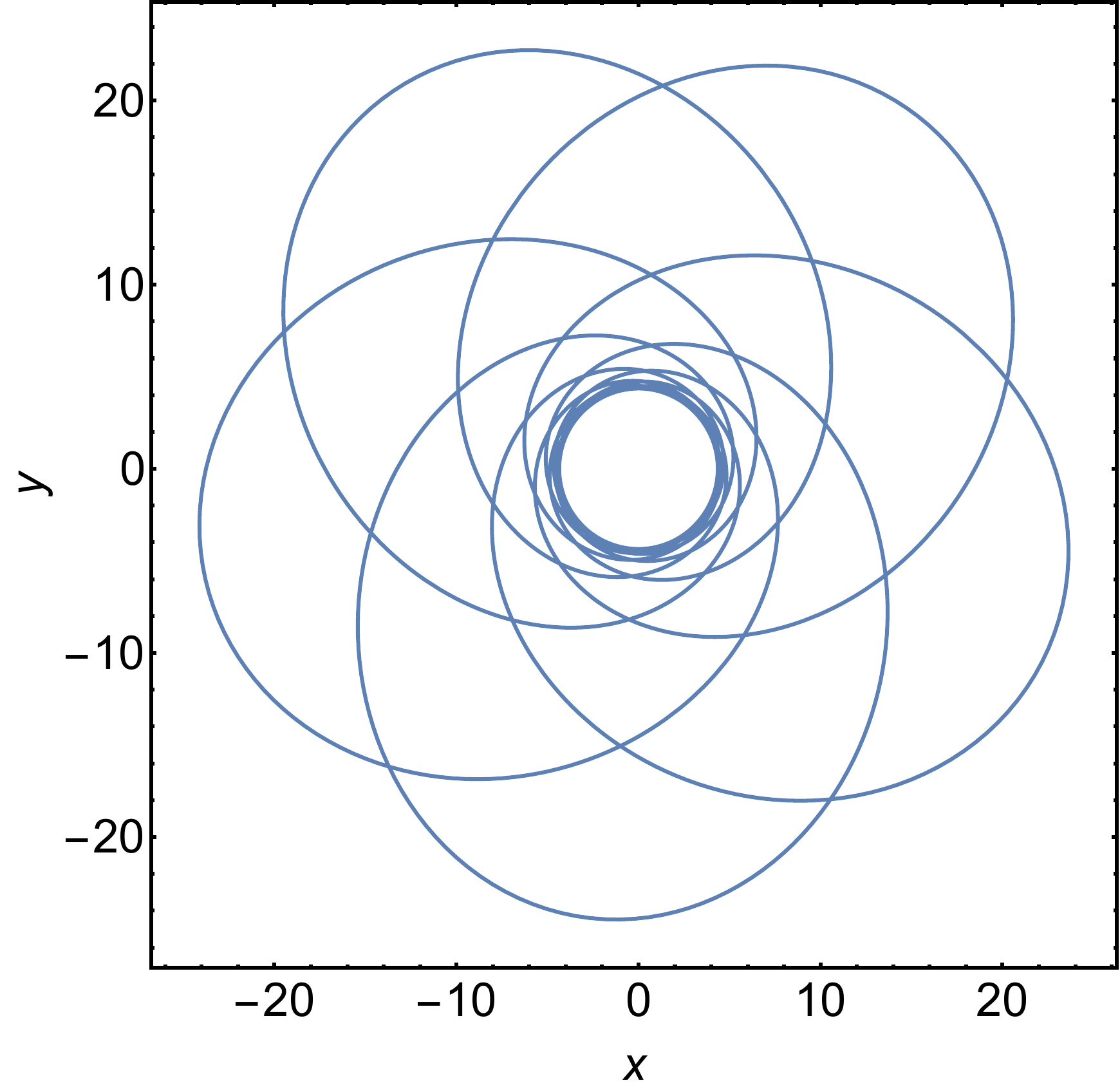}}
\caption{Periodic orbits for different values of $E$ and $(z, w, v)$ with $L=3.7$ and $\alpha=0.8$ $(l=0.011746)$ in \eqref{lcontrol}. Here ${\rm x}$ and ${\rm y}$ have units of meter $\left[m\right]$. }\label{Fig10}
\end{figure*}

In Fig.\,\ref{Fig9} and Fig.\,\ref{Fig10} we show the behaviour of periodic orbits for fixed energy $L=3.7$ which, according to Fig.\,\ref{ELb}, is contained in the suitable range for $E$, taking values for $\alpha=0.7$ and $\alpha=0.8$ in Eq.\,\eqref{lcontrol}. For $l<0$, the orbits have lower energy values compared to orbits of the same taxonomy $(z, w, v)$ for $l>0$ and therefore lower eccentricity.

We point out that the structure of the periodic orbits is the same if we consider the case of Schwarzsichild, but with different amounts of momentum and energy, which can be compared with the results obtained in \cite{TAX}. However, the difference between them can be identified in the shape of the gravitational wave emitted, as we will see in the next section.

\section{GRAVITATIONAL WAVES FROM EMRI: numerical results} \label{secGW}

Encoding the properties of periodic orbits by GWs is fundamental to understanding and studying compact objects \cite{livroPoissonE}. GW signals from EMRIs can be detected by LISA \cite{LISA}, where a compact stellar-mass object orbiting a supermassive black hole is a prominent source of low-frequency gravitational waves. However, such orbits will slowly but steadily lose energy and angular momentum and, therefore, decay towards the black hole, which we consider here is described by the KR metric \eqref{KR0}. We assume the black hole mass to be that of Sgr A* whereas the compact object has a mass of the order of those of the S-stars orbiting it. Furthermore, we shall consider the adiabatic approximation method whose reliability to study the emission of GWs from EMRI systems has been studied elsewhere \cite{Adiabatic, Adiabatic2, Adiabatic3}. In this approach, the energy and orbital angular momentum of the lower-mass object decay very slowly compared to the total energy of the system and its orbital period, which can be considered constant assuming that the orbit is in good geodesic approximation for a certain period and neglecting the influence of gravitational radiation on the motion of the lower-mass object.  

With the results obtained in the previous section regarding the orbits via Eqs.\eqref{rpontoquad} and \eqref{curvaq2}, we use the Kludge \cite{Kludge} method to obtain the GWs  emitted by  periodic orbits in the KR black hole. The method consists of taking the gravitational quadrupole relation to obtain of gravitational waveform up to the quadratic order (for details see Ref.\cite{livroPoissonE})
\begin{eqnarray}
h_{ij}=\frac{4\beta M}{D_L}\left(v_iv_j-\frac{m}{r}n_in_j\right)\,,\label{hij}
\end{eqnarray}
where $M$ is the mass of the supermassive black hole, $m$ is the mass of the stellar mass object, $\beta=Mm/(M+m)^2$ and $D_L$ are the mass ratio and the luminosity distance of the EMRI system, respectively, while $v_{i,j}$ and $n_{i,j}$ are the space velocity and the unit vector of the radial motion of the smaller mass object, respectively. Taking a coordinate system adapted to the gravitational wave detector coinciding with the original coordinate center of the black hole $(x, y, z)$ such that these adapted coordinates are given, as described in \cite{livroPoissonE}, by
\begin{eqnarray}
\mathbf{e}_X&=&[\cos \zeta, -\sin \zeta, 0]\,,\nonumber\\
\mathbf{e}_Y&=&[\cos \iota\sin\zeta, \cos\iota\cos\zeta, -\sin\iota]\,,\nonumber\\
\mathbf{e}_Z&=&[\sin\iota\sin\zeta, \sin\iota\cos\zeta, \cos\iota]\,,
\end{eqnarray}
where $(X, Y, Z)$ are  adapted coordinates of the detector centered on the black hole, $\zeta$ is the latitude of the pericenter measured in the $X-Y$ orbital plane of  smaller object and $\iota$ is the inclination angle of the orbital plane. We can then project Eq.\eqref{hij} onto this adapted coordinate system and obtain the polarization components as
\begin{eqnarray}
h_+&=&-\frac{2\beta M^2}{D_Lr}\left(1+\cos^2\iota\right)\cos\left(2\phi+2\zeta\right)\,,\label{h+} \\
h_\times &=&-\frac{4\beta M^2}{D_Lr}\cos\iota\sin\left(2\phi+2\zeta\right)\,,
\end{eqnarray}
where $\phi$, with implicit dependence on  the affine parameter  $\tau$, is the phase angle linked to the orbital phase, which can be obtained from the geodesic equation \eqref{rponto} after the variable transformation $r(\phi)=1/u(\phi)$.

The adiabatic approximation used to calculate the polarization components $h_+$ and $h_\times$ of the gravitational wave is valid when the radiation reaction takes place over a time scale much longer than the orbital period \cite{Cutler}, which occurs in our model. Furthermore, the adiabatic approximation has been used in conjunction with KR \cite{Bernardo} and bumblebee \cite{Adiabatic} fields, which violate Lorentz symmetry, and studies on gravitational waves are conducted with the KR field \cite{KRparity}.

\begin{figure*}[htb!]
\centering
{\label{rEa}\includegraphics[width=15.75cm]{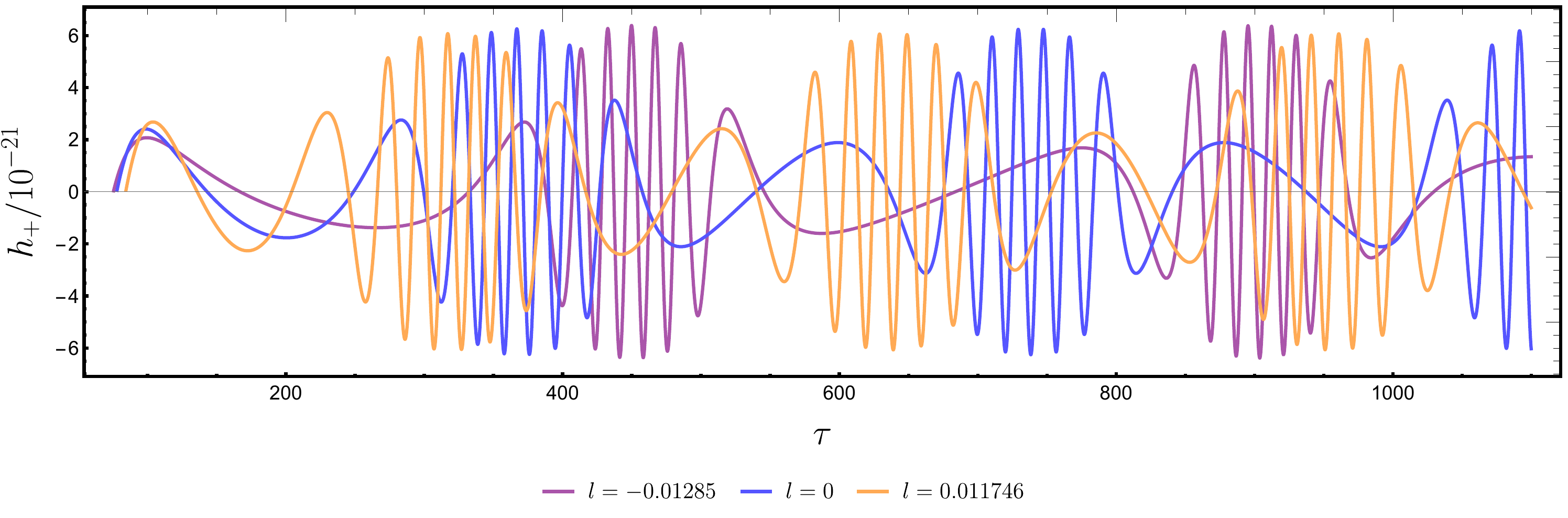} }
{\label{rEb}\includegraphics[width=15.75cm]{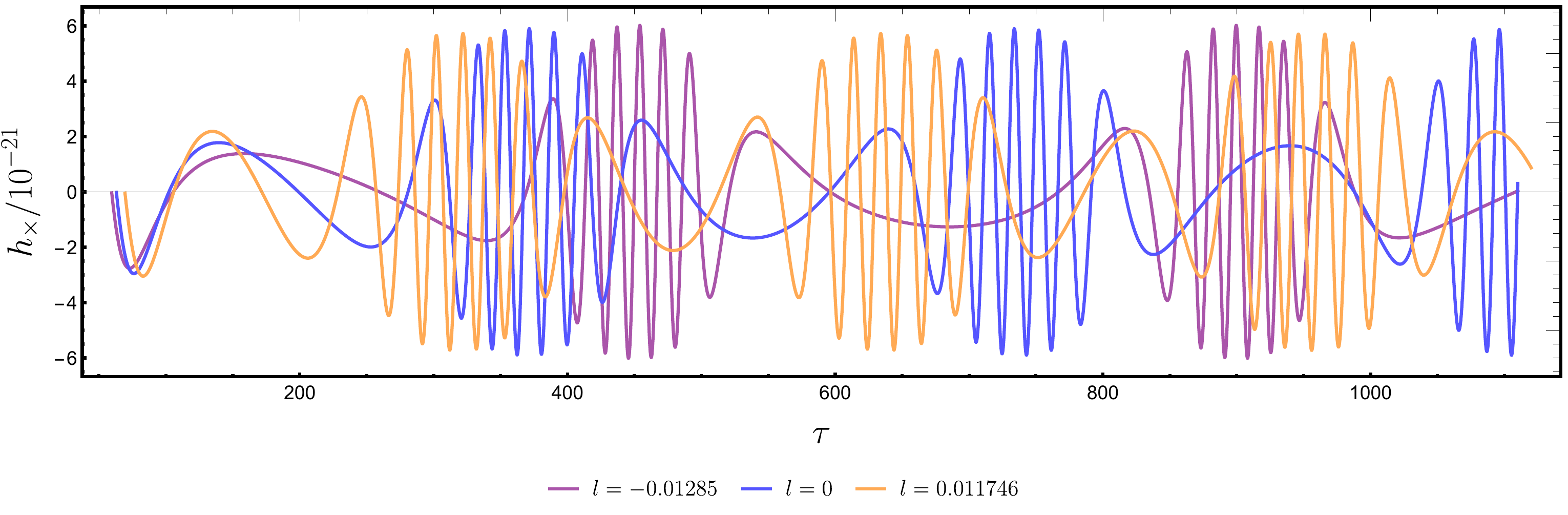} }
\caption{Graphical representation of the polarization modes $h_+$ and $h_\times$ emitted by the EMRI system from the periodic orbit $(3, 2, 2)$ with fixed $E=0.96$ and different values of $l$. Here, $h_+$ and $h_\times$ have units of  $\left[m^{-1}\right]$.}\label{fig11}
\end{figure*}

To analyze how the KR metric parameter $l$ can alter the gravitational waveform, let us consider a fictitious EMRI system composed of the Sgr A* supermassive black hole with mass $M=4\times 10^6 {\rm M}_{\odot}$ and a massive object orbiting it with mass $m=4 {\rm M}_{\odot}$ at a distance $D_L= 7.953 {\rm Kpc}$, inclination   $\iota=4/\pi$ and latitude $\zeta=4/\pi$.  In Fig.\,\ref{fig11} we plot the gravitational waveform from the numerical results of the polarization components $h_+$ and $h_\times$ as a function of the eigentime of the lowest-mass object for an orbit with signature $(3, 2, 2)$ in the parameters $(z,w,v)$ with fixed energy $E=0.96$, to see how the values of $l$ change the gravitational waveform. 
In this plot the blue waveform corresponds to the Schwarzschild space-time while the purple and orange are for the choices of the parameters $l=-0.01285$ and $l=0.011746$, respectively. Such modifications to the geometry slightly alter the amplitude of the wave and significantly its phase, both at the $h_+$ and $h_\times$ polarizations as compared to the $l=0$ case.  The amplitude of the wave is largest when the orbiting object is rotating in the periastron, and as it moves away towards the apoastron the amplitude of the GW decreases, this region representing the zoom number $z$. For $l<0$ the waveform with a smoother amplitude perfectly shows an orbit with a larger eccentricity, while for $l>0$ the waveform has a sharper amplitude and therefore a smaller eccentricity.   This change suggests the possibility of identifying the effects of spontaneous Lorentz symmetry-breaking on the KR metric through the GWs emitted by periodic orbits, since these waves can reveal the zoom and whirl numbers.  

It is important to note that the approach adopted here does not provide complete information on the GW, since we have excluded the multipole contribution beyond the quadratic order. However, the procedure followed so far is sufficient to record and distinguish the gravitational waveforms emitted by the periodic orbit of a massive object around the black hole described by the KR metric. In order to construct GWs adapted to other scenarios, we should extend our analysis to higher orders of multipoles. This is important for future detections of these waves, since EMRI systems are of great interest to the new generation of interferometers 
 such as LISA \cite{LISA}.

\section{Summary and  conclusion}\label{Sec:Conclusion}

In this work, we have analyzed a static, spherically symmetric solution expressed in the form of a Schwarzschild-like metric, which incorporates the effects of the Kalb-Ramond field. The inclusion of this field is implemented through a spontaneous Lorentz symmetry-breaking parameter, denoted by $l$ \cite{KR}. 
Our study primarily focused on examining the orbital dynamics of a particle moving along a time-like geodesic within this modified spacetime. As part of this investigation, we derived the effective potential characterizing the KR spacetime and explored the distinctions it exhibits in comparison to the standard Schwarzschild geometry. 
A notable result of this analysis is that, in the asymptotic limit where $r \to \infty$, the effective potential behaves as $V_{\rm eff}\rightarrow 1/(1-l)$. This asymptotic behavior implies that the maximum energy attainable by a particle in a bound orbit is constrained by the relation $E^2=1/(1-l)$. This value differs from the corresponding energy in the Schwarzschild geometry for any nonzero value of $l \neq 0$. 
In conducting this analysis, we adopted values of the Lorentz symmetry-breaking parameter $l$ constrained within the interval established in our previous work \cite{nosso}, as detailed in Eq.~(\ref{eq:bound}). This ensures consistency with prior findings and provides a well-defined framework for analyzing the impact of the Kalb-Ramond field on orbital behavior.

We computed the fundamental parameters associated with MBOs and the ISCOs using the effective potential. The results are illustrated in Fig.\,\ref{fig1}, which presents the quantities $E_{\rm ISCO}$, $L_{\rm ISCO}$, $L_{\rm MBO}$, $r_{\rm ISCO}$, and $r_{\rm MBO}$ in the context of the KR geometry and compares them to their counterparts in the Schwarzschild spacetime. Our findings reveal that as the Lorentz symmetry-breaking parameter $l$ increases within the interval $l_{\rm min} = -0.185022$ to $l_{\rm max} = 0.060938$, the angular momentum ($L_{\rm ISCO}$, $L_{\rm MBO}$) and the radial positions ($r_{\rm ISCO}$, $r_{\rm MBO}$) systematically decrease. Conversely, the energy at the ISCO ($E_{\rm ISCO}$) exhibits an increasing trend with $l$. 
Interestingly, similar to the Schwarzschild geometry, the KR geometry also satisfies the relation $r_{\rm MBO} = L_{\rm MBO}$. This behavior in the radial position and angular momentum is consistent with patterns identified in other modified spacetime scenarios, such as polymer black holes within the framework of Loop Quantum Gravity \cite{LQC}. 

Since bound orbits are constrained to lie between MBOs and ISCOs, we further analyzed their existence for different values of $l$, as depicted in Fig.\,\ref{Fig3}. Notably, at the extrema of $l$, there are no bound orbits for a specific chosen energy value, $E = 0.96$. This particular energy was selected as it allows the existence of bound orbits for both positive and negative values of $l$ for certain values of $\alpha$ in Eq.~\eqref{lcontrol}. This behavior is depicted in Fig.\,\ref{Fig4}, which displays the $E$-$L$ region where bound orbits are permitted. As $l$ approaches its extreme values, there is a discernible shift in the region of allowed bound orbits, effectively precluding their coexistence for fixed energy or angular momentum values. 
The existence of bound orbits is fundamentally tied to the presence of at least two distinct roots of the equation $\dot{r}^2 = 0$. To investigate this further, we numerically computed the roots of Eq.~\eqref{rpontoquad} for various values of $L$ and $E$, systematically exploring the dependence on $l$. Additionally, the condition $\dot{r}^2 > 0$ was utilized to examine the behavior of periodic orbits, offering insights into the impact of the symmetry-breaking parameter $l$ on the orbital dynamics.

Subsequently, we employed the method proposed in \citep{TAX} to characterize a periodic orbit using a triplet $(z, w, v)$, which defines a rational number $q$ encapsulating the unique properties of each orbit. Using the information derived from the region between MBOs and ISCOs, we numerically solved Eq. \eqref{curvaq2} in Sec.\,\ref{PO}. This was achieved by adopting parameterized coordinate systems $(x, y) = (r\cos\phi, r\sin\phi)$, which allowed for the precise computation of periodic orbits of a massive particle in the vicinity of a Kalb-Ramond (KR) black hole. 
The results of this analysis, illustrated in Figs.\,\ref{Fig7} and \ref{Fig8}, reveal that for a fixed energy value $E = 0.96$, periodic orbits of the same taxonomy exhibit a higher angular momentum when $l < 0$ compared to $l > 0$. This behavior corresponds to orbits with greater eccentricity in the $l < 0$ case. Similarly, when fixing the angular momentum at $L = 3.7$ and varying $l$, we observed that for $l < 0$, the energy of the orbit is lower, which in turn implies a lower eccentricity compared to orbits of the same taxonomy for $l > 0$. 
These results highlight the influence of the parameter $l$ in altering the spacetime structure near the black hole. This modification is significant enough to distinguish the dynamics around KR black holes from those around Schwarzschild black holes, as evidenced by the motion of a massive particle in their respective gravitational fields. Our results are similar to those obtained for the Schwarzschild solution in \cite{TAX} and \cite{TAXKeer2, TAXKeer3, TAXKeer4, TAXKeer5, TAXRN, TAXQC, SCNovo, P1, P3} when we take $l=0$, and therefore confirm the consistency with GR.
 
In order to explore the gravitational waveforms emitted by a KR black hole, in Sec.\,\ref{secGW} we analyzed an EMRI system. Specifically, we considered a black hole with a mass equivalent to that of Sgr A*, orbited by a secondary object with a mass on the order of S-stars \cite{EHT1,EHT2,EHT3,EHT4,EHT5,EHT6}. Due to the slow variation in the energy and angular momentum of the system over time, we employed the adiabatic approximation to simplify the analysis. Using the Kludge method \cite{Kludge}, we computed the gravitational waveforms, denoted as $h_+$ and $h_\times$, which are presented in Fig.\,\ref{fig11}. Our study focused on gravitational wave emission from an orbit with fixed energy $E = 0.96$ and taxonomy $(3, 2, 2)$, comparing the cases of $l < 0$ and $l > 0$ with the Schwarzschild geometry as a reference. The results reveal distinct characteristics in the waveform. During the ``zooming'' phase, which corresponds to regions of the orbit with higher eccentricity, the amplitude of the emitted waves is significantly lower. In contrast, during the "whirling" phase, where the orbit involves multiple close turns near the black hole, the wave amplitude is considerably higher. These phases align with the number of turns in the orbit.
The Lorentz symmetry-breaking parameter $l$ plays a crucial role in altering the phase of the emitted waves and has a slight effect on their amplitude. For $l < 0$, we observe an increase in the period of periodic orbits, consistent with the increase in angular momentum compared to the $l > 0$ case, as depicted in Figs.\,\ref{Fig5} and \ref{Fig6}. Notably, this behavior persists across all orbits, implying that the effect of $l$ is uniform for any triplet $(z, w, v)$. These findings emphasize the significant influence of the parameter $l$ on the gravitational wave emission from KR black holes, offering a potential observational signature to distinguish them from their Schwarzschild counterparts.

The results of this analysis demonstrate that the gravitational waveforms produced by KR black holes, within the framework of our approach, effectively encapsulate the characteristic features of periodic orbits. This highlights their utility in exploring the effects of spontaneous Lorentz symmetry-breaking on the dynamics of stellar-mass objects orbiting this class of black holes. Specifically, the modifications induced in the gravitational wave polarizations $h_+$ and $h_\times$ by the parameter $l$ are substantial, offering a promising avenue for distinguishing KR black holes from their Schwarzschild counterparts.
Thus, the parameter $l$, which embodies the symmetry-breaking mechanism, imparts observable deviations in the waveform structure, particularly in the phase and, to a lesser extent, the amplitude of the emitted gravitational waves. These deviations, tied to the underlying spacetime geometry, provide a distinctive signature of KR black holes that could be exploited in astrophysical observations. Consequently, the study of such waveforms offers a pathway to probing the fundamental effects of Lorentz symmetry-breaking in a strong-gravity regime, with potential implications for testing modifications to general relativity.

However, to take full advantage the potential of this approach, it is essential to extend the analysis to encompass more general scenarios. One key direction for future research is the generalization of this framework to rotating black holes within KR gravity. Such an extension would enable the investigation of gravitational waves emitted during the inspiral and merger of binary KR black holes, a process expected to yield a wealth of information about the influence of the parameter $l$ in dynamical and highly non-linear gravitational regimes. 
By addressing these challenges, future studies could provide a comprehensive understanding of the astrophysical and theoretical implications of KR gravity, further establishing its observational differences from Schwarzschild and Kerr geometries. This broader perspective would also contribute to advancing our ability to detect and interpret gravitational wave signals as unique probes of modified theories of gravity.

\section*{Acknowledgements}

MER thanks Conselho Nacional de Desenvolvimento Cient\'ifico e Tecnol\'ogico - CNPq, Brazil, for partial financial support. DRG is supported by the Spanish Agencia Estatal de Investigación Grant No. PID2022-138607NB-I00, funded by MCIN/AEI/10.13039/501100011033, FEDER, UE, and ERDF A way of making Europe. This study was financed in part by the Coordena\c{c}\~{a}o de Aperfei\c{c}oamento de Pessoal de N\'{i}vel Superior - Brasil (CAPES) - Finance Code 001.
FSNL acknowledges support from the Funda\c{c}\~{a}o para a Ci\^{e}ncia e a Tecnologia (FCT) Scientific Employment Stimulus contract with reference CEECINST/00032/2018, and funding through the research grants UIDB/04434/2020, UIDP/04434/2020 and PTDC/FIS-AST/0054/2021.



\end{document}